\documentclass[aps,prd,preprint,eqsecnum,nofootinbib,showpacs,floatfix]{revtex4}

\usepackage[dvips]{graphicx}
\usepackage{multirow}
\usepackage{amsmath}
\usepackage{placeins}
\usepackage{mathtools}
\newcommand{\BlackHat}{{\sc BlackHat}}

\newcommand{\SHERPA}{{\sc SHERPA}}
\newcommand{\MCFM}{{\sc MCFM}}

\newcommand{\COMIX}{{\sc COMIX}}
\newcommand{\SISCone}{{\sc SISCone}}
\newcommand{\ntuple}{{$n$-tuple}}
\newcommand{\ntuples}{{$n$-tuples}}

\newif\ifdraft
\draftfalse
\newif\ifpreprint
\preprinttrue

\def\fig#1{fig.~{\ref{#1}}}

\def\sect#1{section~{\ref{#1}}}
\def\eqn#1{eq.~(\ref{#1})}

\def\eqns#1#2{eqs.~(\ref{#1}) and~(\ref{#2})}

\def\tab#1{table~{\ref{#1}}}
\def\Tab#1{Table~{\ref{#1}}}

\def\nn{\nonumber}
\def\y{\gamma}
\def\qb{\bar q}

\def\e{\epsilon}
\def\eps{\epsilon}

\def\Wjj{$W\,\!+\,2$}
\def\Wjjj{$W\,\!+\,3$}

\def\Wjjjjj{$W\,\!+\,5$}

\def\Wjnp1{$W\,\!+\,(n\!+\!1)$}

\def\Zjn{$Z\,\!+\,n$}

\def\YYjj{$\gamma\gamma\,\!+\,2$}
\def\YYj{$\gamma\gamma\,\!+\,1$}
\def\YYz{$\gamma\gamma\,\!+\,0$}

\def\gjn{$\gamma\,\!+\,n$}
\def\jet{{\rm jet}}
\def\pT{p_{\rm T}}
\def\kT{k_{\rm T}}

\def\root{{\sc root}}

\def\ET{E_{\rm T}}
\def\HTpartonic{{\hat H}_{\rm T}}
\def\HTpartonicp{{\hat H}_{\rm T}'}

\def\MLO{d\sigma^{(0)}}
\def\MNLO{d\sigma_V^{(1)}}
\def\MOLS{d\sigma_{\rm OLS}}
\def\hatMNLO{ {\widehat{d\sigma}_V}^{(1)}}

\newbox\charbox
\newbox\slabox
\def\s#1{{      
        \setbox\charbox=\hbox{$#1$}
        \setbox\slabox=\hbox{$/$}
        \dimen\charbox=\ht\slabox
        \advance\dimen\charbox by -\dp\slabox
        \advance\dimen\charbox by -\ht\charbox
        \advance\dimen\charbox by \dp\charbox
        \divide\dimen\charbox by 2
        \raise-\dimen\charbox\hbox to \wd\charbox{\hss/\hss}
        \llap{$#1$}
}}

\begin{document}

\def\pspacer{\hfil\rule{1cm}{0cm}\hfil}
\title{%
\ifpreprint
\vskip -1 cm 
\hbox to \textwidth{
\small\rm UCLA/13/TEP/111\pspacer
SLAC--PUB--15870\pspacer
SB/F/434--13 \pspacer
IPhT--T13/270}\vspace{-4mm}
\hbox to \textwidth{
\small\rm IPPP/13/1\pspacer
FR-PHENO-2014-001
}\fi

Next-to-Leading Order \YYjj-Jet Production at the LHC}

\author{Z.~Bern${}^a$, L.~J.~Dixon${}^b$, F.~Febres Cordero${}^c$, S.~H{\" o}che${}^b$, H.~Ita${}^{d}$, D.~A.~Kosower${}^{e}$, N.~A.~Lo Presti${}^{e}$ and
D.~Ma\^{\i}tre${}^{f}$
\\
$\null$
\\
${}^a$Department of Physics and Astronomy, UCLA, Los Angeles, CA
90095-1547, USA \\
${}^b$SLAC National Accelerator Laboratory, Stanford University,
             Stanford, CA 94309, USA \\
${}^c$Departamento de F\'{\i}sica, Universidad Sim\'on Bol\'{\i}var, 
 Caracas 1080A, Venezuela\\
${}^d${Physikalisches Institut, Albert-Ludwigs-Universit\"at Freiburg,
       D--79104 Freiburg, Germany}\\
${}^e$Institut de Physique Th\'eorique, CEA--Saclay,
          F--91191 Gif-sur-Yvette cedex, France\\
${}^f$Department of Physics, University of Durham, Durham DH1 3LE, UK\\
}

\begin{abstract}
We present next-to-leading order QCD predictions for cross sections
and for a comprehensive set of distributions in \YYjj-jet
production at the Large Hadron Collider.  We consider the contributions
from loop amplitudes for two photons and four gluons, but we neglect top quarks.  
We use \BlackHat\ together with \SHERPA\ to carry out
the computation.   We use a Frixione cone isolation for the photons.
We study standard sets of cuts on the jets and the photons,
and also sets of cuts appropriate for studying backgrounds
to Higgs-boson production via vector-boson fusion.
\end{abstract}

\pacs{12.38.-t, 12.38.Bx, 13.87.-a, 14.70.Bh \hspace{1cm}}

\maketitle

\section{Introduction}

Reliable theoretical predictions for Standard-Model processes at the
Large Hadron Collider (LHC) are
important to ongoing searches for new physics.  They are also
important to the increasingly precise studies of the recently discovered
Higgs-like boson~\cite{HiggsBosonExpt}, of the top quark,
and of vector boson self-interactions.  Uncovering hints of new physics
beyond the Standard Model requires 
a good quantitative understanding of the Standard-Model backgrounds and their
uncertainties.  

Predictions for background rates at the LHC
rely on perturbative QCD, which enters all aspects of
short-distance collisions at a hadron collider.  Leading-order (LO)
predictions in QCD suffer from a strong dependence on the unphysical
renormalization and factorization scales.  This dependence gets
stronger with increasing jet multiplicity.  Next-to-leading order (NLO)
results generally reduce this dependence dramatically, typically to a
10--15\% residual sensitivity.  Thus they offer the first quantitatively
reliable order in perturbation theory.

Photon pairs are a key decay channel for detecting and
measuring the Higgs-like boson.  A good
understanding of prompt photon-pair background is important for
precision measurements of its properties and for uncovering
deviations from Standard Model expectations. In particular, when
the photon pair is produced in association with two hadronic jets, the
process is an important background to Higgs-like boson production via
vector-boson fusion (VBF).  We study this background in the present
paper, both for standard cuts on the jets and the photons, as
well as for other sets of cuts designed to isolate the VBF region of phase space.

Inclusive photon-pair production was studied at NLO by a number of
groups~\cite{InclusiveDiphoton,MCFMPhoton}.  Gluon-initiated
subprocesses, which arise only at one loop, account for an important
fraction of the cross section.  Studying these subprocesses to their
NLO requires two-loop amplitudes~\cite{TwoLoopPhoton}, which have been 
applied to photon-pair production~\cite{TwogammaMC}.  More
recently, NNLO results for inclusive di-photon production
have been presented by Catani, Cieri, de Florian,
Ferrera and Grazzini~\cite{RecentNNLODiPhoton}.
NLO predictions for the production of a photon pair in association with a
single jet were given some time ago~\cite{PhotonOneJet,GGH1}.  Here we
present predictions for inclusive photon-pair production in
association with two jets at NLO.  This process has also been studied recently
by Gehrmann, Greiner, and Heinrich (GGH)~\cite{GGH2}, and by
Badger, Guffanti and Yundin~\cite{BGY}.  (The latter paper also provides
NLO results for photon-pair production in association with three jets.)
We study three pairs of cuts.  Each pair consists of a standard jet cut,
and a cut appropriate for isolating Higgs bosons formed from vector boson
fusion.  The second and third pairs of cuts are oriented toward
specific experimental analyses by the ATLAS and CMS collaborations.

In the present paper we use on-shell methods as implemented in
numerical form in the \BlackHat{} software library~\cite{BlackHatI}.
This library, together with the \SHERPA{} package~\cite{Sherpa}, has
previously been used to make NLO predictions for a variety of vector
boson plus multi-jet production processes~\cite{W3jDistributions,
  BlackHatZ3jet, Wpolarization, W4j,Z4j}, most recently for
\Wjjjjj-jets~\cite{W5j}, and for four-jet production~\cite{FourJets}.
It has also been used to compute \gjn-jet to \Zjn-jet ratios for
assessing theoretical uncertainties~\cite{PhotonZ,PhotonZ3} in the CMS
searches~\cite{PhotonZexp} for supersymmetric particles.  The ATLAS
collaboration has also used results from \BlackHat{} computations with
\SHERPA{} for Standard-Model studies of electroweak vector-boson
production in association with three or more jets~\cite{NTupleUse}.
Other programs that use on-shell methods are described in
refs.~\cite{OtherOnShellPrograms,GoSam}.

\SHERPA{} is used to manage the numerous partonic subprocesses
entering the calculation, to integrate over phase space, to construct
physical distributions, and to output \root{}~\cite{ROOT} $n$-tuples.
We use the \COMIX{} package~\cite{Comix} to compute Born and
real-emission matrix elements, along with the corresponding
Catani--Seymour~\cite{CS} dipole subtraction terms.  Rather than
repeating the entire computation for each scale and for each parton
distribution function (PDF) set, we store intermediate results in
$n$-tuple format, recording momenta for all partons in an event, along
with the coefficients of various scale- or PDF-dependent functions in
the event weight~\cite{NTuples}.
The \ntuple{} storage makes it possible to evaluate
cross sections and distributions for different scales and PDF error
sets.  We have generated two sets of \ntuple{}s, one corresponding to
loose standard jet cuts, and another adding VBF cuts.  We are then
able to study modifications of each of these cuts without the
time-consuming recomputation of matrix elements.
The \ntuple{}s generated for the present study are available 
in the format of ref.~\cite{NTuples} with process directories
{\tt YY2j\/} and {\tt YY2j\_VBF\/}.

This paper is organized as follows.  In \sect{BasicSetUpSection} we 
summarize the basic setup of the computation.  In \sect{ResultsSection}
we present our results for cross sections, ratios and distributions. 
We summarize and give our conclusions in \sect{ConclusionSection}.
Tables for distributions are in three appendices.  A fourth appendix contains
matrix elements at a point in phase space.

\section{Basic Setup}
\label{BasicSetUpSection}

\begin{figure}[t]
\includegraphics[clip,scale=0.65]{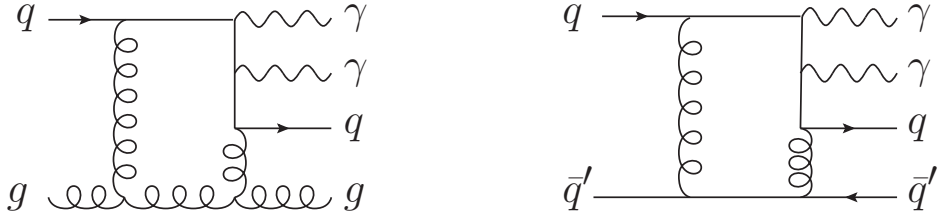}
\caption{Examples of six-point loop diagrams for the processes
$q g \rightarrow \gamma\gamma q g $ and
$q \qb' \rightarrow \gamma\gamma  q\qb'$.}
\label{lcdiagramsFigure}
\end{figure}

\begin{figure}[t]
\includegraphics[clip,scale=0.65]{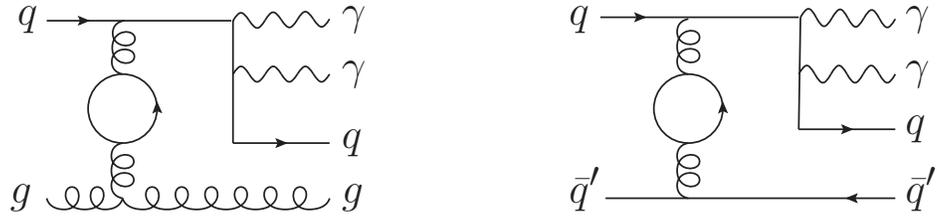}
\caption{Examples of six-point fermion-loop diagrams for the processes
$q g \rightarrow \gamma\gamma q g $ and
$q \qb' \rightarrow \gamma\gamma  q\qb'$. These diagrams have a closed
quark loop, but the photons do not couple directly to it.}
\label{nfdiagramsFigure} \end{figure}

\begin{figure}[t]
\null\hskip -3mm
\includegraphics[clip,scale=0.65]{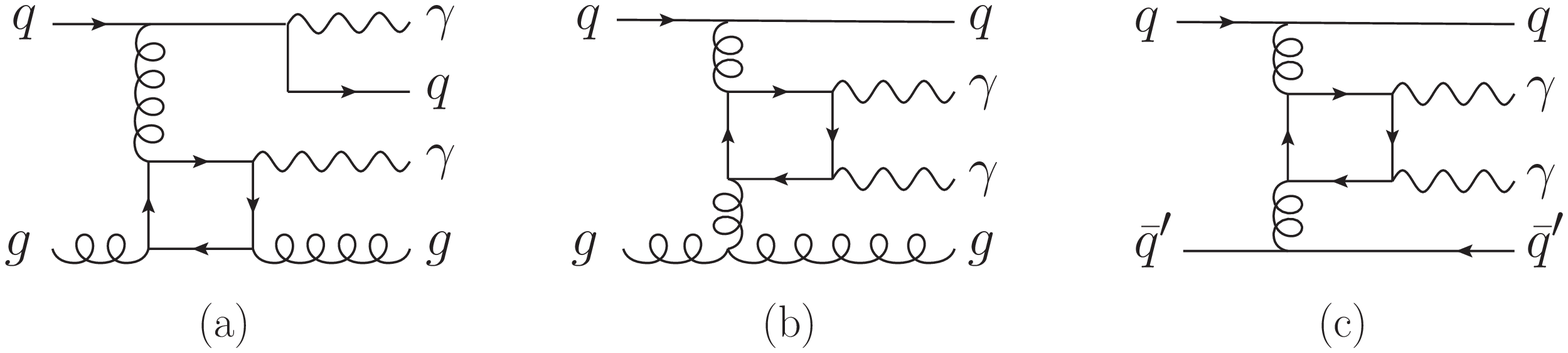}
\vskip -.4 cm 
\caption{Examples of six-point fermion-loop diagrams for the processes
$q g \rightarrow \gamma\gamma q g $ and
$q \qb' \rightarrow \gamma\gamma  q\qb'$. These diagrams have a closed
quark loop. In (a), one
photon couples directly to the quark loop, whereas in (b) and~(c), both photons couple to the
quark loop.}
\label{vectdiagramsFigure} \end{figure}

\begin{figure}[t]
\includegraphics[clip,scale=0.65]{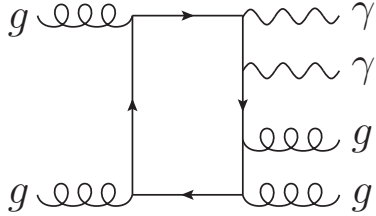}
\vskip -.2 cm 
\caption{Example of a six-point one-loop diagram for the process
$g g \rightarrow \gamma\gamma g g $.  This one-loop amplitude is finite
because the corresponding tree-level amplitude vanishes.}
\label{gluediagramsFigure}
\end{figure}

\begin{figure}[t]
\includegraphics[clip,scale=0.65]{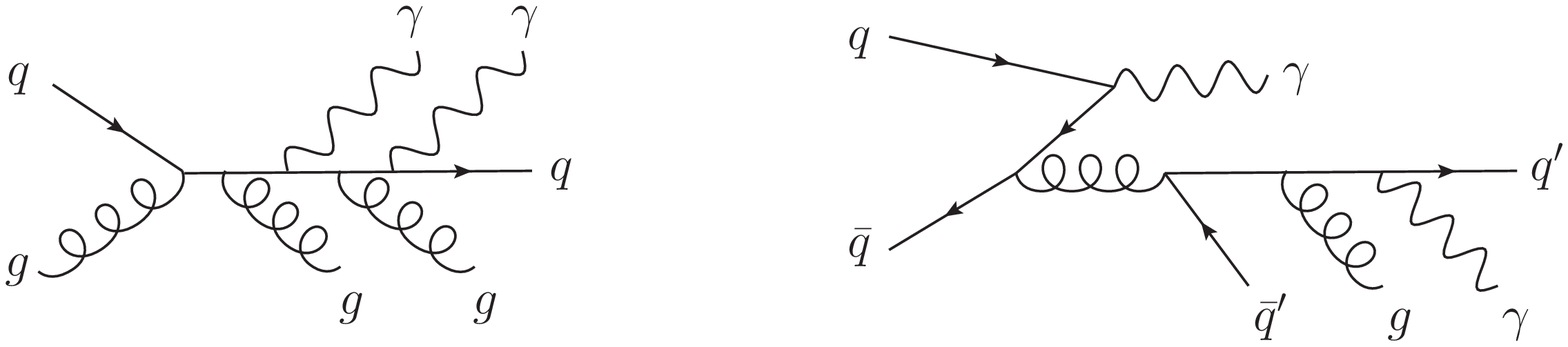}
\vskip -.4 cm 
\caption{Examples of seven-point real-emission diagrams for the
processes $q g \rightarrow \gamma\gamma  q ggg$ and 
$q \qb \rightarrow \gamma\gamma  q' \qb' g$.
}
\label{treediagramsFigure}
\end{figure}

In this paper we compute the \YYjj-jet processes at NLO in QCD,
\begin{eqnarray}
&& pp \, \longrightarrow \gamma\gamma+2\, \hbox{ jets} \,.
\end{eqnarray}
These processes receive contributions from several partonic subprocesses.
At leading order, and in the virtual NLO contributions,
the subprocesses are all obtained from
\begin{eqnarray}
 && q \qb gg\rightarrow \gamma\gamma\,,\nn \\
 && q \qb q' \qb' \rightarrow \gamma\gamma\,,
\label{subprocesses}
\end{eqnarray}
by crossing two of the initial-state partons into the final state. 
 We illustrate the virtual contributions
with one or two external quark pairs in
figs.~\ref{lcdiagramsFigure}--\ref{vectdiagramsFigure}, although we do
not need any of the diagrams explicitly, as our calculation uses
on-shell methods rather than Feynman diagrams. There
are additional `pure-gluon' scattering processes that we may consider,
\begin{eqnarray}
 && g g \rightarrow \gamma\gamma g g\,,
\label{puregluon-subprocess}
\end{eqnarray}
which have no external quark legs.  This process,
illustrated in \fig{gluediagramsFigure}, vanishes
at tree level and appears only at one loop.  Accordingly, the amplitude is
finite at one loop, and it appears in the squared matrix element only at
relative order $\alpha_s^2$, as a one-loop squared contribution.  In
\YYz-jet or \YYj-jet production, these processes contribute at a
significant or noticeable level (respectively), because there is no
tree-level process with a $gg$ initial state.  The large value of the
gluon distribution can compensate for the additional two powers of
$\alpha_s$, so these subprocesses must be taken into account.  In
\YYjj-jet production, in contrast, one crossing of the first
subprocess in \eqn{subprocesses}, in which the quark pair is moved
to the final state, does give a tree-level contribution with a
$gg$ initial state.  We might then expect the contribution of
the pure-gluon subprocess to be genuinely suppressed by two powers
of $\alpha_s$, relative to this other $gg$ initial-state contribution
(although it does have a different dependence on the quark electric charges).
We shall test this expectation by including the matrix
element for the $gg\rightarrow\gamma\gamma gg$ subprocess explicitly
in the NLO calculation.  While we will find that its contribution is small
in the total, it is not as small compared to the tree-level $gg$ initial-state
contribution as this argument would suggest.  (We do not include similar
contributions from the squaring of finite one-loop helicity amplitudes
in the $g g\rightarrow \gamma\gamma q\qb$ subprocess, which are
expected to give smaller contributions.)

In our computation, we obtain amplitudes with multiple identical quark
flavors by appropriate antisymmetrization of amplitudes for distinct
flavors.  The virtual contributions to any given subprocess can be divided
into gauge-invariant subparts.  For example, the contributions to
amplitudes with a closed quark loop form a gauge-invariant subset.
The quark-loop contributions can be split up further, depending on
the lines to which the external photons couple. 
Terms in which neither photon couples to the closed quark loop, but only to
the open quark lines (lines that connect to external states), 
as shown in \fig{nfdiagramsFigure}, give a contribution
proportional to $n_{\!f}$, the number of quark flavors.  Terms with one photon
coupling to the closed quark loop and one to an open quark line, as shown
in \fig{vectdiagramsFigure}(a), give a contribution
proportional to the flavor sum of quark charges, $\sum_f Q_f$.
Finally, terms in which both photons couple
directly to the closed quark loop, as shown in \fig{vectdiagramsFigure}(b),
give a contribution proportional to the flavor sum of squared quark charges,
$\sum_f Q_f^2$.  The pure-gluon subprocesses~(\ref{puregluon-subprocess}),
shown in \fig{gluediagramsFigure}, are likewise proportional to this latter
flavor sum.

Calculations to NLO in QCD also require real-emission matrix elements,
corresponding to contributions with an additional parton in the final state.
We obtain the required subprocesses by crossing three initial-state partons
into the final state in one of the two basic processes,
\begin{eqnarray}
 && q \qb ggg\rightarrow \gamma\gamma\,,\nn \\
 && q \qb q' \qb' g\rightarrow \gamma\gamma\,.
\label{subprocesses-RE}
\end{eqnarray}
We illustrate these processes in \fig{treediagramsFigure}.

In our calculation, the five lightest quarks, $u,d,c,s,b$, are all
treated as massless.  We do not include contributions to the
amplitudes from real or virtual top quarks; we expect this omission
to affect our results only at the percent level.

\subsection{Photon Isolation}
\label{PhotonIsolationSection}
\def\egamma{\epsilon_\gamma}

Photon measurements make use of an isolation criterion in order to
suppress backgrounds from photons arising from hadrons.  
From an
experimental point of view, the isolation requirement is necessary to reduce
an important background, consisting of jets with a $\pi^0$ or
$\eta$ meson carrying most of the jet's energy,
which is then misidentified as an isolated photon because it decays 
to a nearly-collinear photon pair.
Experimental collaborations typically use an isolation
criterion (see {\it e.g.\/} refs.~\cite{ATLASPhoton,CMSPhotonNote}), imposing a
limit on the hadronic energy in a cone around the photon.  This limit may be
applied after subtractions to account for detector noise, the effects of the
underlying event or of pile-up of other $pp$ collisions, and possible
adjustments for photon energy not captured within the cone.  As a result, the
hadronic energy within the cone may even be negative; along with the
accounting for underlying event or pile-up activity, this weakens the 
link with a purely perturbative implementation of a fixed-cone isolation
criterion.

A fixed-cone isolation criterion requires the use of nonperturbative photon
fragmentation functions in order to obtain theoretical predictions.  The use of
fragmentation functions requires additional work, and in any case it
would limit the precision attainable, because the fragmentation
functions are not that well constrained by experimental data.
Furthermore, unlike the case of the parton distribution functions, no
error sets are available that would allow us to estimate the
uncertainties due to the fragmentation functions.  
These issues weaken the motivation for using a fixed-cone
isolation in a theoretical calculation, compared to possible alternative
isolation procedures.

Frixione proposed such
an alternative photon isolation procedure, which avoids the need for
fragmentation-function contributions~\cite{Frixione} by suppressing the
region of phase space where photons are collinear with jets.  It still
allows soft radiation arbitrarily close to the photon, ensuring that
it is infrared safe. We use this procedure, requiring that the
partons obey,
\begin{equation}
\label{iso}
\sum\limits_i E_{{\rm T}i}\, \Theta \left(\delta - R_{i\gamma}\right)
   \leq E (\delta) \,,
\end{equation}
for all $\delta=R(\phi,\eta;\phi_\gamma,\eta_\gamma)
\leq
\delta_0$, where 
\begin{equation}
    R(\phi_1,\eta_1; \phi_2,\eta_2)=[(\phi_1-\phi_2)^2+(\eta_1-\eta_2)^2]^{1/2}\,,
\end{equation}
is the usual longitudinally boost-invariant angular distance measure.  
In the inequality (\ref{iso}), $R_{i\gamma}=
R(\phi_i,\eta_i; \phi_\gamma,\eta_\gamma)$ is the distance of parton $i$ from
the photon, $E_{{\rm T}i}$ is the transverse energy of the $i$th parton and the
restricting function $E(\delta)$ is given by 
\begin{equation} 
    E (\delta) = \ET^\gamma \, \egamma \left( \frac{ 1 - \cos \delta }{1 - \cos
    \delta_0 } \right)^n \, .  \label{photoniso}
\end{equation} 
The restriction is scaled by the parameter $\egamma$ to the photon transverse
energy $\ET^\gamma$.  
The inequality (\ref{iso}) 
constrains the hadronic energy in a cone of fixed half-angle $\delta_0$
around the photon axis.  
The restricting function has the property that it
vanishes as $\delta \rightarrow 0$ and thus suppresses collinear
configurations, but allows soft radiation arbitrarily close to the photon.  We
will use the Frixione cone, with, 
\begin{equation} 
    \egamma = 0.5\,,\quad  \delta_0 = 0.4\,, \quad \mbox{ and}\quad n = 1\,.
\end{equation} 
An earlier study~\cite{PhotonZ} of inclusive photon production found that the
difference in cross sections between Frixione cone and fixed-cone isolation,
with similar parameters to the present study, was less than 1\% at large photon
transverse momenta.   Although the Frixione isolation has not been applied
directly in experiments, here at least, we do not expect the discrepancy
to be large.

Our implementation follows the standard \SHERPA{} one; the photon
isolation and jet algorithm using the FastJet~\cite{FastJet} library
are applied independently, with no special treatment for partons
inside the photon cone.  After isolation and jet-finding, we apply an
additional angular separation criterion to photon--jet pairs and to
the pair of photons.

\subsection{Formalism and Software}
\label{FormalismSubsection}

Several ingredients enter into an NLO calculation:
the Born cross section, the virtual  (one-loop)
corrections, and the radiative (real-emission) corrections.
The computation of the latter requires tree-level matrix elements with an
additional parton in the final state compared to the Born process.
The virtual corrections have explicit divergences in the dimensional
regulator $\epsilon = (4-D)/2$, whereas the canceling divergences in
the real-emission contributions arise only after integration over
$D$-dimensional phase space.  
We use the Catani--Seymour dipole subtraction scheme~\cite{CS}
in order to implement these
cancellations in a numerical calculation.  This scheme
adds and subtracts contributions to the evaluation of the NLO cross
section; schematically, we decompose it as,
\begin{equation}
 \sigma_n^{{\rm NLO}}\,\,=\,\,
   \int_n \sigma_n^{{\rm born}}
   \,+\,
 \int_n \sigma_n^{{\rm virt}}
    \,+\, \int_n\Sigma_n^{{\rm subtr}}
   \,+\,
 \int_{n+1} \left(\sigma_{n+1}^{{\rm real}}
    \,-\, \sigma_{n+1}^{{\rm subtr}}\right)\,.
\end{equation}
Here the subscripts on the integrals denote the number of final-state
partons, and $\Sigma_n^{{\rm subtr}}$ is the result of integrating
$\sigma_{n+1}^{{\rm subtr}}$ analytically over a one-particle unresolved
phase space.  Other subtraction methods in current use include the
FKS approach~\cite{FKS} and antenna subtraction~\cite{TwoLoopAntenna};
the former has been automated~\cite{MadFKS}.  We use the SHERPA
package~\cite{Sherpa} to manage the partonic subprocesses, to
integrate over phase space, and to output \root{}~\cite{ROOT}
\ntuples. 

The techniques we use for computing virtual contributions are
collectively known as on-shell methods, and are reviewed in
refs.~\cite{OnShellReviews}.  These methods rely on underlying
properties of amplitudes --- factorization and unitarity --- in order
to express them in terms of simpler, on-shell amplitudes of lower
multiplicity, reducing the swell of terms.  Early application of the
unitarity method~\cite{UnitarityMethod} to collider physics was to
computing analytically the one-loop matrix elements for $q\bar{q}gg
g$, $q\bar{q}gg V$ and $q \qb q'\qb' V$ ($V = W$ or $Z$)
processes~\cite{Zqqgg}.  The latter matrix elements are used, for
example, in the NLO program MCFM~\cite{MCFM} as well as in studies at
$e^+ e^-$ colliders.  In recent years, on-shell methods have been
implemented in a more flexible numerical form.  These methods scale
well as the number of external legs
increases~\cite{OtherOnShellPrograms,GoSam,GenHel,GZ,EMZW3j,W3jPRL,%
  W3jDistributions,NLOttjj,NLOVVjj,W4j,Z4j}.  There have also been
important advances in computing virtual corrections with more
traditional methods~\cite{HeavyQuarkFeynman,BDDP}.

One-loop amplitudes in QCD with massless quarks may be expressed as a sum
over three different types of Feynman integrals (boxes, triangles, and
bubbles) with additional `rational' terms.  These latter terms
are rational functions of spinor variables associated to the external momenta.
The integrals' coefficients are also rational functions of these variables.
The integrals are
universal and well-tabulated; the aim of the calculation is to compute their
coefficients as well as the rational terms.  In an on-shell approach,
the integral coefficients may be computed using four-dimensional
generalized unitarity~\cite{UnitarityMethod, Zqqgg, NewUnitarity},
while the rational terms may be computed either by a loop-level
version~\cite{GenHel} of on-shell recursion~\cite{BCFW} or using
$D$-dimensional unitarity~\cite{DDimUnitarity}.  We use a numerical
version~\cite{BlackHatI} of Forde's method~\cite{Forde} for the
integral coefficients, and subtract box and triangle
integrands along the lines of the Ossola--Papadopoulos--Pittau
procedure~\cite{OPP}, improving the numerical
stability.  To compute the rational terms, we use a numerical
implementation of Badger's massive continuation method~\cite{Badger},
which is related to $D$-dimensional unitarity.

These algorithms are implemented in the \BlackHat{} software
library~\cite{BlackHatI,W3jPRL}.  \BlackHat{} organizes the
computation of the amplitudes in terms of elementary gauge-invariant
``primitive amplitude'' building blocks~\cite{TwoQuarkThreeGluon,Zqqgg}.  The
primitive amplitudes are then assembled into partial amplitudes, which
are the kinematic coefficients of the different color tensors that can
appear in the amplitude.  The complete virtual cross section is
obtained by interfering the one-loop partial amplitudes with the
tree-level amplitude and summing over spins and color indices.  A
given primitive amplitude can appear in multiple partial amplitudes
and does not have to be recomputed for each one.

This approach also allows for a straightforward separation of leading-
and subleading-color contributions.  The subleading-color
contributions are much smaller, yet more computationally costly (10 times
slower per phase-space point for \YYjj-jet production), but
using the separation we can evaluate them at far fewer
phase-space points than the leading-color contributions, while
obtaining comparable absolute statistical uncertainties.
Similarly to the production of a $W$ boson in association with 
three~\cite{W3jDistributions} or four jets~\cite{ItaOzeren}, the
subleading-color terms in the virtual contributions are small.  The magnitude
of these subleading-color contributions depends strongly on the cuts.
With standard cuts, they are typically 2\% of the leading-color virtual terms,
and about 0.2\% of the cross section.  With VBF cuts applied in addition,
these percentages increase to 5\% of the virtual, and 2\% of the cross
section. Our results are based on event
samples of $7\cdot 10^6$ leading-color virtual events, 
and $6\cdot 10^5$ subleading color ones.

As explained earlier in this section, there are four distinct types of
contributions to an NLO calculation: Born, virtual,
integrated-subtraction, and subtracted real-emission.  We perform the
phase-space integration of each type independently, using adaptive
Monte-Carlo integration~\cite{Vegas}.  We use an efficient
hierarchical phase-space generator based on QCD antenna
structures~\cite{AntennaIntegrator}, as incorporated into
\SHERPA{}~\cite{Comix}.
For each integration, the code adapts a grid during an initial phase;
the grid is then frozen, and used in the next, high-statistics phase,
which provides an estimate of the integration result
and associated statistical uncertainties.  \SHERPA's integrator
adjusts the relative number of evaluations between different
subprocesses during grid generation, in order to optimize the
statistical uncertainties of the computed cross section with a fixed
number of matrix-element evaluations.

For the virtual contributions, we use the associated Born matrix elements
to adapt and refine the integration grid.  For the pure-gluon terms,
we cannot do this, because the corresponding tree-level amplitudes vanish
identically; instead, we use antenna functions for this purpose.
The choice of the antenna functions is somewhat arbitrary, but the choice
will affect only how quickly the final phase-space integration converges,
and not the result itself.
We choose the antenna functions to incorporate most of the singularities
present in the one-loop amplitudes squared.
As an example, consider the integration of the squared matrix element
$M\big(g_1,g_2,g_3,\gamma_4,\gamma_5\big)$ in more detail.
We compute an antenna function using a combination of
color-ordered gluon tree amplitudes symmetrized over $g_4$ and $g_5$: 
$A^{\rm sym} \equiv (A^{\rm tree}\big(g_1,g_2,g_3,g_4,g_5\big)
+A^{\rm tree}\big(g_1,g_2,g_3,g_5, g_4\big))$.
The antenna function is then the squared matrix element $|A^{\rm sym}|^2$,
summed over colors and helicities.

The NLO result also requires real-emission corrections to the LO
process, which arise from tree-level amplitudes with one additional
parton; illustrative diagrams are shown in
\fig{treediagramsFigure}.  We use the~\COMIX{} library~\cite{Comix}, included in
the \SHERPA{} framework~\cite{Sherpa}, to compute these contributions,
including the Catani--Seymour dipole subtraction terms~\cite{CS}.  The \COMIX{}
code is based on a color-dressed form~\cite{CDBG} of the Berends-Giele
recursion relations~\cite{BG}, making it very efficient for processes with high
multiplicities.  

In the results described in the present article, we restrict attention
to one PDF set, and one jet algorithm.  We do use several correlated
values of the renormalization and factorization scales in order to
estimate the scale-dependence bands at LO and at NLO.  In addition, we
study several different choices for the experimental cuts.  In
general, however, we might need to compute the same physical distributions
for a collection of PDF error sets, and for different jet algorithms,
in addition to different renormalization and factorization scales.  We
organize the computation so that the matrix elements do not have to be
reevaluated anew for each choice of PDF, of scales, or of
jet-algorithm parameters (within a limited
set)~\cite{NTuples}.  We do this by storing intermediate information
in \root{}-format \ntuple{} files~\cite{ROOT}.
This format has also
been used by the experimental collaborations to compare results from
\BlackHat{} $+$ \SHERPA{} to experimental data~\cite{NTupleUse}.

\subsection{Checks}
\label{ChecksSection}

We have performed a number of consistency checks on the virtual
amplitudes and on integrated cross sections.  We have checked the
factorization properties of primitive amplitudes. 
As checks on our diphoton setup, at isolated
phase-space points, we have checked the $gg\gamma\gamma$ amplitude
against \MCFM{}~\cite{MCFM}; the $q\bar{q}\gamma\gamma$ amplitude
against HELAC-1LOOP~\cite{Helac}; the $ggg\gamma\gamma$ amplitude
against GoSam~\cite{GoSam}; the $q\bar{q}g\gamma\gamma$ amplitude against older
analytic results obtained from various permutations of the $q\bar{q}ggg$
amplitudes in ref.~\cite{TwoQuarkThreeGluon}
and against GoSam; and a selection of \YYjj-jet amplitudes against
GoSam.  We have also compared the cross section for \YYz-jet production
with \MCFM, that for \YYj-jet production with the results of
Gehrmann, Greiner and Heinrich (GGH)~\cite{GGH1}, and that for \YYjj-jet
production with the GGH results~\cite{GGH2}. When we use their cuts and choice of
central scale, we find agreement for the total cross section.

\subsection{Kinematics and Observables}
\label{KinematicsSection}

In our study, we consider the inclusive process
$p p \rightarrow\,\gamma\gamma\,+\,2\,$ jets at an LHC center-of-mass
energy of $\sqrt{s} = 8$ TeV, applying the following cuts:
\begin{equation}
\begin{aligned}
    & \pT^{\gamma_1} > 50 \hbox{ GeV} \,, \hskip 1.2cm
&&\pT^{\gamma_2} > 25 \hbox{ GeV} \,, \hskip 1.2 cm 
&&|\eta_{\gamma}| < 2.5\,, \hskip 1.2 cm
&&R_{\gamma\gamma} > 0.45\,, \\
& \pT^{\jet_1} > 40 \hbox{ GeV}\,, \hskip 1.2 cm 
   &&\pT^{\jet_2} > 25 \hbox{ GeV}\,, \hskip 1.2 cm 
 &&|\eta_{\jet}| < 4.5\,, \hskip 1.2 cm 
&&R_{\gamma,\jet} > 0.4\,.
\end{aligned}
\label{BasicCuts}
\end{equation}
We will call these the `basic' set of cuts.
In these expressions, $R$ is the usual longitudinally boost-invariant
angular distance,
$R_{ab} = [\Delta\phi_{ab}^2+\Delta\eta_{ab}^2]^{1/2}$.
We define jets using the anti-$k_T$ algorithm~\cite{antikT} with
parameter $R = 0.4$.  The jets are ordered in transverse momentum $\pT$,
and are labeled numerically in order of decreasing $\pT$, with jet $1$ being
the leading (hardest) jet. 

In addition, we also consider further cuts, which select the kinematic region
for VBF production of the Higgs-like boson, with the boson decaying into
two photons. We will call these the VBF cuts,
\begin{equation}
 M_{jj} > 400  \hbox{ GeV}\,, \hskip 1.5 cm 
    |\Delta\eta_{jj}| > 2.8 \,,
\label{BasicVBFCuts}
\end{equation}
where $M_{jj}$ is the invariant mass of the subsystem made up of the two
hardest jets, and $\Delta\eta_{jj}$ is the difference in pseudorapidity 
between these two jets. We will show distributions both with and without
VBF cuts.

For the central renormalization and factorization scale in our calculation, 
we use the dynamical scale $\HTpartonic/2$, where
\begin{equation}
\HTpartonic \,\equiv\, \pT^{\gamma_1} + \pT^{\gamma_2} + \sum_m p_{\rm T}^m \,.
\label{HT}
\end{equation}
The sum runs over all final-state partons $m$, whether or not they are
inside jets that pass the cuts.  This means that modifications to the
experimental cuts will not affect the value of the matrix element at a
given point in phase space. We note in passing that because the photons
are massless, $\HTpartonic$ in this calculation has the same value as the
$\HTpartonicp$ variable the \BlackHat{} collaboration has employed
previously for studies of $W$ or $Z$ production accompanied by jets.
(In $\HTpartonicp$, the transverse momentum of a boson with mass $M$
is replaced by the transverse energy $\ET = \sqrt{\pT^2 + M^2}$.)
We quote scale variation bands corresponding to varying the scales
simultaneously up and down by a factor of two, taking the maximum and
minimum of differential cross sections at the five scales
$\HTpartonic/2\times (1/2,1/\sqrt{2},1, \sqrt{2},2)$.

We also study the effect of an additional set of cuts, suggested by the
ATLAS collaboration, which selects a window on the diphoton invariant
mass centered around the Higgs-like boson mass,
\begin{equation}
\begin{aligned}
&\mathrlap{122~{\rm GeV} \leq m_{\gamma\gamma}\leq 130~{\rm GeV}\,,}\\
& \pT^{\gamma_1} > 0.35\,m_{\gamma\gamma}  \,, \hskip 10 mm
&&\pT^{\gamma_2} > 0.25\,m_{\gamma\gamma} \,, \hskip 10 mm 
&&|y_{\gamma}| < 2.37\,, \hskip 10 mm
&&R_{\gamma\gamma}>0.45\,,\hskip 10mm\\
& \pT^{\jet} > 30 \hbox{ GeV}\,, \hskip 10 mm  
&&R_{\gamma,\jet} > 0.4\,,\hskip 10 mm
&&|y_{\jet}| < 4.4\,.
\end{aligned}
\label{ATLASCuts}
\end{equation}
We will call these the `ATLAS' cuts.
The additional VBF cuts here are the same as those in \eqn{BasicVBFCuts}.

Finally, we study a set of cuts suggested by the CMS collaboration,
\begin{equation}
\begin{aligned}
&\mathrlap{100~{\rm GeV}\,\leq m_{\gamma\gamma}\leq\,180~{\rm GeV}\,,}\\
&\pT^{\gamma_1} > m_{\gamma\gamma}/2\,,\hskip 10mm
&&\pT^{\gamma_2} >  25 \,{\rm GeV}\,,\hskip 10mm
&&|\eta_{\gamma}|<2.5\,,\hskip 10mm\\
&\pT^{\jet} >  30 \,{\rm GeV}\,,\hskip 10mm
&&R_{\gamma\gamma}>0.45\,,\hskip 10mm
&&|\eta_{\rm jet}|<4.7\,,\hskip 10mm\\
&R_{\gamma, \jet}>0.5\,,\hskip 10mm
&&|\phi_{jj} - \phi_{\gamma\gamma}|> 2.6\,,\hskip 3mm
&&|\eta^*| < 2.5\,.
\end{aligned}
\label{CMSCuts}
\end{equation}
In these inequalities, $\phi_{jj}$ and $\phi_{\gamma\gamma}$ 
denote the azimuthal angle of the dijet and diphoton systems, respectively.\;
and $\eta^*$ denotes the relative diphoton pseudorapidity (as introduced by
Rainwater, Szalapski, and Zeppenfeld~\cite{RSZ}),
\begin{equation}
   \eta^*=\eta_{\gamma\gamma}-\frac1{2}({\eta_{\jet_1} + \eta_{\jet_2}})\,.
\end{equation}
In this equation, the pseudorapidity $\eta_{\gamma\gamma} =
-\ln\tan(\theta_{\gamma\gamma}/2)$, where $\theta_{\gamma\gamma}$ is
the polar angle in the lab frame for the diphoton momentum vector.
The jet algorithm used here is anti-$\kT$ with $R=0.5$.  We will call
these the `CMS' cuts.

The additional VBF cuts in this case are,
\begin{equation}
M_{jj}> 500\,{\rm GeV}\,,\hskip 10mm
|\Delta\eta_{jj}|>3\,.
\label{CMSVBFCuts}
\end{equation}

The calculation proceeds in two phases: generation of \ntuples, and analysis.
In the first phase, we generate two sets of \root{}~\cite{ROOT}
format \ntuples{} using a looser set of cuts,
\begin{equation}
\begin{aligned}
& \pT^{\gamma_1} > 25 \hbox{ GeV} \,, \hskip 10mm
&&\pT^{\gamma_2} > 25 \hbox{ GeV} \,, \hskip 10mm 
&&|\eta_{\gamma}| < 2.5\,, \hskip 10mm 
&&R_{\gamma\gamma} > 0.2\,, \\
& \pT^{\jet_1} > 25 \hbox{ GeV}\,, \hskip 10mm 
  && \pT^{\jet_2} > 25 \hbox{ GeV}\,, \hskip 10mm 
 &&|\eta_{\jet}| < 4.8\,, \hskip 10 mm 
&&R_{\gamma,\jet} > 0.4\,,
\end{aligned}
\label{GenerationCuts}
\end{equation}
where the $R_{\gamma,\jet}$ cut at generation level is applied
only to the leading two jets,
with the second set also imposing VBF cuts that
 are looser than those of \eqn{BasicVBFCuts},
\begin{equation}
M_{jj} > 300  \hbox{ GeV}\,, \hskip 1.5 cm 
    |\Delta\eta_{jj}| > 2.0 \,. 
\label{GenerationVBFCuts}
\end{equation}
In principle, if we had sufficient statistics in the first set,
generated with the cuts of \eqn{GenerationCuts}, we would not need a
second, more targeted set, in order to study the effect of VBF cuts.
These cuts push us into a small corner of phase space, however,
reducing the cross section by a factor of roughly 20.  Adequate
statistics in the first set would thus be 400 times larger than would
be needed for studies without VBF cuts.  It is much more efficient to
generate a second set of \ntuples{} in order to obtain reasonable
statistical uncertainties for the latter cuts.  The first
set of \ntuple{}s are in the process directory {\tt YY2j\/},
and the second in {\tt YY2j\_VBF\/}.  The location of the directory may
be found in 
{\tt http://blackhat.hepforge.org/trac/wiki/Location\/}.

In the second, analysis, phase of
our calculation we impose the following six sets of cuts:
\begin{eqnarray}
 &&\hbox{Basic:\hskip2cm     cuts of \eqn{BasicCuts}}\nonumber\\
 &&\hbox{Basic+VBF:\hskip0.8cm  cuts of \eqn{BasicCuts} and \eqn{BasicVBFCuts}}
\nonumber\\
 &&\hbox{ATLAS:\hskip1.6cm     cuts of \eqn{ATLASCuts}}\nonumber\\
 &&\hbox{ATLAS VBF:\hskip0.6cm cuts of \eqn{ATLASCuts} and \eqn{BasicVBFCuts}}
\nonumber\\
 &&\hbox{CMS:\hskip2.1cm     cuts of \eqn{CMSCuts}}\nonumber\\
 &&\hbox{CMS VBF:\hskip1.1cm cuts of \eqn{CMSCuts} and \eqn{CMSVBFCuts}}
\nonumber
\label{Allsixsets}
\end{eqnarray}
We compute the
cross section for each set of cuts, as well as various kinematical
distributions.

The \ntuples{} we have generated are also valid for anti-$k_T$, $k_T$ and
\SISCone{} algorithms~\cite{antikT,JetAlgorithms} for $R=0.4, 0.5, 0.6, 0.7$,
as implemented in the FASTJET package~\cite{FastJet}.  In the \SISCone{}
case the merging parameter $f$ is chosen to be $0.75$. 

In addition to distributions in transverse momenta, invariant masses,
rapidities, and azimuthal angles,
we will also study a distribution in $\cos\theta^*$,
the cosine of the polar angle of the photon pair with 
respect to the $z$ axis of the Collins--Soper frame~\cite{CollinsSoper}.
This variable can also be expressed as,
\begin{equation}
|\cos\theta^*|\,=\,
 \frac{|\sinh(\Delta\eta_{\gamma\gamma})|}
{\sqrt{1+(\pT^{\gamma\gamma}/m_{\gamma\gamma})^2}}
 \frac{2\pT^{\gamma_1}\pT^{\gamma_2}}{m_{\gamma\gamma}^2}\,.
\label{CosTheta}
\end{equation}
It has been used by the ATLAS~\cite{ATLASSpin} and CMS~\cite{CMSDiphoton}
collaborations in their studies of the diphoton decays of the Higgs-like boson.

In our study, we use the MSTW2008 LO and NLO PDFs~\cite{MSTW2008} at
the respective orders.  We use the five-flavor running $\alpha_s(\mu)$
and the value of $\alpha_s(M_Z)$ supplied with the parton distribution
functions.
As explained in
ref.~\cite{PhotonZ} (see also refs.~\cite{Alpha}), we use the
zero-momentum-squared value, $\alpha_{\rm EM}(0) = 1/137$ (to our required
precision), for the
electromagnetic coupling.

We perform our fixed-order NLO computation at the parton level.  We do
not apply a parton shower, or corrections due to non-perturbative effects
such as those induced by the underlying event or hadronization.
For comparisons to experiment it is important to incorporate these
effects or at least estimate their size.  

\section{Results}
\label{ResultsSection}
\subsection{Scale dependence}

\begin{figure}[t]
\includegraphics[clip,scale=0.65]{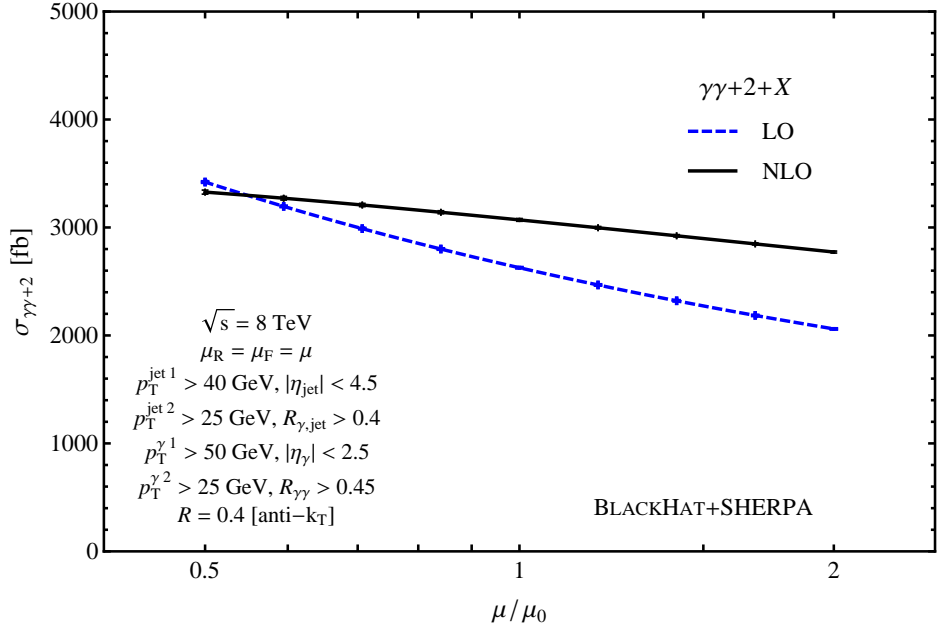}
\caption{ 
The renormalization-scale dependence of the cross section for
\YYjj-jet production using a dynamical central scale of $\mu_0=\HTpartonic/2$.  
The renormalization and factorization scales are kept equal
and varied simultaneously.  The LO result is given by the dashed (blue)
line, and the NLO one by the solid (black) line.  The error bars indicate
the numerical integration uncertainties.}
\label{ScalesTotXSFigure}
\end{figure}

We expect perturbative results to be more stable under variation of
the renormalization and factorization scales as the perturbative order
is increased.  The residual variability has been used as a proxy for
the expected uncertainty due to higher-order corrections beyond the
calculated order.  As an example, we saw that in studies of $W$ in
association with several jets~\cite{W3jDistributions,BlackHatZ3jet,W4j,W5j},
the variability increases substantially with a growing number of jets
at LO, but stabilizes at under 20\% at NLO (for a range of
scales between half and twice the central value).  In
\fig{ScalesTotXSFigure}, we show how the cross section for \YYjj-jet production
varies with a common renormalization and factorization scale,
$\mu_R = \mu_F = \mu$.   We vary the common scale up
and down by a factor of 2 at both LO and NLO, around a central choice
of $\HTpartonic/2$.  The NLO variation is under 10\% of the central value.

The kinematical distributions we study have a large dynamic range, and
$\HTpartonic/2$ is a suitable event-by-event scale, matching typical
energy scales individually rather than merely on average.   
In~\sect{CrossSectionsSection}, we plot a variety of distributions.
The bands in the plots all correspond to varying the scales up and down
by a factor of 2 around the central value.  Other authors have suggested
alternate choices of dynamical scale~\cite{OtherScaleChoices,BDDP}; GGH
have used such an alternate dynamical scale~\cite{GGH2}. 


\subsection{Dependence on the Frixione-Cone Energy Fraction}

\begin{figure}[t]
\includegraphics[clip,scale=0.65]{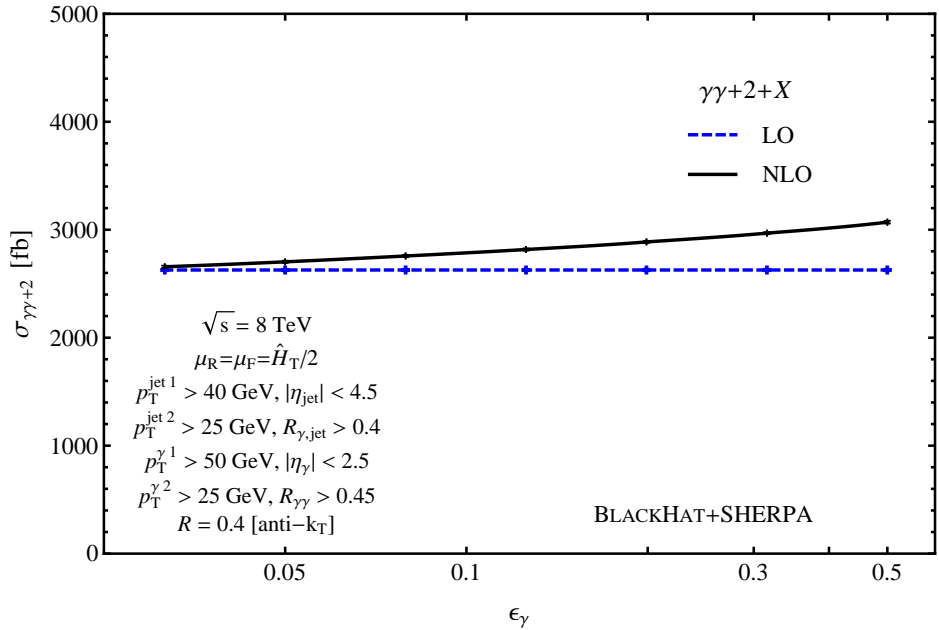}
\caption{ The dependence of the cross section on the $\epsilon_\gamma$
parameter in the Frixione-cone photon isolation.  The LO result is given
by the dashed (blue) line, and the NLO one by the solid (black) line.
The error bars indicate the numerical integration uncertainties.}
\label{EpsilonGammaTotXSFigure}
\end{figure}
\FloatBarrier
In a previous study of single-photon production in association
with jets, we observed that the NLO cross section depended only weakly on
the parameters used for the Frixione-cone isolation of the photons.  We
have examined the dependence on one of these parameters, the energy fraction
$\epsilon_\gamma$, in the present study.  The results are shown in
\fig{EpsilonGammaTotXSFigure}.  The LO result is of course independent of
the parameter, as there is no additional radiation that could enter the
photon cone; this result is shown for comparison in the figure.  The NLO
cross section is only weakly dependent on this parameter in the range 
$0.03<\epsilon_\gamma<0.5$. 

\subsection{Cross Sections and Distributions}
\label{CrossSectionsSection}

\newcommand\Tstrut{\rule{0pt}{2.4ex}}       
\newcommand\Bstrut{\rule[-1.1ex]{0pt}{0pt}} 

\begin{table}[t]
\vskip .4 cm
\centering
\begin{tabular}{||c||c|c|c||}
\hline
Cuts &  LO  & NLO & $gg\rightarrow \gamma\gamma gg$ \\  \hline
\Tstrut\Bstrut Basic &~$2627(3)^{+794}_{-567}$~&~$3070(13)^{+257}_{-298}$~&~$48(3)$~\\ \hline
\Tstrut\Bstrut Basic+VBF &~$136.0(0.2)^{+52.6}_{-34.9}$~&~$155(1)^{+14}_{-18}$~&~$2.75(0.05)$~\\ \hline
\Tstrut\Bstrut ATLAS &~$89.3(0.5)^{+26.6}_{-19.1}$~&~$100(2)^{+7}_{-9}$~&~$1.46(0.05)$~\\ \hline
\Tstrut\Bstrut ATLAS+VBF &~$3.91(0.03)^{+1.53}_{-1.01}$~&~$4.6(0.1)^{+0.5}_{-0.6}$~&~$0.075(0.004)$~\\ \hline
\Tstrut\Bstrut CMS &~$574(1)^{+170}_{-122}$~&~$596(3)^{+21}_{-43}$~&~$7.82(0.08)$~\\ \hline
\Tstrut\Bstrut CMS+VBF &~$11.84(0.05)^{+4.68}_{-3.09}$~&~$14.7(0.2)^{+2.0}_{-2.0}$~&~$0.34(0.01)$~\\ \hline

\end{tabular}
\caption{Total cross sections in femtobarns for \YYjj-jet production with
various sets of cuts:
basic (\protect\eqn{BasicCuts}), VBF (\protect\eqns{BasicCuts}{BasicVBFCuts}), 
ATLAS (\protect\eqn{ATLASCuts}),
ATLAS VBF (\protect\eqns{ATLASCuts}{BasicVBFCuts}),
CMS (\protect\eqn{CMSCuts}), and CMS VBF (\protect\eqns{CMSCuts}{CMSVBFCuts}).
The numerical integration uncertainty is given in parentheses, and the scale
dependence is quoted in superscripts and subscripts.  The contribution of
the $gg\rightarrow \gamma\gamma gg$ subprocess, shown separately in the last
column, is small but is included in the NLO value.}
\label{CrossSectionTable}
\end{table}

In \Tab{CrossSectionTable}, we present the LO and NLO parton-level cross
sections for inclusive diphoton production accompanied by two jets.
We consider the six different sets of cuts discussed in \sect{KinematicsSection}.
We list separately the contributions from the $gg\rightarrow\gamma\gamma gg$
subprocess (this contribution is also included in the NLO prediction).  

The pure-gluon process starts only at one loop, and is therefore
suppressed by two powers of $\alpha_s$.  As discussed earlier, we
might expect it to be genuinely suppressed compared to the tree-level
$gg$ initial-state contribution.  We find that the pure-gluon
subprocess does give only a small contribution, as shown in
\Tab{CrossSectionTable}: it contributes less than 2.5\% of the NLO
result in all cases.  However, it is not as suppressed as one might have naively expected,
compared to the LO $gg$ initial-state contribution, which is
approximately 5\% of the LO cross section for both Basic and Basic+VBF
cuts.

\def\leftmarg{\null\hskip -5mm}
\def\tighten{\vspace{-7mm}}
\def\tightcaption{\baselineskip=13pt}
\begin{figure}[tb]
\begin{center}
\begin{minipage}[b]{1.03\linewidth}
\leftmarg
\includegraphics[clip,scale=0.29]{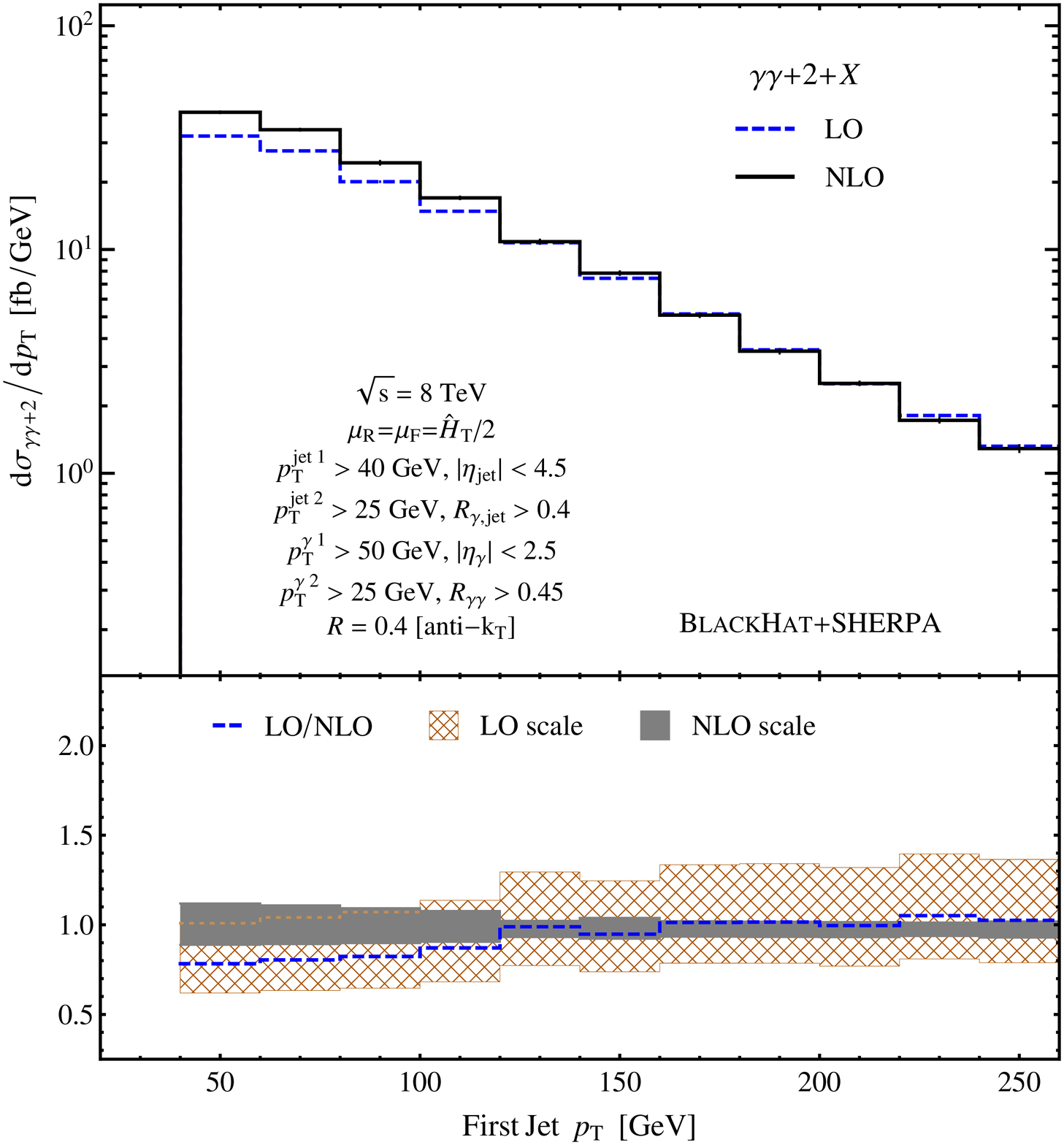} 
\includegraphics[clip,scale=0.29]{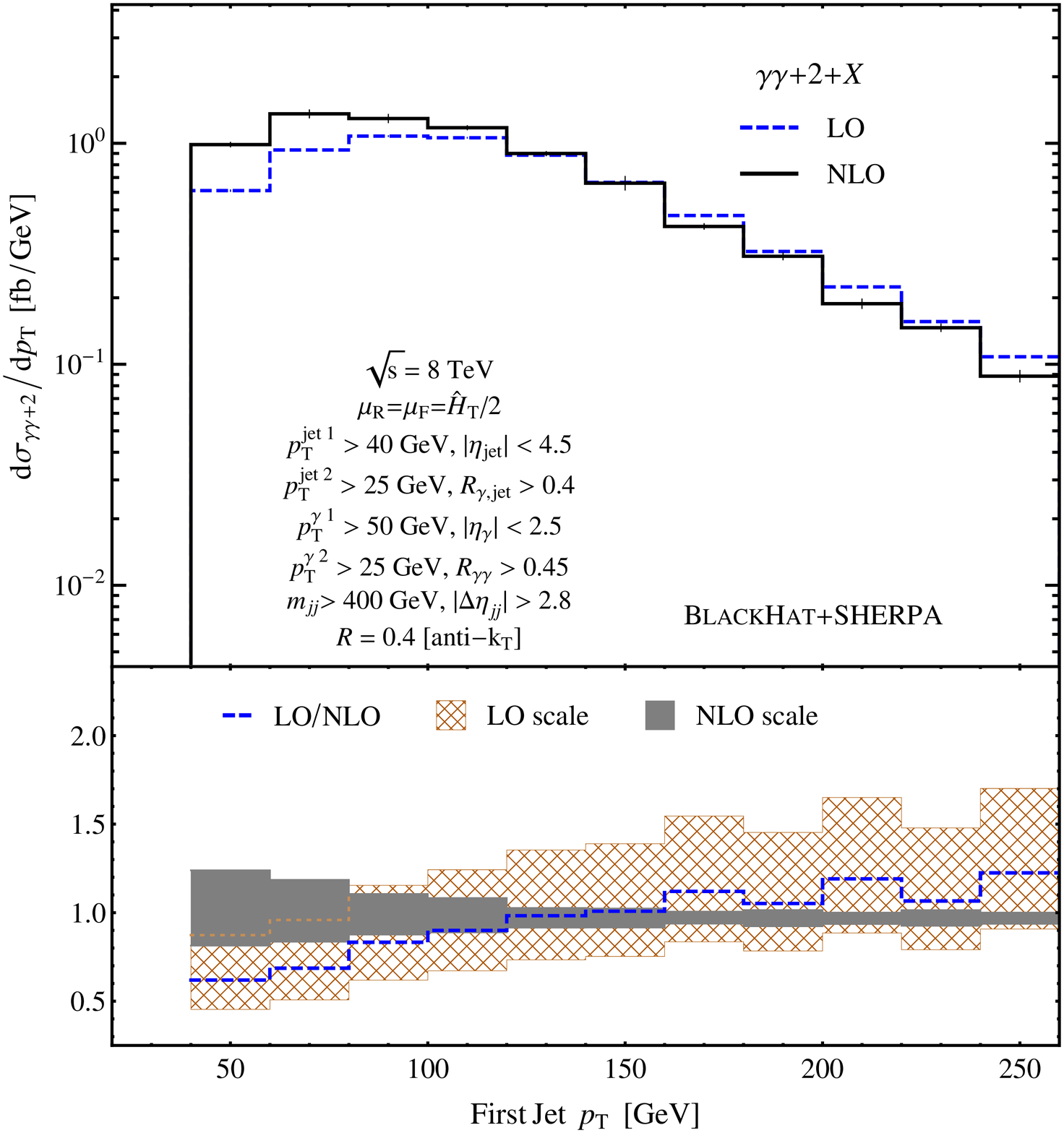}
\end{minipage}
\end{center}
\tighten
\caption{\tightcaption The leading-jet transverse-momentum
  distribution in \protect\YYjj-jet production.
  The left plot shows the distribution for the basic cuts of
  \protect\eqn{BasicCuts}, and the right plot with the VBF cuts of
  \protect\eqn{BasicVBFCuts} in addition.  The upper panels show the
  LO (dashed blue) and NLO (solid black) distributions, while the
  lower panels show the ratios to the NLO prediction, including the LO
  (hatched brown) and NLO (gray) scale-dependence bands.  The thin
  vertical lines at the center of each bin (where visible) indicate
  the numerical integration errors for the bin.}
\label{Jet1pTFigure}
\end{figure}

\begin{figure}[bt]
\begin{center}
\begin{minipage}[b]{1.03\linewidth}
\leftmarg
\includegraphics[clip,scale=0.29]{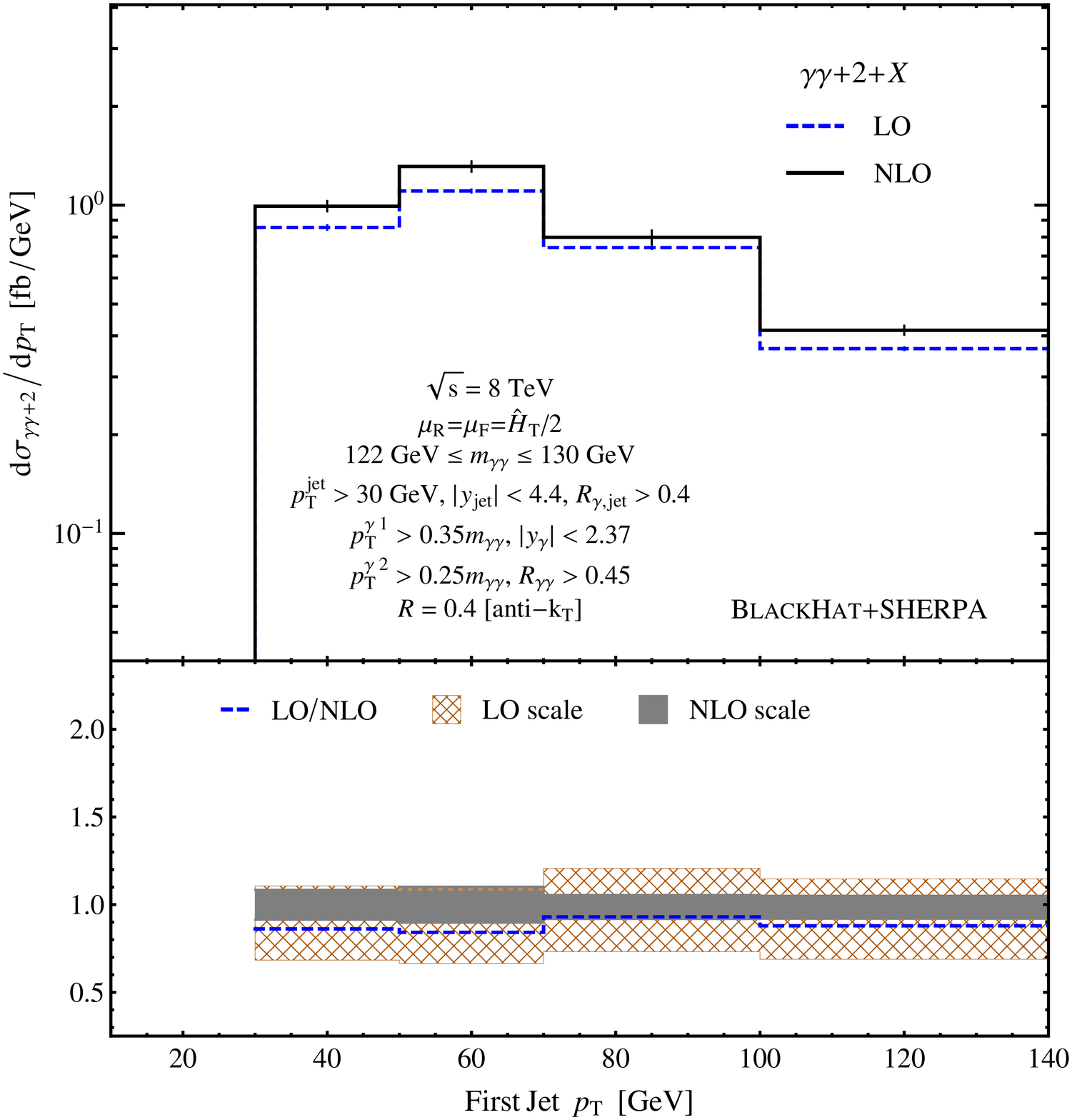} 
\includegraphics[clip,scale=0.29]{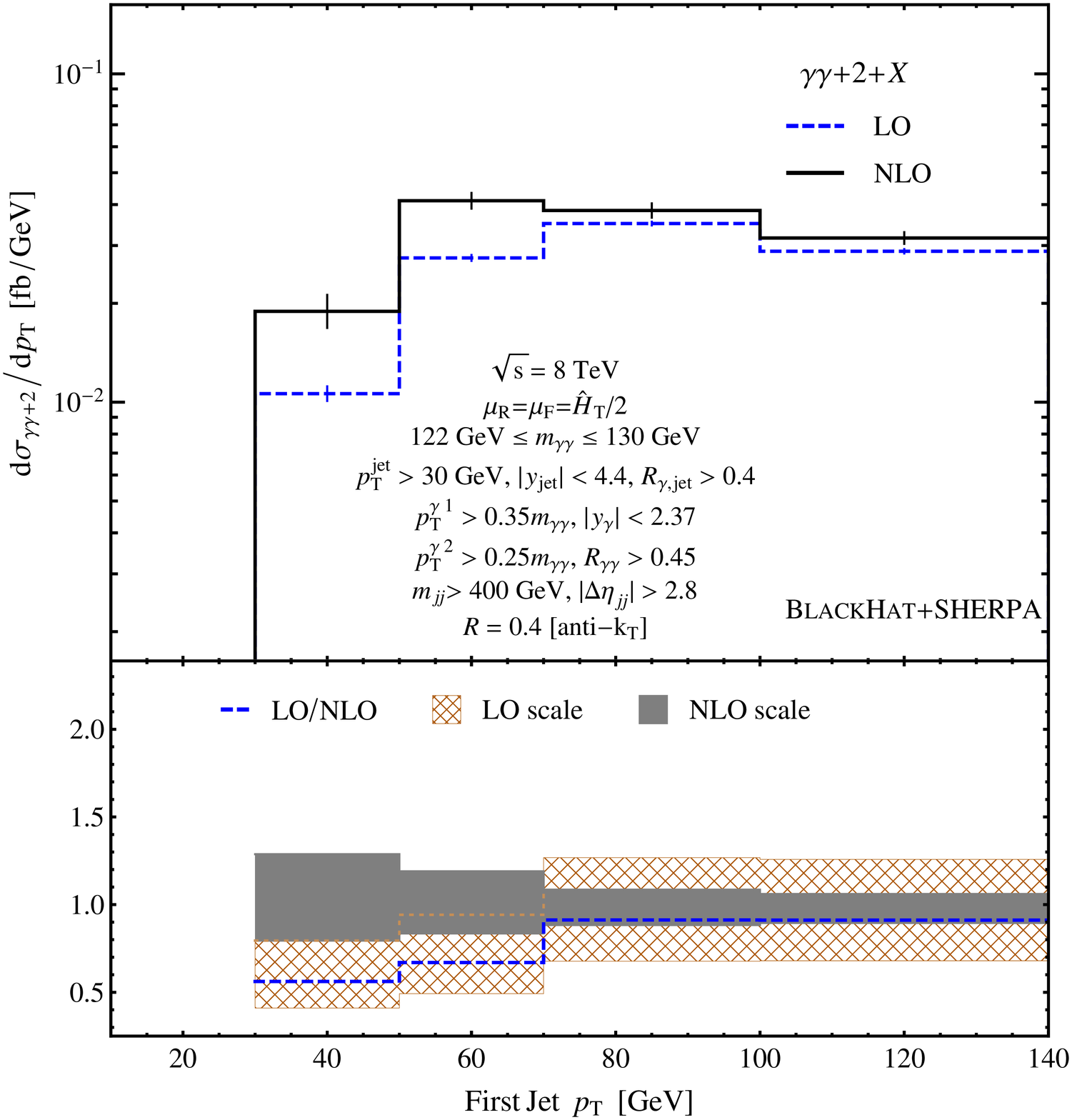}
\end{minipage}
\end{center}
\tighten
\caption{\tightcaption 
The leading-jet transverse-momentum distribution in \YYjj-jet production.
The left plot shows the distribution for the ATLAS cuts of
\protect\eqn{ATLASCuts}, and the right plot with the VBF cuts of
\protect\eqn{BasicVBFCuts} in addition.  The panels, curves, and bands are
as in \protect\fig{Jet1pTFigure}.}
\label{Jet1pT-ATLASFigure}
\end{figure}

\FloatBarrier

We also present predictions for a number of distributions.
In \fig{Jet1pTFigure}, we show the distribution in the transverse momentum 
of the leading jet for the cuts of \eqn{BasicCuts}, and also with the
addition of the VBF cuts of \eqn{BasicVBFCuts}.  In \fig{Jet1pT-ATLASFigure},
we show the same distribution with the ATLAS cuts of \eqn{ATLASCuts} as well
as with the additional VBF cuts of \eqn{BasicVBFCuts}.  We provide detailed
tables of our results in appendices~\ref{BasicTablesAppendix},
\ref{ATLASTablesAppendix}, and~\ref{CMSTablesAppendix}.

\begin{figure}[tb]
\begin{center}
\vspace{-2mm}
\begin{minipage}[b]{1.03\linewidth}
\leftmarg
\includegraphics[clip,scale=0.29]{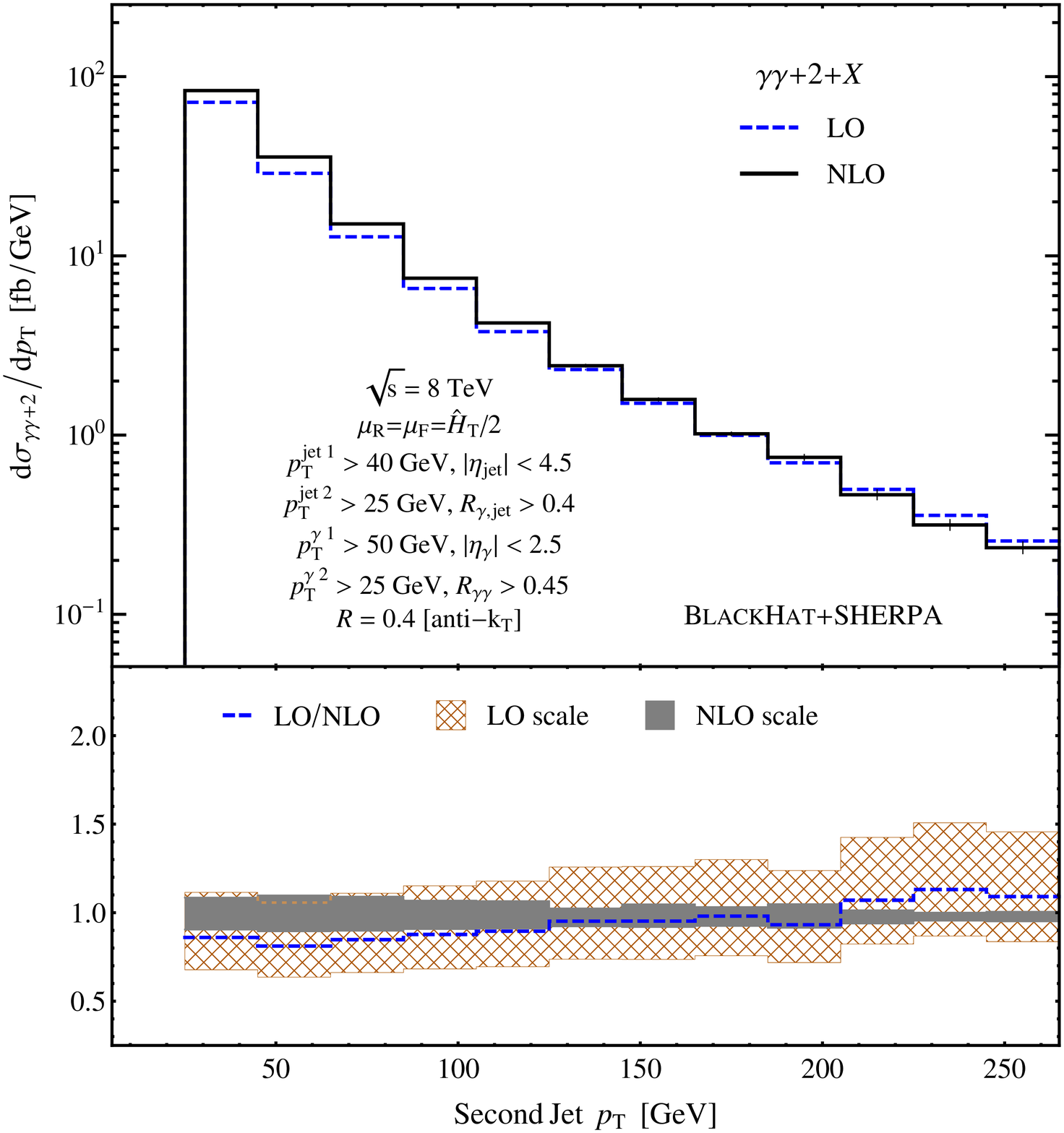} 
\includegraphics[clip,scale=0.29]{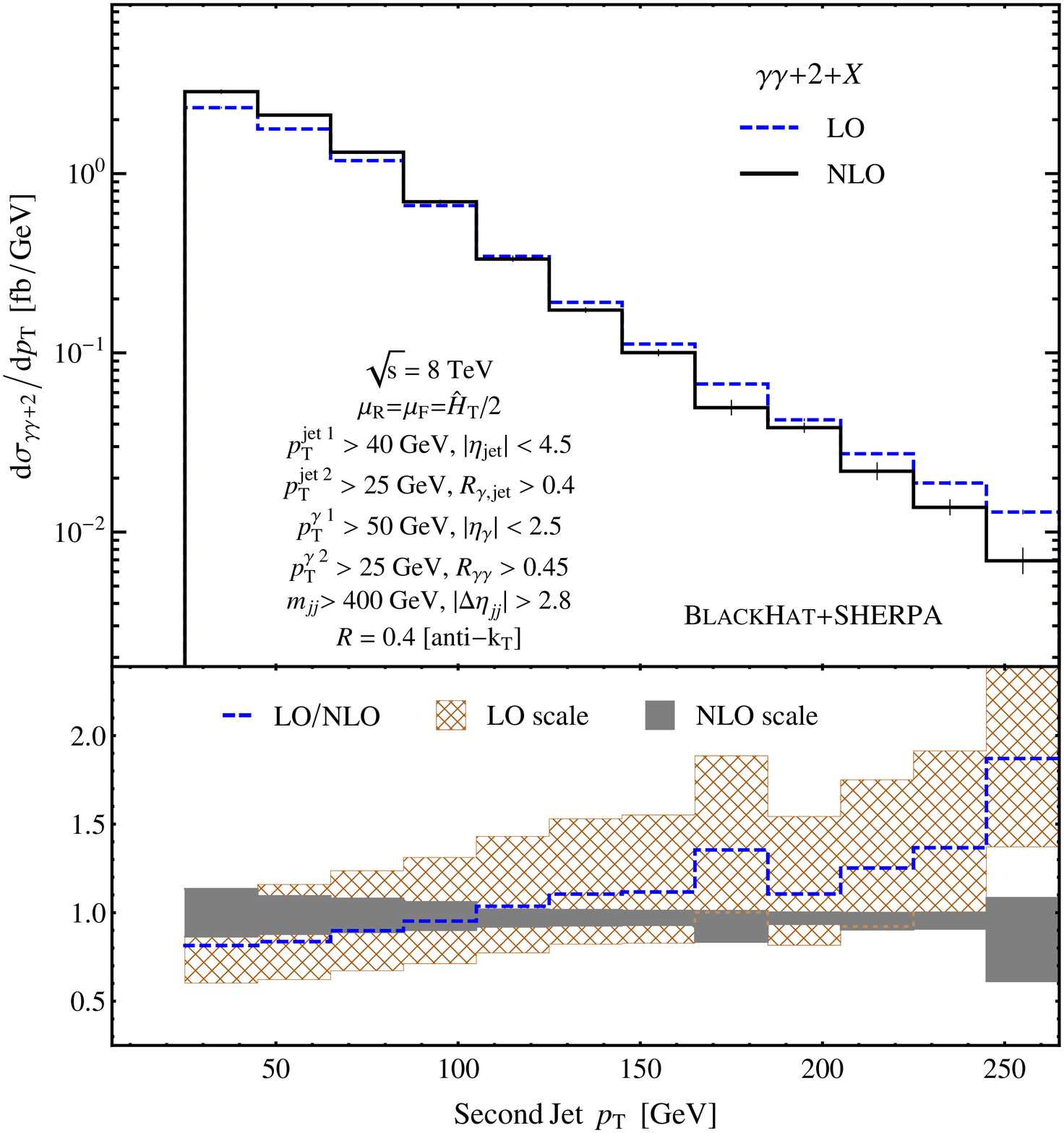}
\end{minipage}
\end{center}
\vspace{-2mm}\tighten
\caption{\tightcaption
The second jet transverse-momentum distribution in \YYjj-jet production.
The left plot shows the distribution for the basic cuts of
\protect\eqn{BasicCuts}, and the right plot with the VBF cuts of
\protect\eqn{BasicVBFCuts} in addition.  The panels, curves, and bands are
as in \protect\fig{Jet1pTFigure}.}
\label{Jet2pTFigure}
\end{figure}

\begin{figure}[tbh]
\begin{center}
\vspace{-2mm}
\begin{minipage}[b]{1.03\linewidth}
\leftmarg
\includegraphics[clip,scale=0.29]{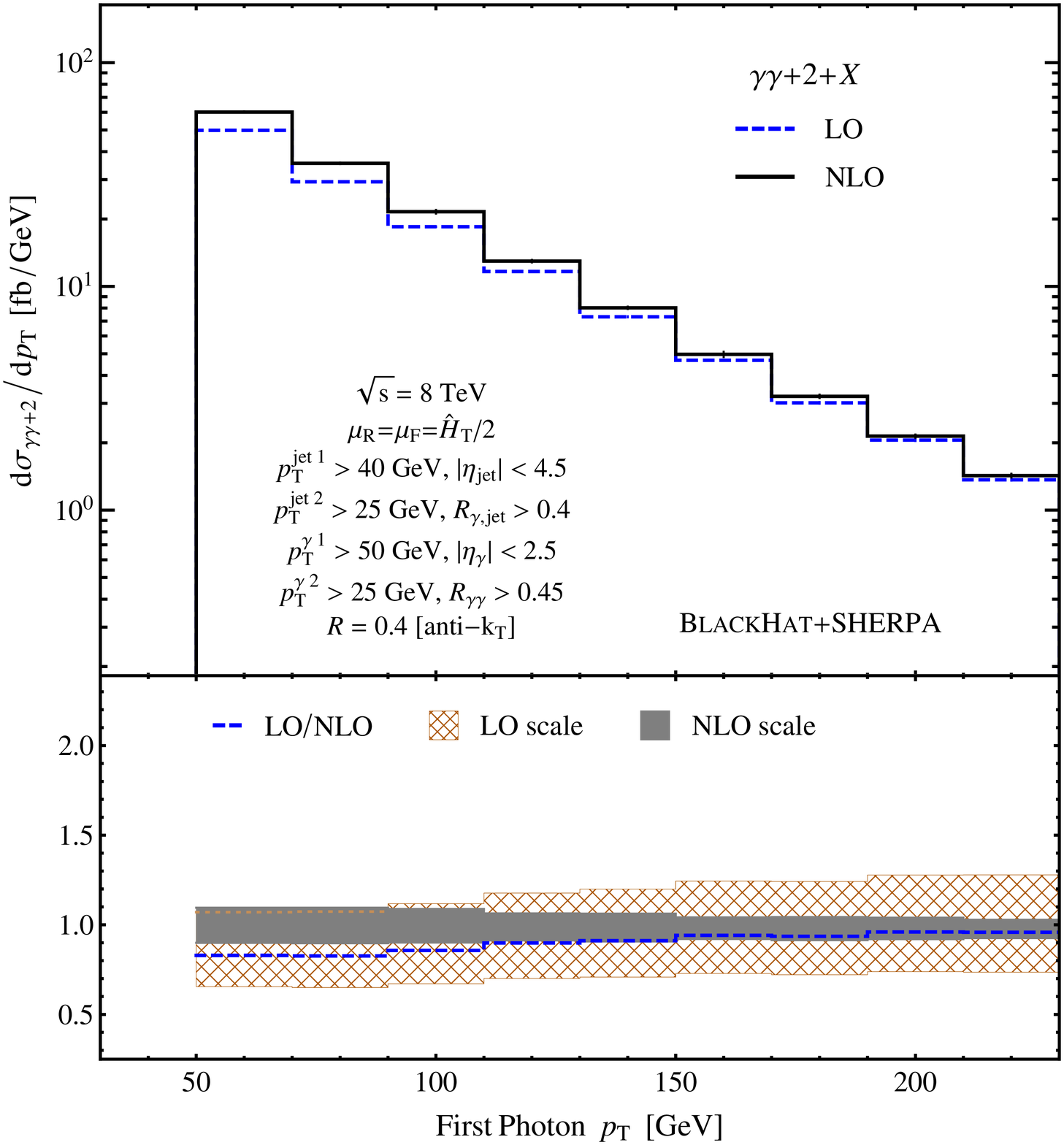} 
\includegraphics[clip,scale=0.29]{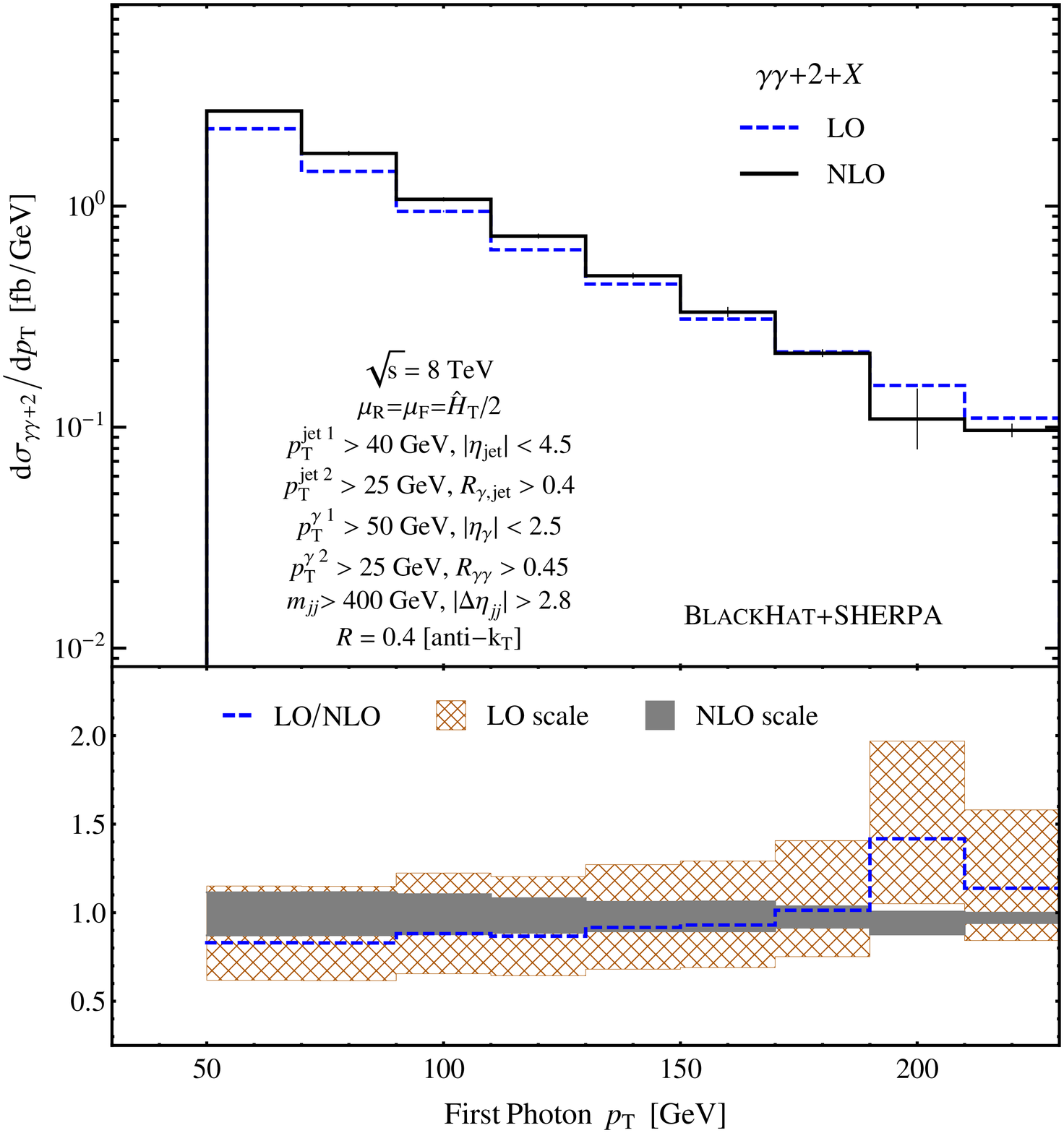}
\end{minipage}
\end{center}
\vspace{-2mm}\tighten
\caption{\tightcaption
The leading-photon transverse-momentum distribution in \YYjj-jet production.
The left plot shows the distribution for the basic cuts of
\protect\eqn{BasicCuts}, and the right plot with the VBF cuts of
\protect\eqn{BasicVBFCuts} in addition.  The panels, curves, and bands are
as in \protect\fig{Jet1pTFigure}.}
\label{Photon1pTFigure}
\end{figure}

\begin{figure}[!tbhp]
\begin{center}
\begin{minipage}[b]{1.03\linewidth}
\leftmarg
\includegraphics[clip,scale=0.29]{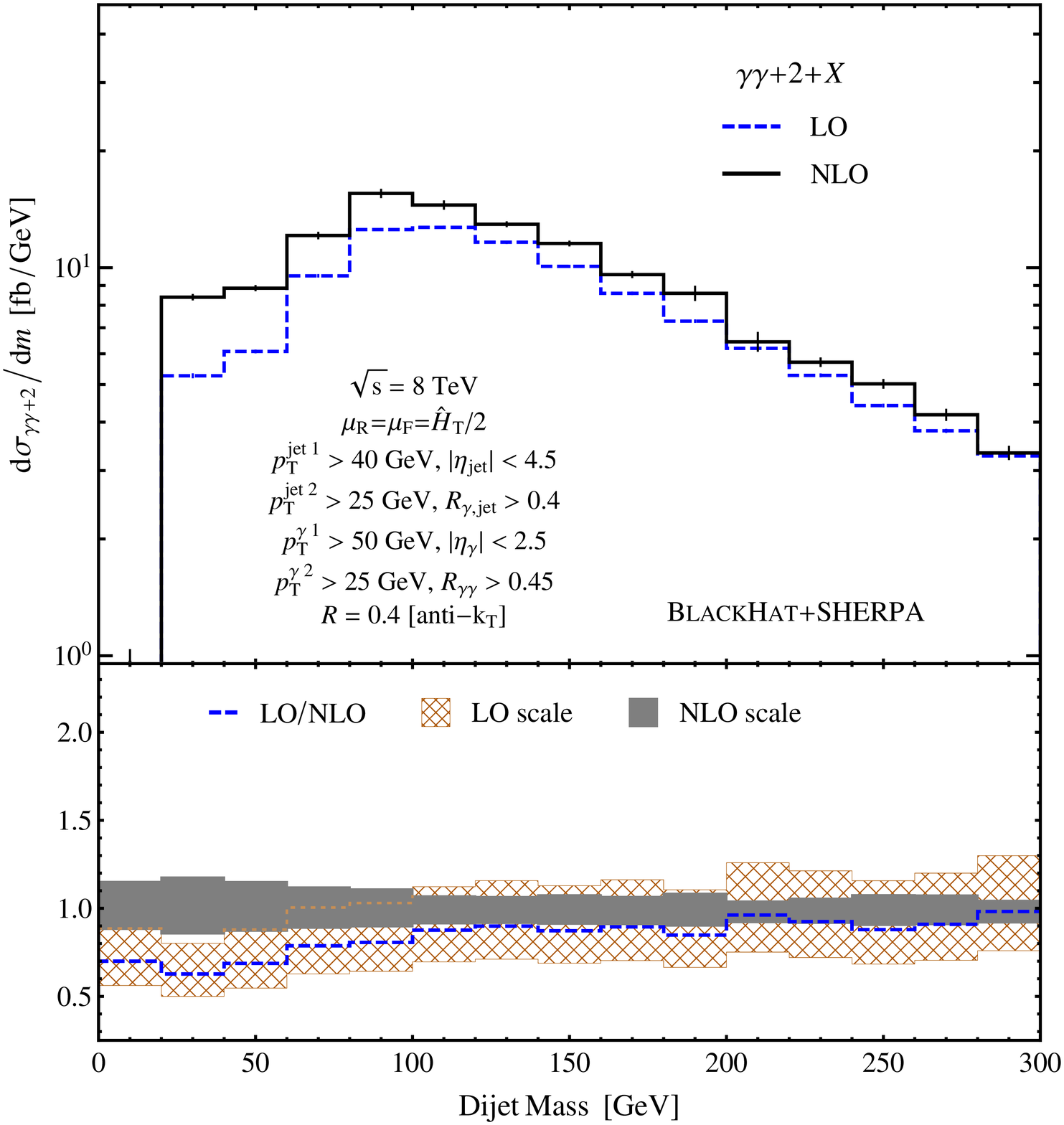} 
\includegraphics[clip,scale=0.29]{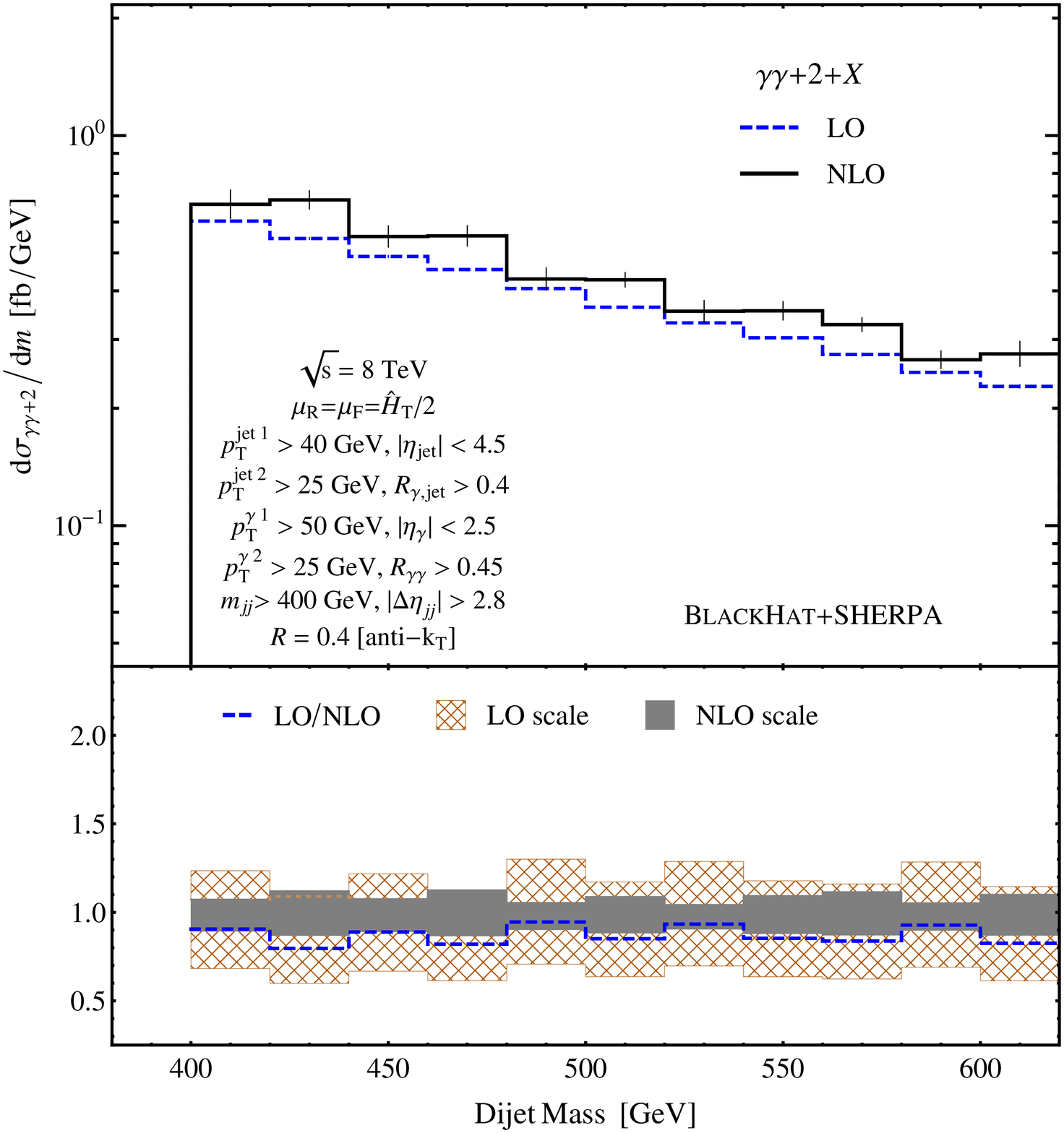}
\end{minipage}
\end{center}
\tighten
\caption{\tightcaption
The dijet invariant-mass distribution in \YYjj-jet production.  The left plot
shows the distribution for the basic cuts of \protect\eqn{BasicCuts}, and the
right plot with the VBF cuts of \protect\eqn{BasicVBFCuts} in addition.
The panels, curves, and bands are as in \protect\fig{Jet1pTFigure}.}
\label{DijetMassFigure}
\end{figure}

\begin{figure}[!tbhp]
\begin{center}
\begin{minipage}[b]{1.03\linewidth}
\leftmarg
\includegraphics[clip,scale=0.29]{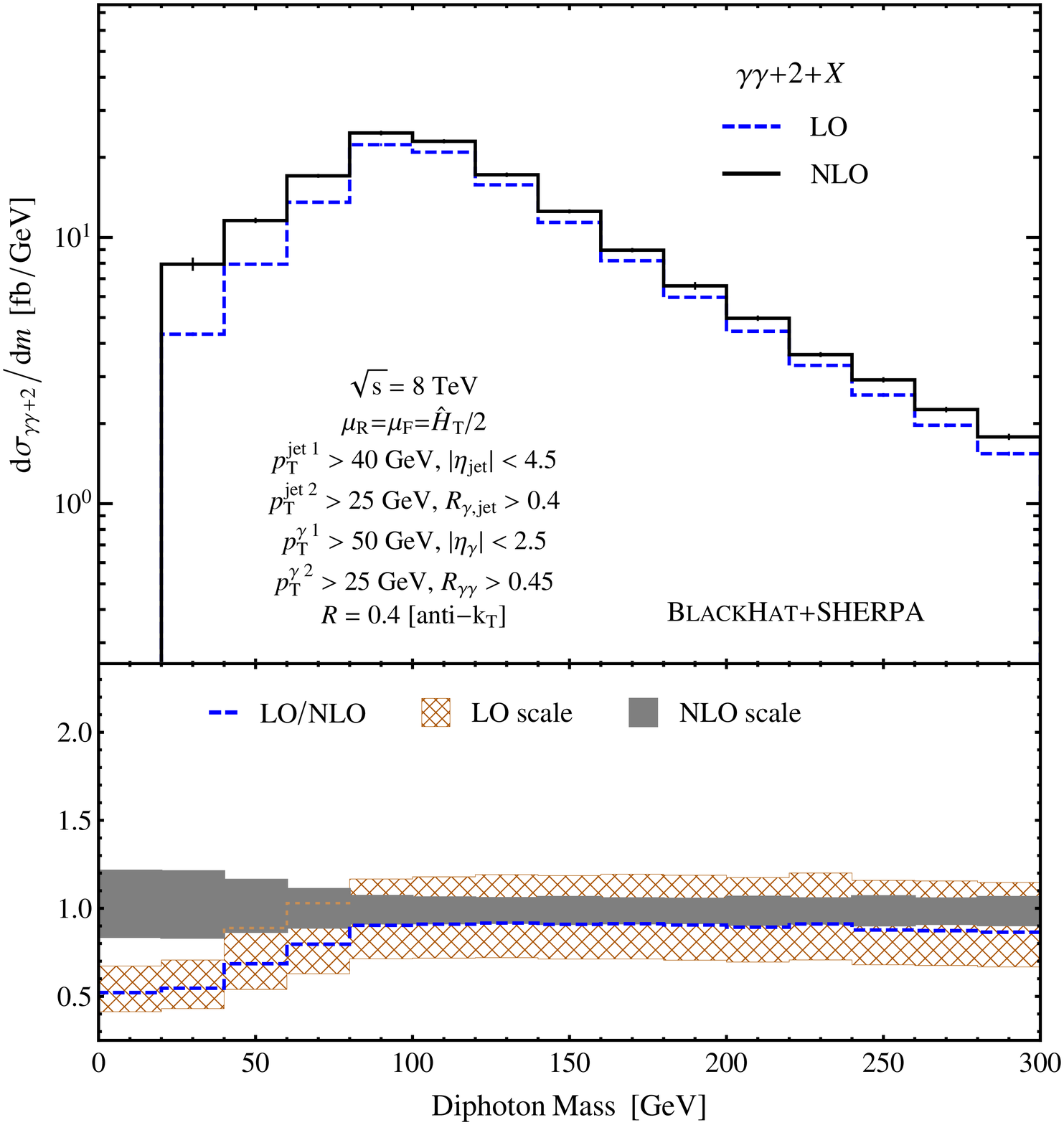} 
\includegraphics[clip,scale=0.29]{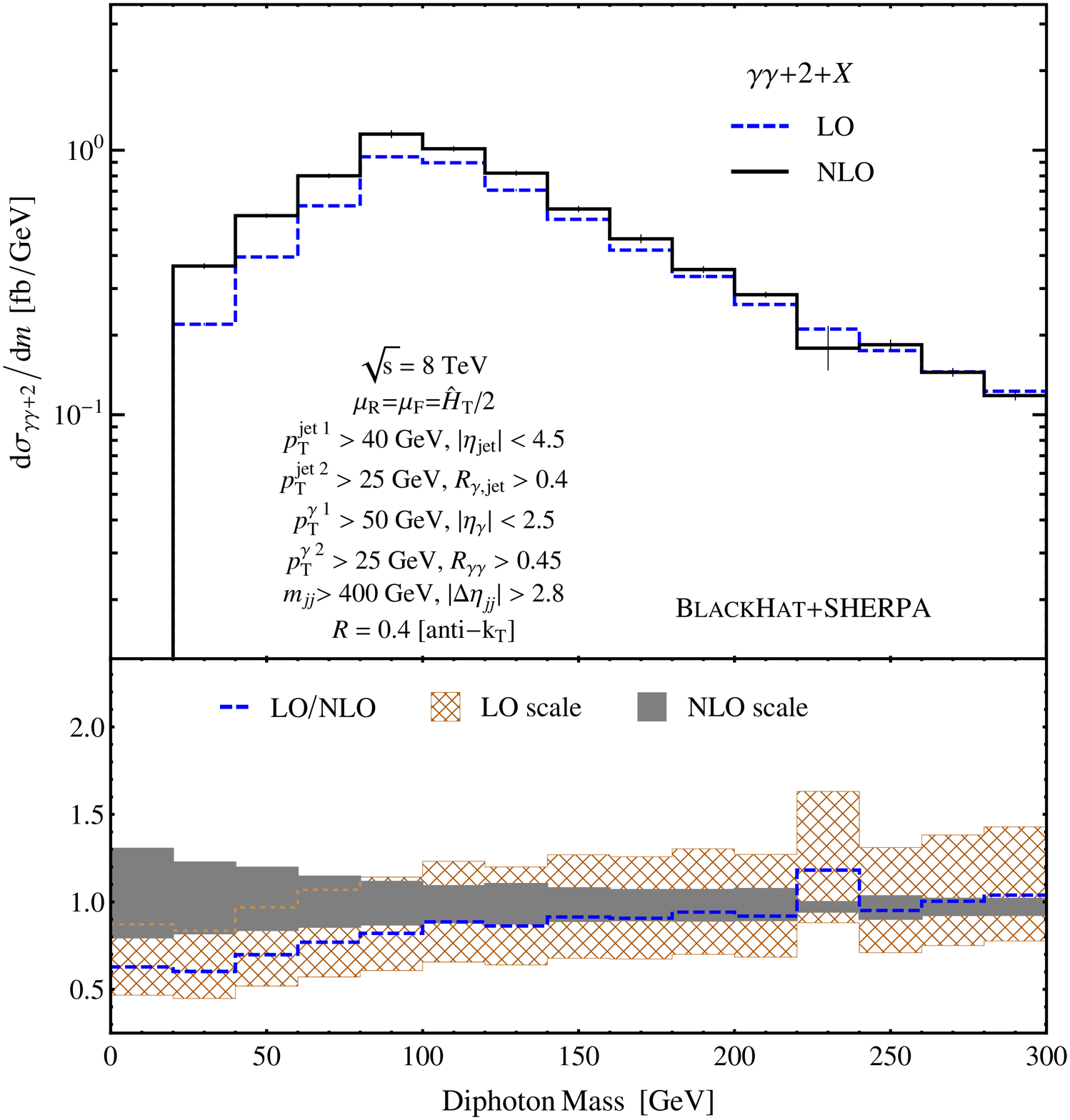}
\end{minipage}
\end{center}
\tighten
\caption{\tightcaption
The photon-pair invariant-mass distribution in \YYjj-jet production.  The left
plot shows the distribution for the basic cuts of \protect\eqn{BasicCuts},
and the right plot with the VBF cuts of \protect\eqn{BasicVBFCuts} in addition.
The panels, curves, and bands are as in \protect\fig{Jet1pTFigure}.}
\label{DiphotonMassFigure}
\end{figure}

In figs.~\ref{Jet2pTFigure}--\ref{DiphotonMassFigure}, we show a series of
distributions side-by-side for the cuts of \eqn{BasicCuts} and for the
same cuts with the addition of the VBF cuts of \eqn{BasicVBFCuts}:
in \fig{Jet2pTFigure}, the transverse momentum of the second jet;
in \fig{Photon1pTFigure}, the transverse momentum of the leading photon;
in \fig{DijetMassFigure}, the dijet invariant mass; and
in \fig{DiphotonMassFigure}, the photon-pair invariant mass.
\clearpage

\begin{figure}[!tbhp]
\begin{center}
\begin{minipage}[b]{1.03\linewidth}
\null\hskip -4mm
\includegraphics[clip,scale=0.29]{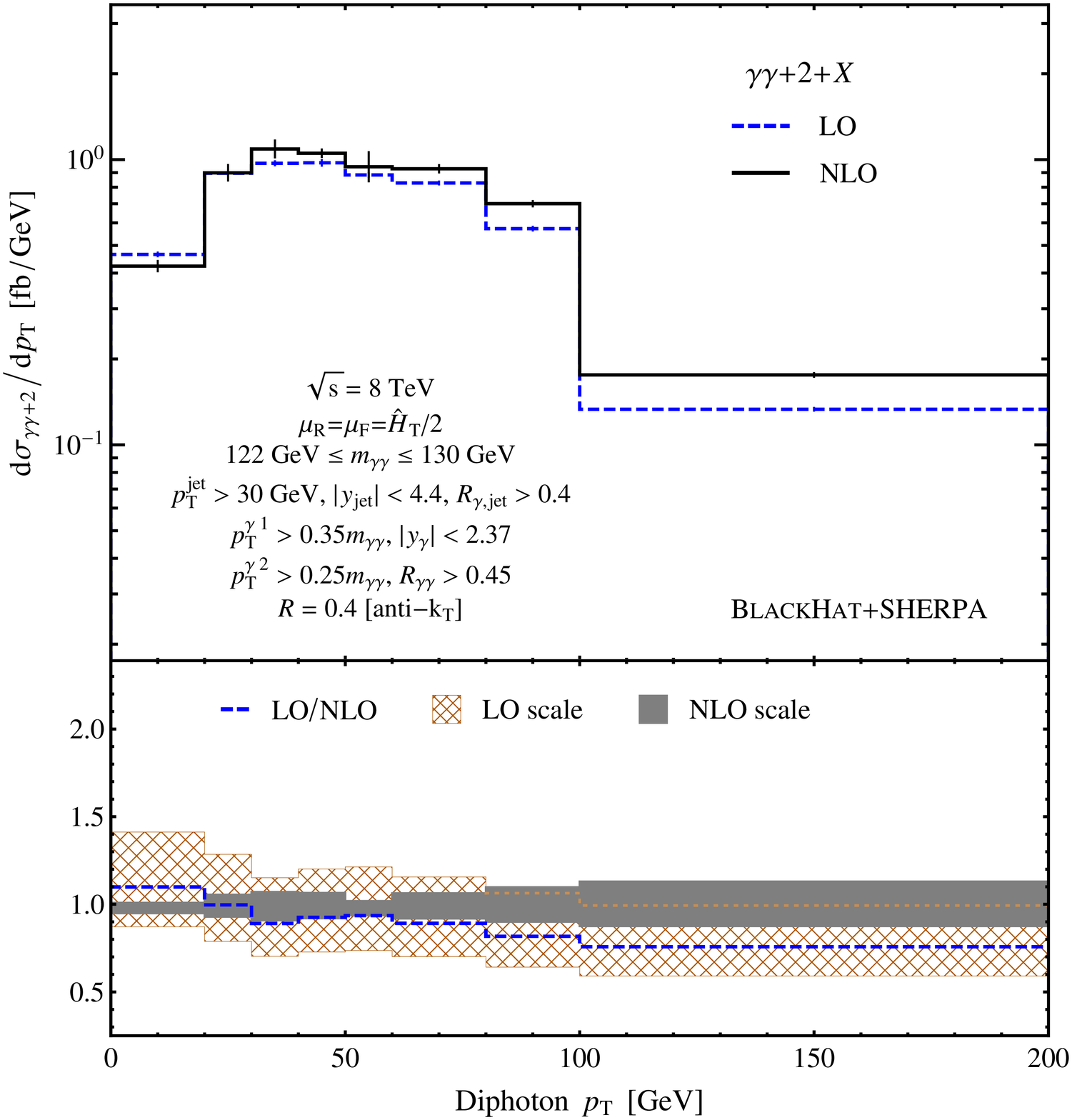} 
\includegraphics[clip,scale=0.29]{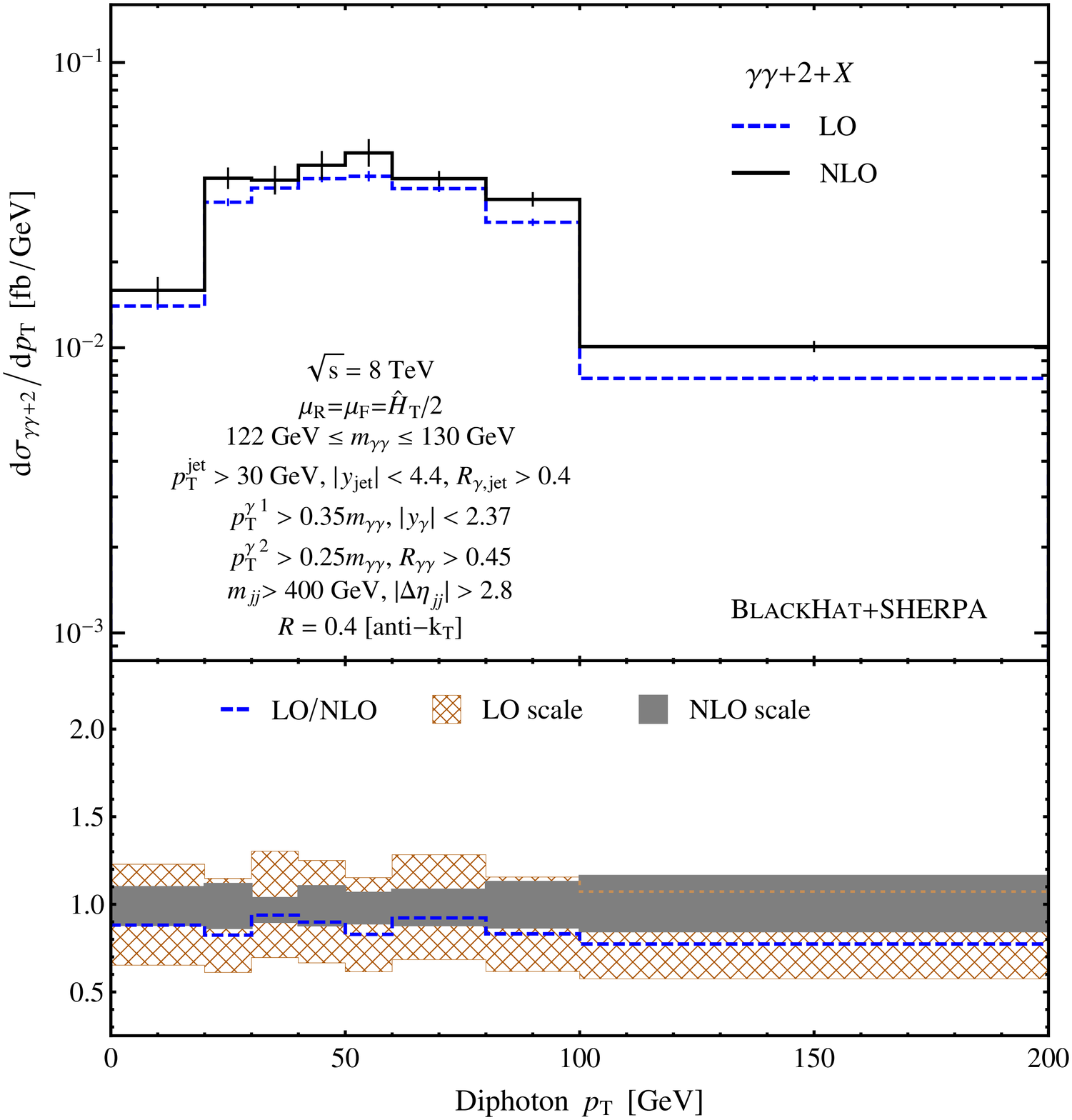}
\end{minipage}
\end{center}
\tighten
\caption{\tightcaption
The diphoton transverse-momentum distribution in \YYjj-jet production.
The left plot shows the distribution for the ATLAS cuts of
\protect\eqn{ATLASCuts}, and the right plot with the VBF cuts of
\protect\eqn{BasicVBFCuts} in addition.  The panels, curves, and bands are
as in \protect\fig{Jet1pTFigure}.}
\label{DiphotonPT-ATLASFigure}
\end{figure}

\begin{figure}[!tbhp]
\begin{center}
\begin{minipage}[b]{1.03\linewidth}
\null\hskip -4mm
\includegraphics[clip,scale=0.29]{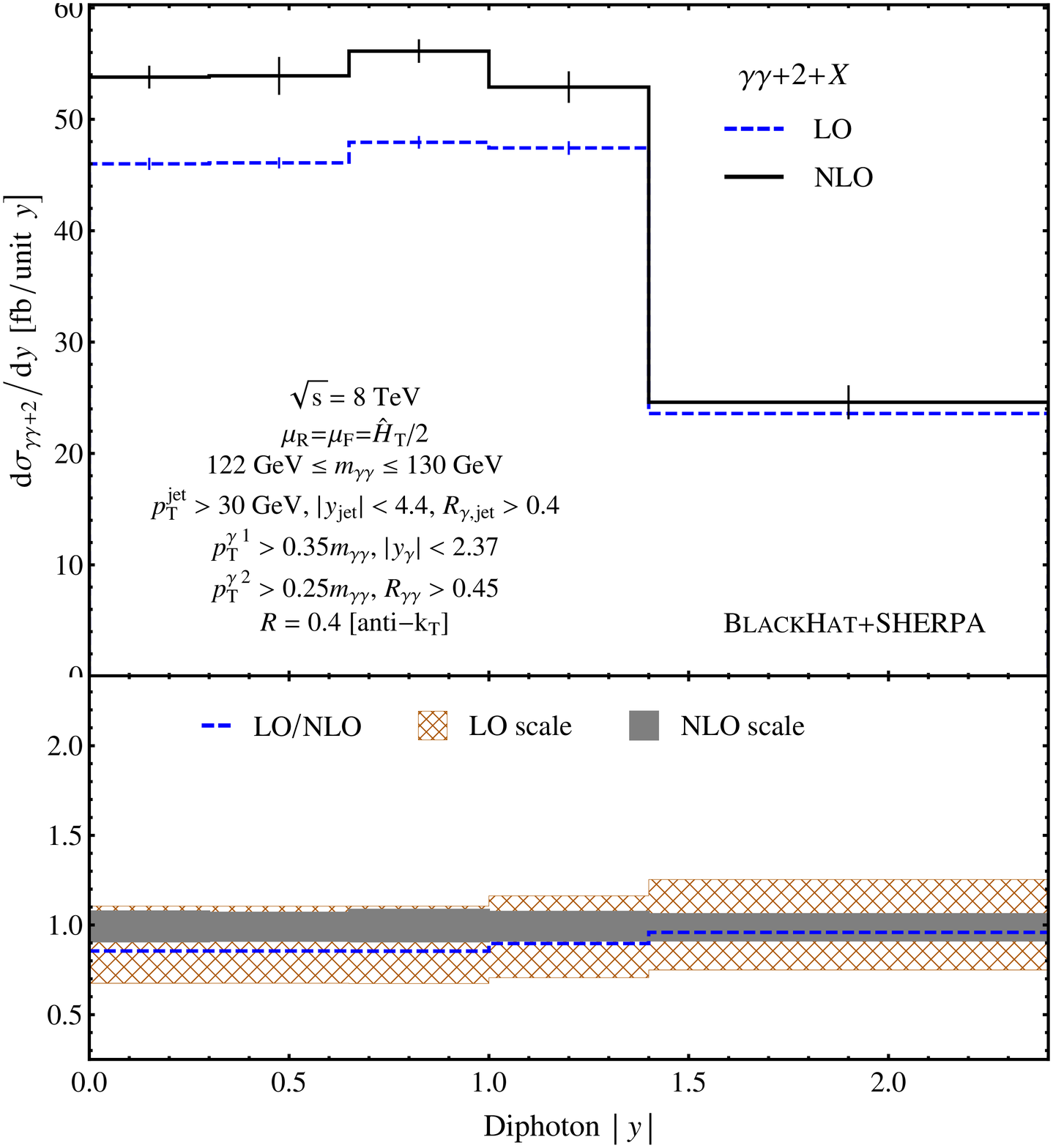} 
\includegraphics[clip,scale=0.29]{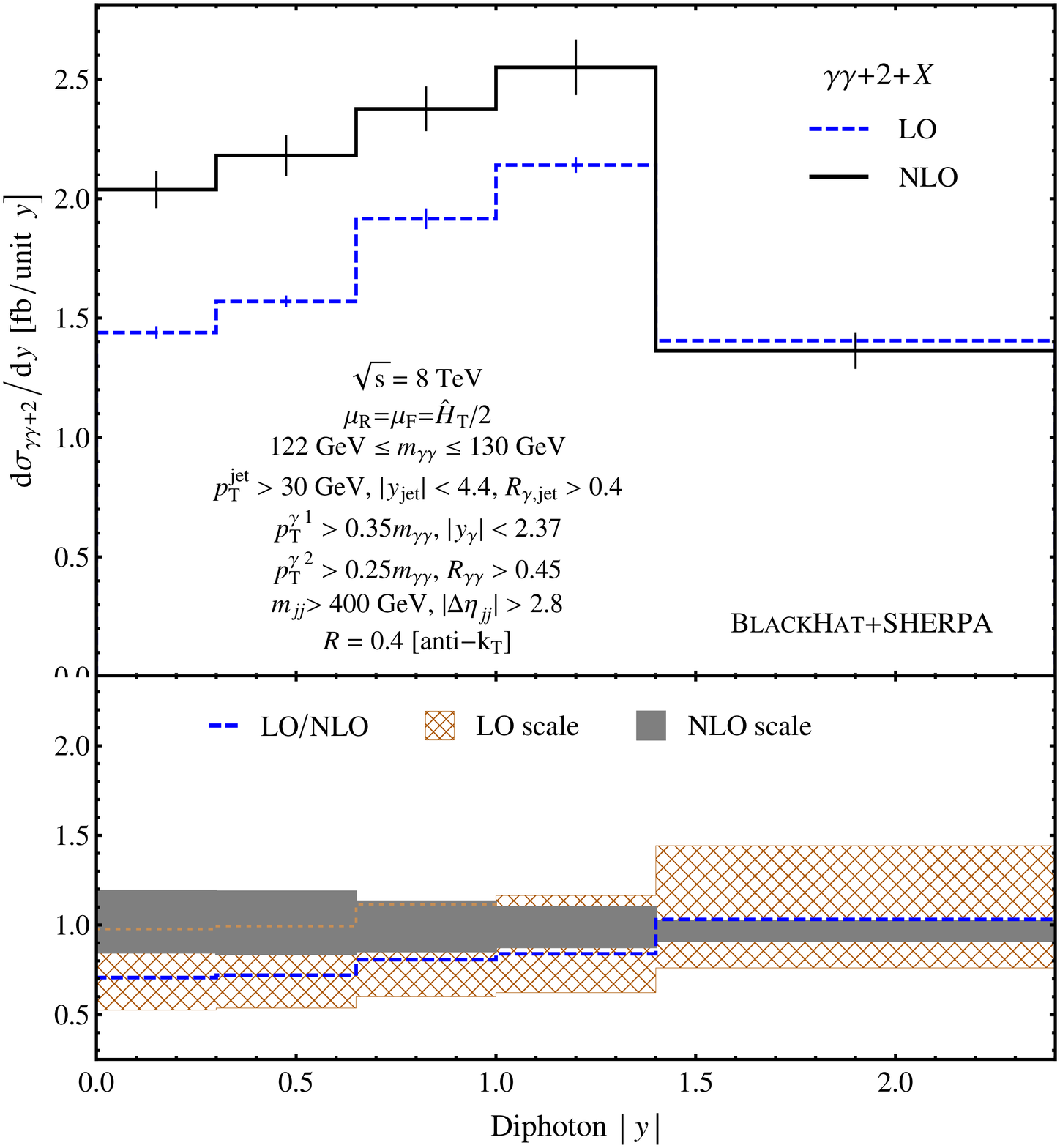}
\end{minipage}
\end{center}
\tighten
\caption{\tightcaption
The distribution of the absolute value of the diphoton rapidity in \YYjj-jet
production.  The left plot shows the distribution for the ATLAS cuts of
\protect\eqn{ATLASCuts}, and the right plot with the VBF cuts of
\protect\eqn{BasicVBFCuts} in addition.  The panels, curves, and bands
are as in \protect\fig{Jet1pTFigure}.}
\label{DiphotonY-ATLASFigure}
\end{figure}

\begin{figure}[!tbhp]
\begin{center}
\begin{minipage}[b]{1.03\linewidth}
\null\hskip -4mm
\includegraphics[clip,scale=0.29]{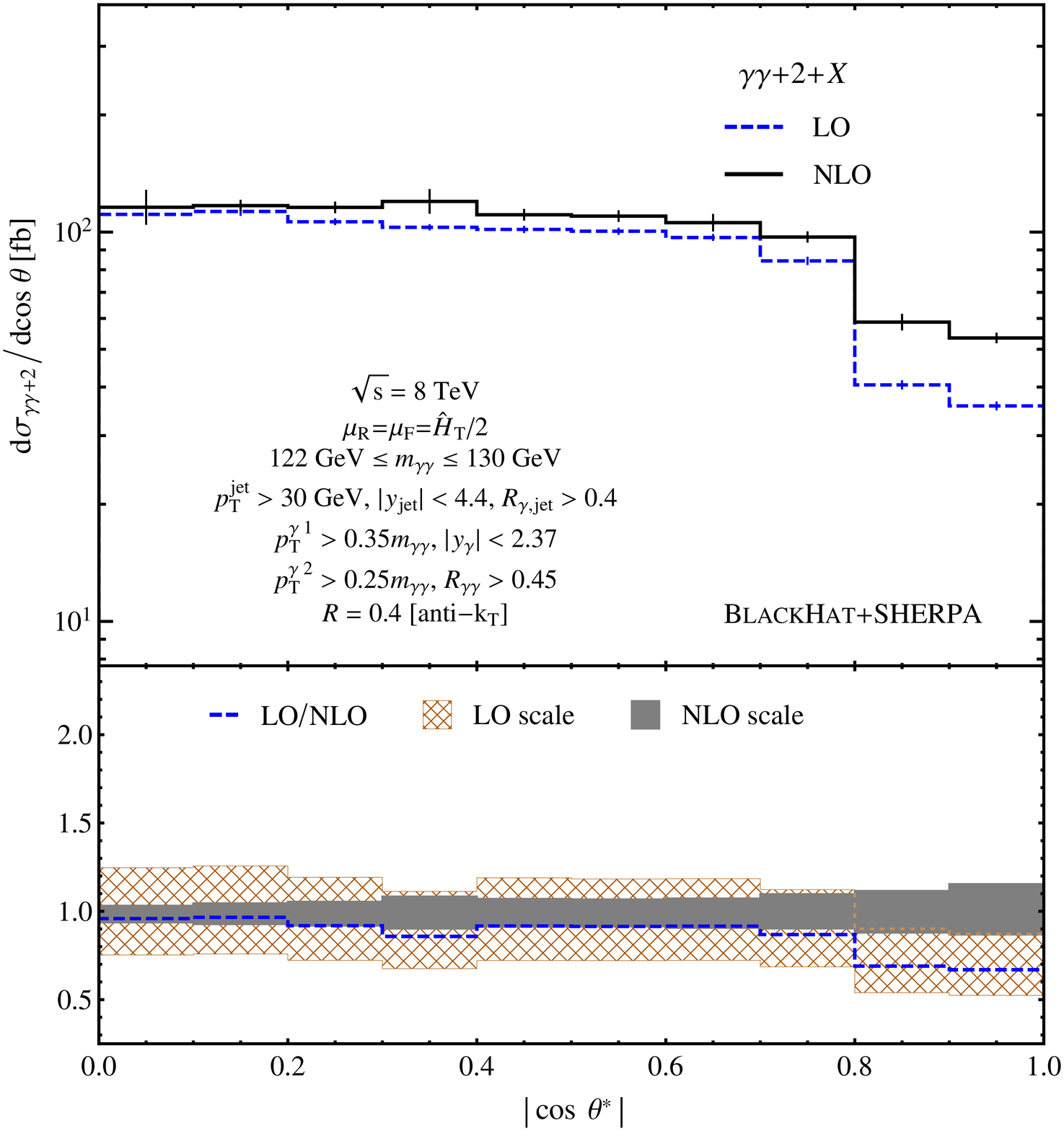} 
\includegraphics[clip,scale=0.29]{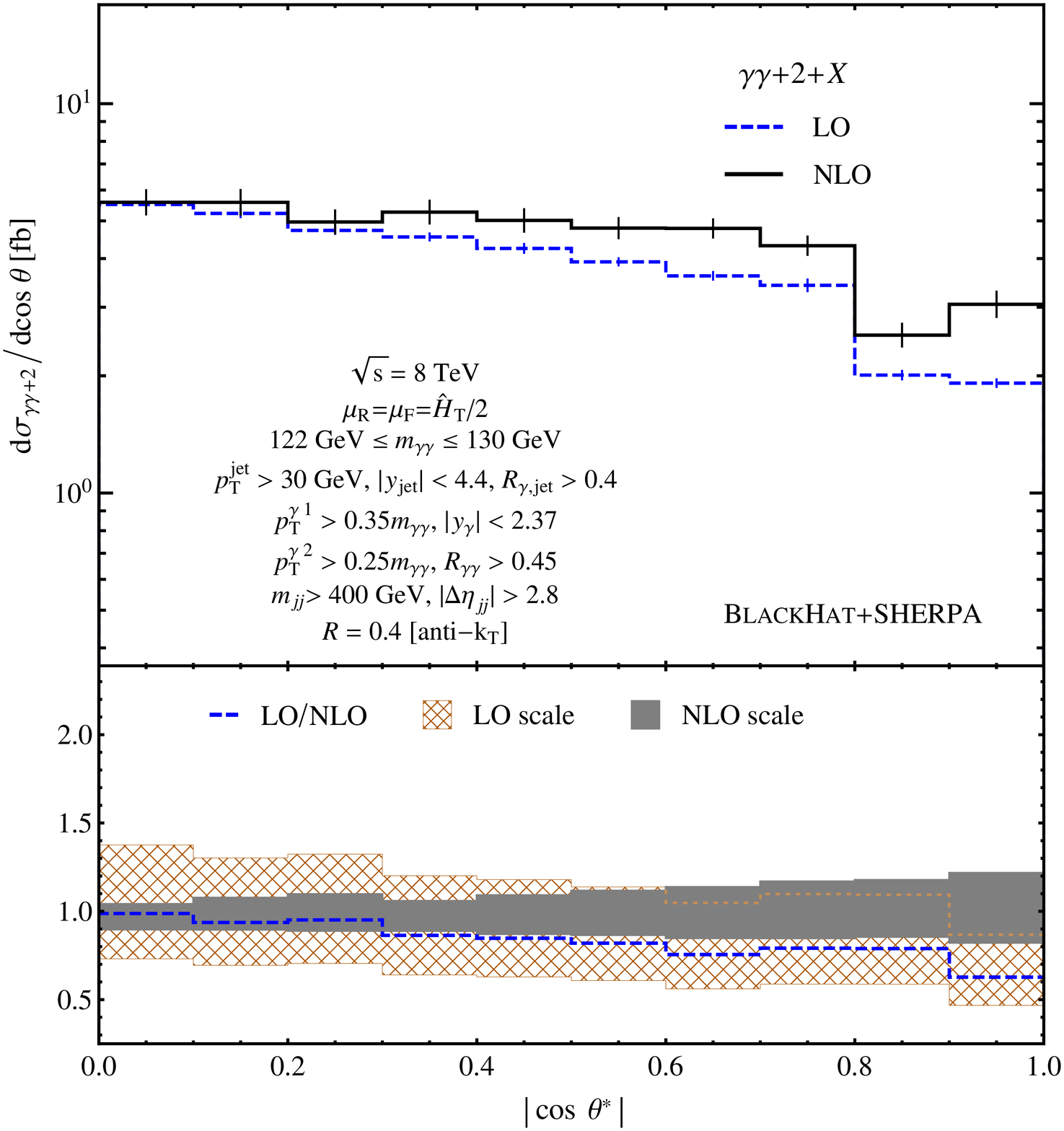}
\end{minipage}
\end{center}
\tighten
\caption{\tightcaption
The distribution of $|\cos\theta^*|$, as defined in \protect\eqn{CosTheta},
in \YYjj-jet production.  The left plot shows the distribution for the ATLAS
cuts of \protect\eqn{ATLASCuts}, and the right plot with the VBF cuts of
\protect\eqn{BasicVBFCuts} in addition.  The panels, curves, and bands are
as in \protect\fig{Jet1pTFigure}.}
\label{CosTheta-ATLASFigure}
\end{figure}

\begin{figure}[!tbhp]
\begin{center}
\begin{minipage}[b]{1.03\linewidth}
\null\hskip -4mm
\includegraphics[clip,scale=0.29]{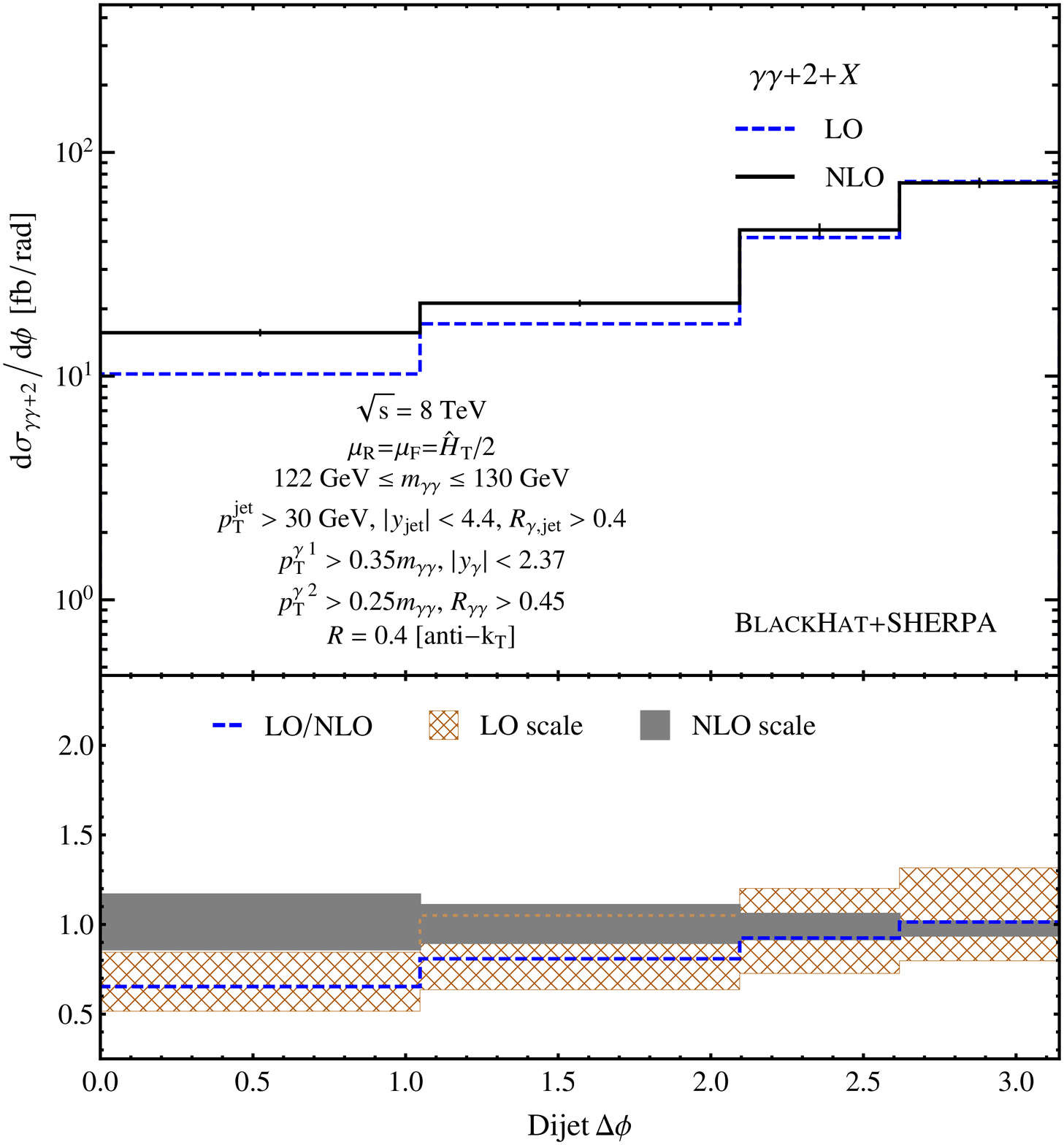} 
\includegraphics[clip,scale=0.29]{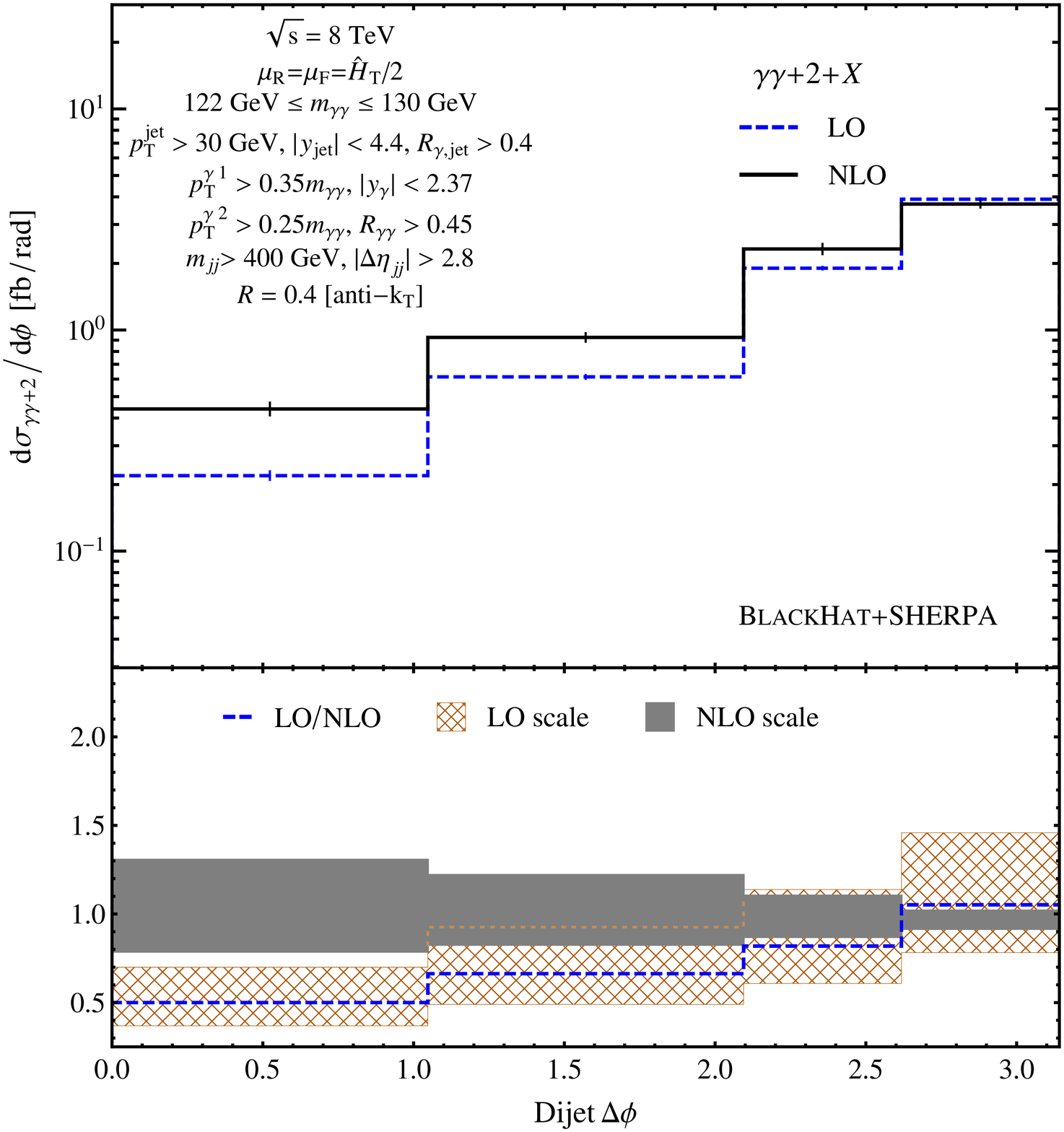}
\end{minipage}
\end{center}
\tighten
\caption{\tightcaption
The distribution of the azimuthal angle between the two leading jets in
\YYjj-jet production.  The left plot shows the distribution for the ATLAS
cuts of \protect\eqn{ATLASCuts}, and the right plot with the VBF cuts of
\protect\eqn{BasicVBFCuts} in addition.  The panels, curves, and bands are
as in \protect\fig{Jet1pTFigure}.}
\label{DijetAngle-ATLASFigure}
\end{figure}
\FloatBarrier

In figs.~\ref{DiphotonPT-ATLASFigure}--\ref{DijetAngle-ATLASFigure}, we
show a series of distributions side-by-side for the ATLAS cuts of
\eqn{ATLASCuts} and for the same cuts with the addition of the VBF cuts of
\eqn{BasicVBFCuts}:
in \fig{DiphotonPT-ATLASFigure}, the transverse momentum of the photon pair;
in \fig{DiphotonY-ATLASFigure}, the absolute value of the rapidity of the
photon pair;
in \fig{CosTheta-ATLASFigure}, the absolute value of $\cos\theta^*$,
as defined in \eqn{CosTheta};
and in \fig{DijetAngle-ATLASFigure}, the azimuthal angle difference between
the leading two jets.
\clearpage

\begin{figure}[tbh]
\begin{center}
\begin{minipage}[b]{1.03\linewidth}
\leftmarg
\includegraphics[clip,scale=0.29]{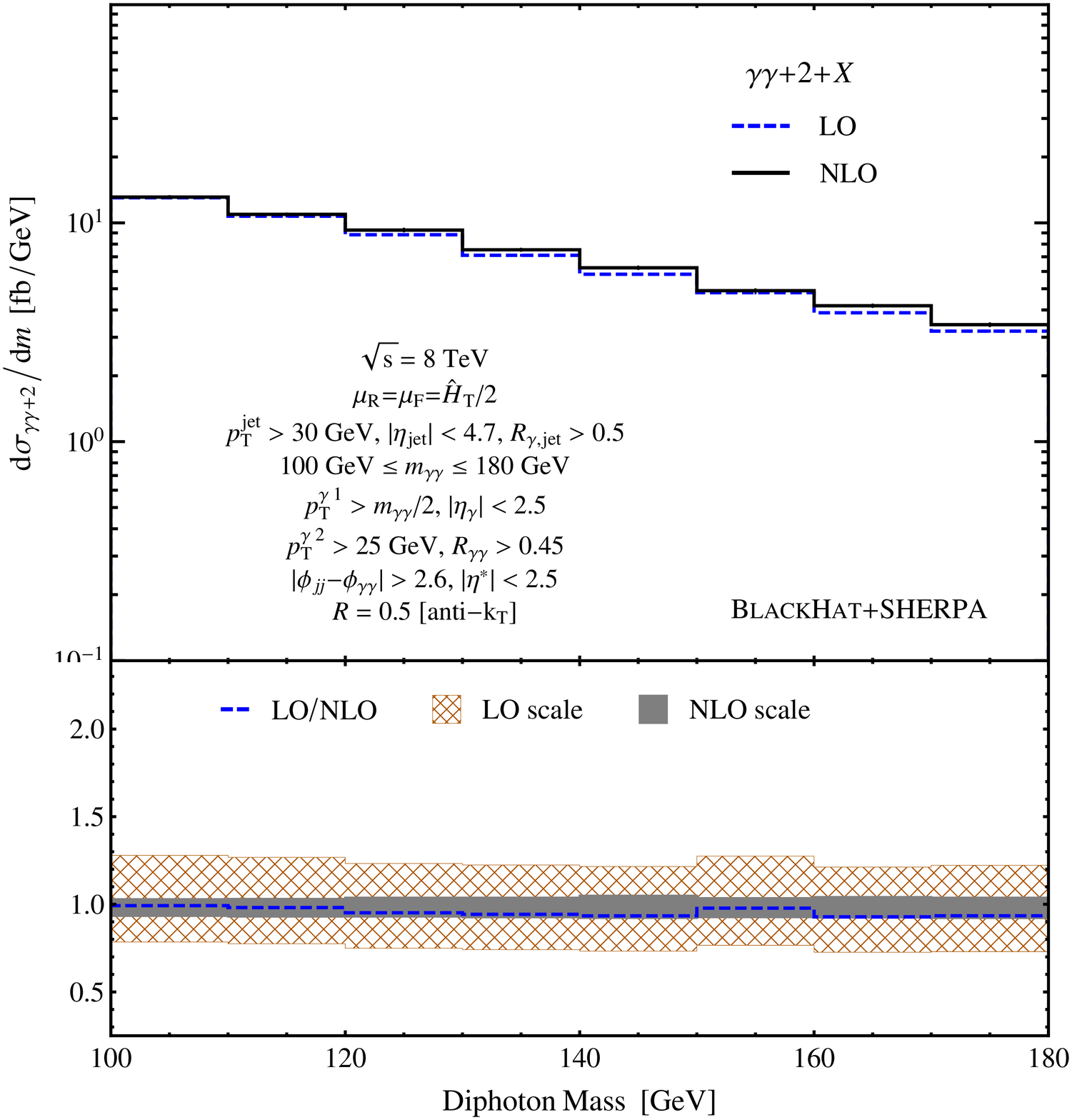} 
\includegraphics[clip,scale=0.29]{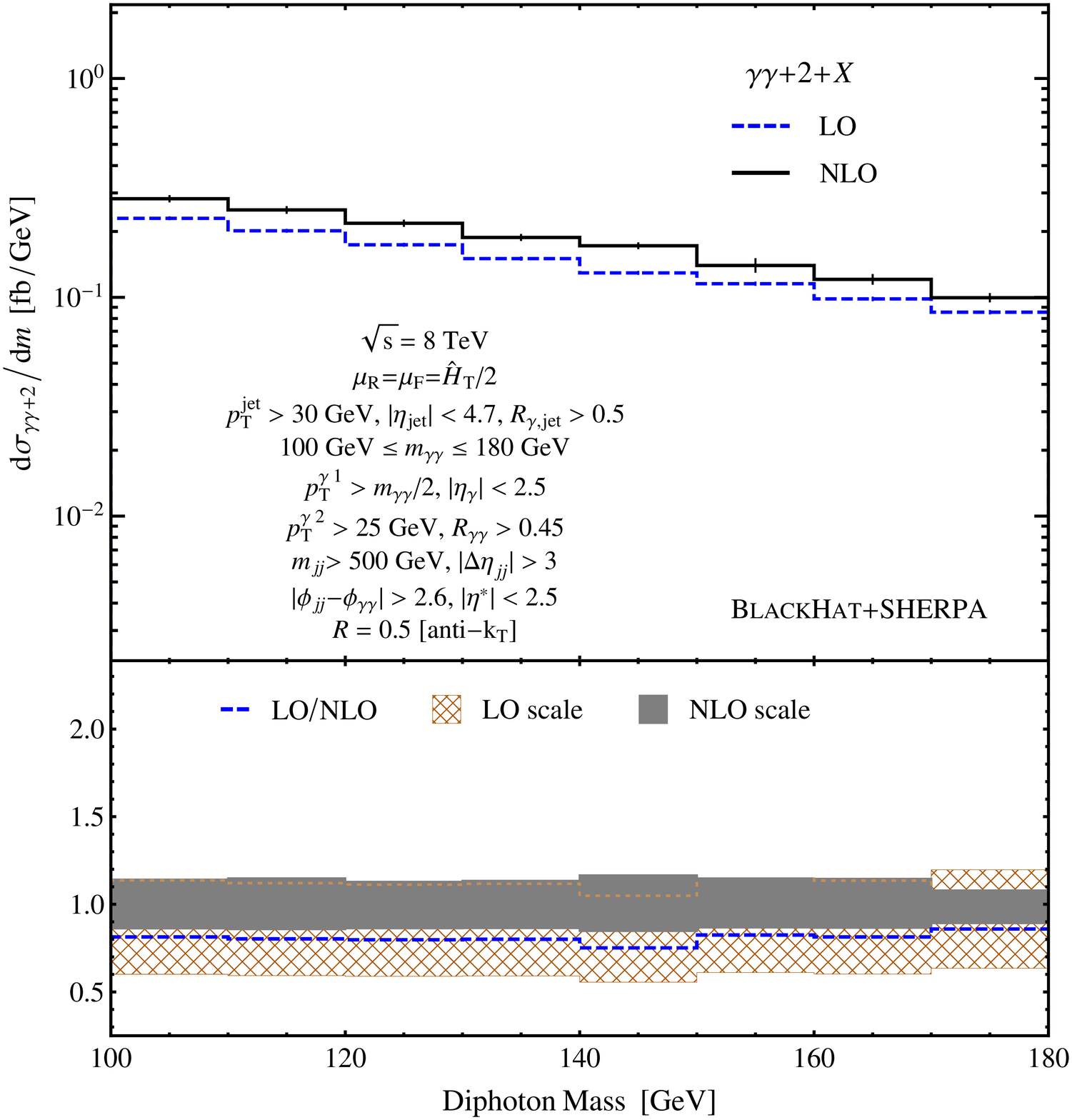}
\end{minipage}
\end{center}
\tighten
\caption{\tightcaption
The photon-pair invariant-mass distribution in \YYjj-jet production.
The left plot shows the distribution for the CMS cuts of
\protect\eqn{CMSCuts}, and the right plot with the VBF cuts of
\protect\eqn{CMSVBFCuts} in addition.  The panels, curves, and bands are
as in \protect\fig{Jet1pTFigure}.}
\label{DiphotonMass-CMSFigure}
\end{figure}

\begin{figure}[tbh]
\begin{center}
\begin{minipage}[b]{1.03\linewidth}
\null\hskip -4mm
\includegraphics[clip,scale=0.29]{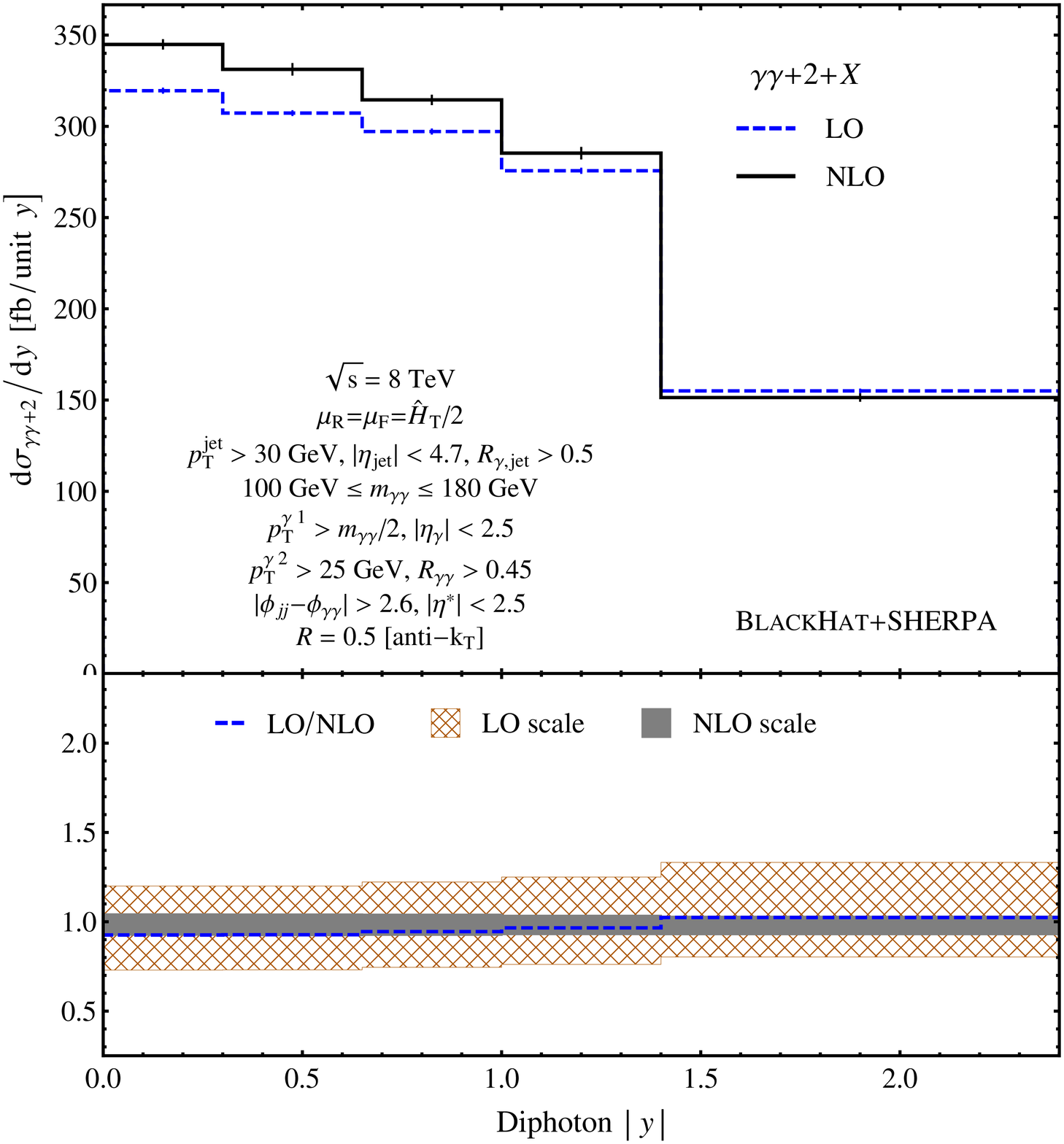} 
\includegraphics[clip,scale=0.29]{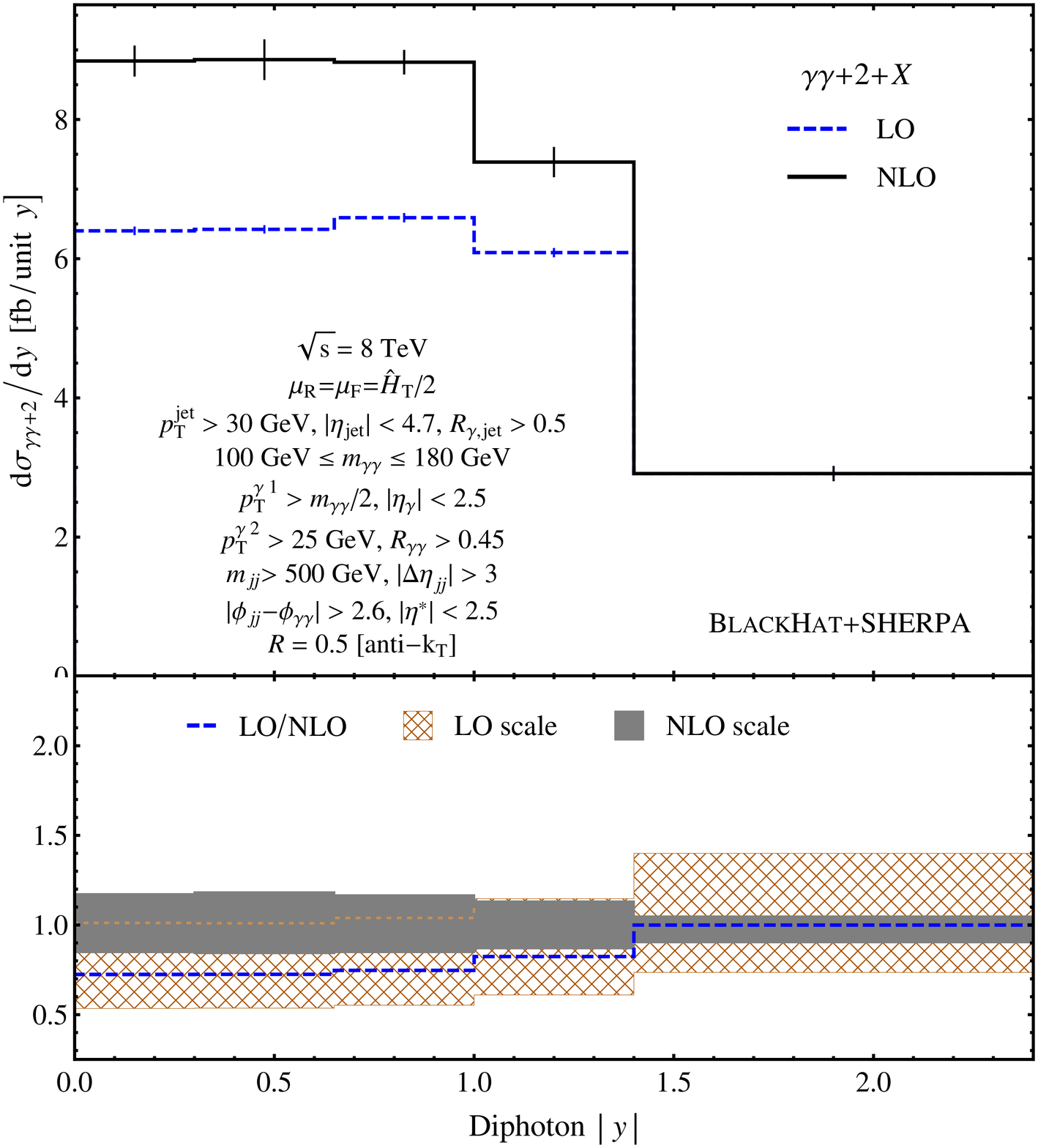}
\end{minipage}
\end{center}
\tighten
\caption{\tightcaption
The distribution of the absolute value of the diphoton rapidity in \YYjj-jet
production.  The left plot shows the distribution for the CMS cuts of
\protect\eqn{CMSCuts}, and the right plot with the VBF cuts of
\protect\eqn{CMSVBFCuts} in addition.  The panels, curves, and bands are
as in \protect\fig{Jet1pTFigure}.}
\label{DiphotonY-CMSFigure}
\end{figure}

\begin{figure}[tbh]
\begin{center}
\begin{minipage}[b]{1.03\linewidth}
\null\hskip -4mm
\includegraphics[clip,scale=0.29]{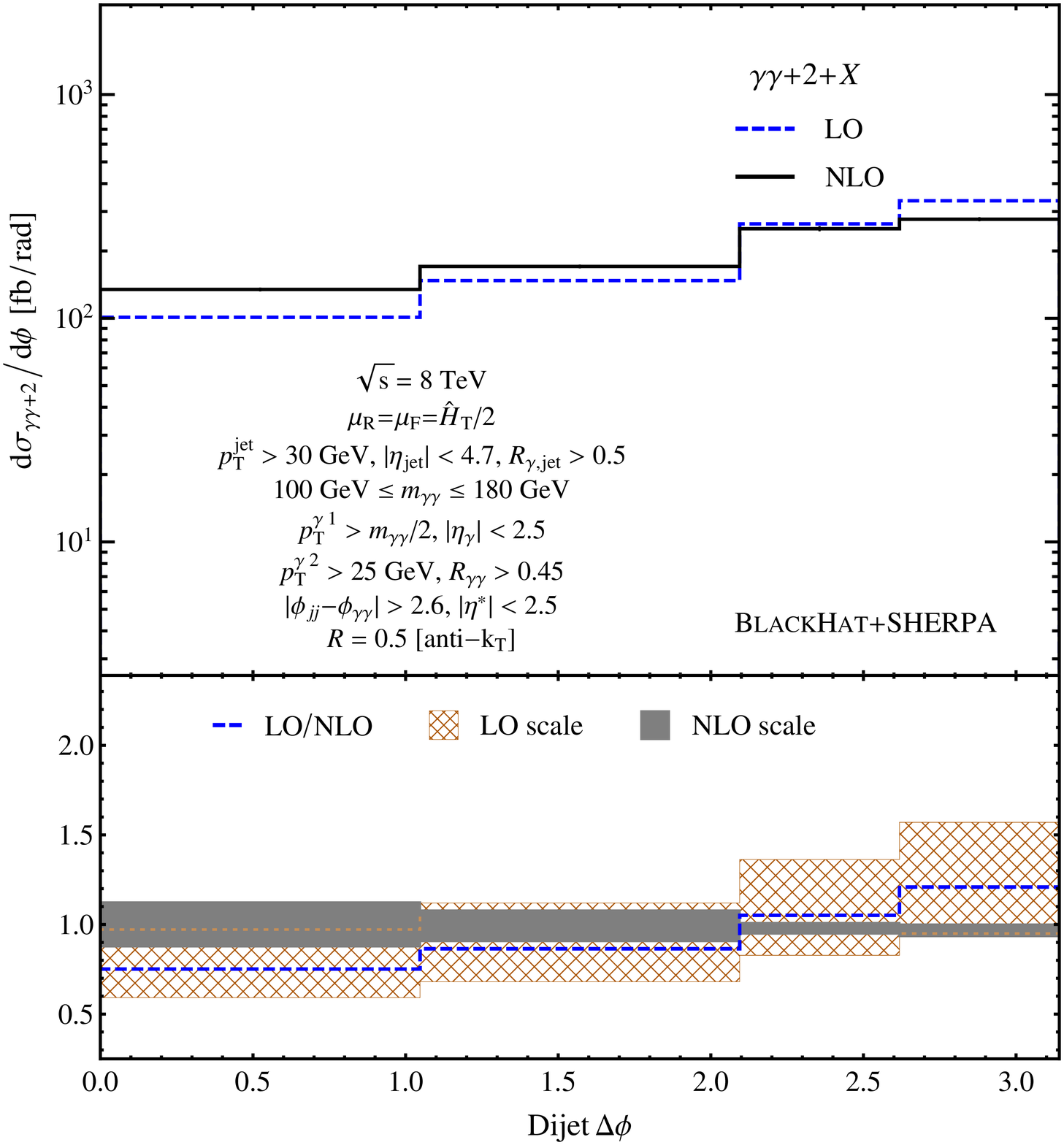} 
\includegraphics[clip,scale=0.29]{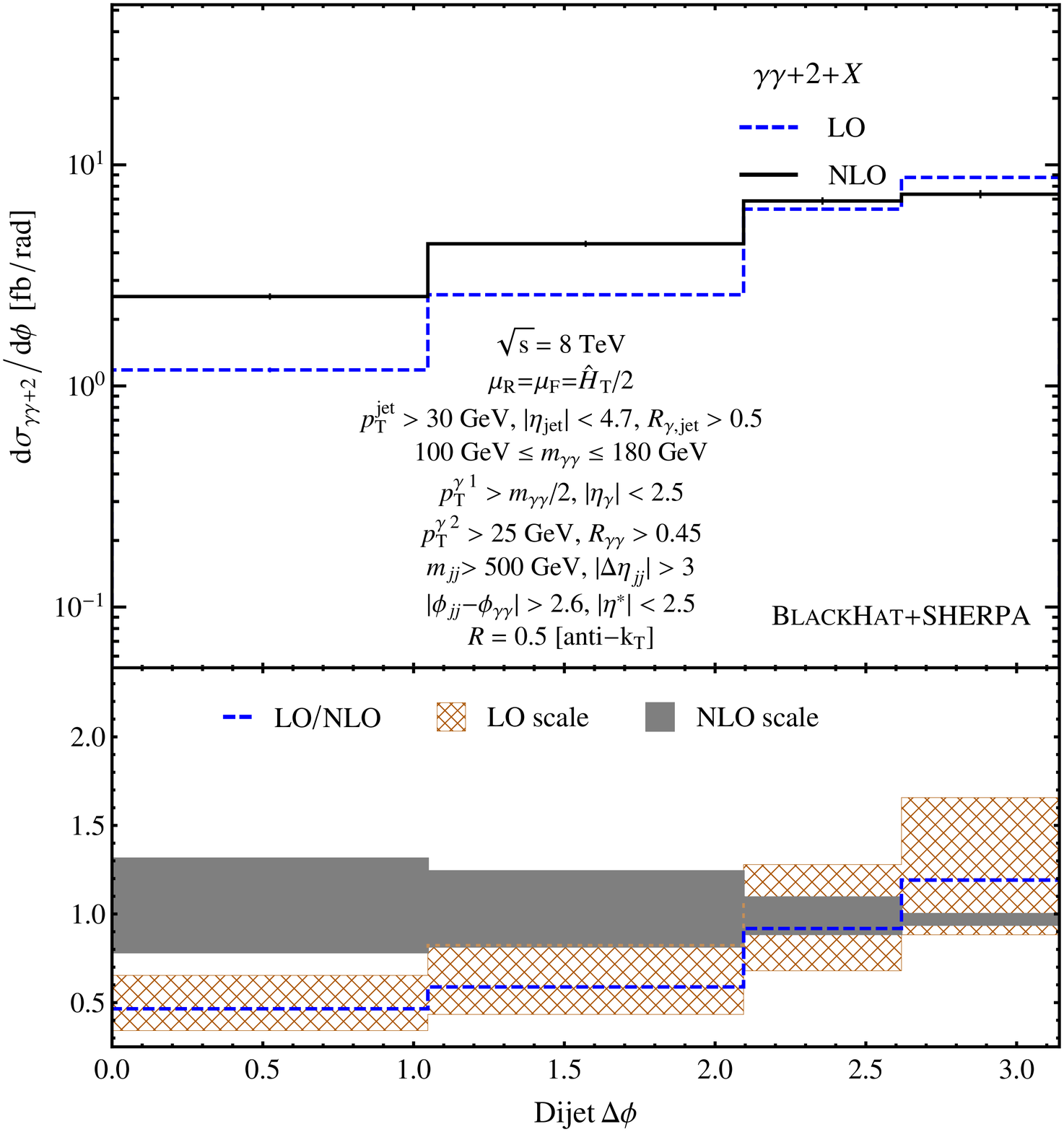}
\end{minipage}
\end{center}
\tighten
\caption{\tightcaption
The distribution of the azimuthal angle between the two leading jets in
\YYjj-jet production.  The left plot shows the distribution for the CMS cuts
of \protect\eqn{CMSCuts}, and the right plot with the VBF cuts of
\protect\eqn{CMSVBFCuts} in addition.  The panels, curves, and bands are
as in \protect\fig{Jet1pTFigure}.}
\label{DijetAngle-CMSFigure}
\end{figure}

\FloatBarrier

In figs.~\ref{DiphotonMass-CMSFigure}--\ref{DijetAngle-CMSFigure}, 
we show three distributions side-by-side for
the CMS cuts of \eqn{CMSCuts} and for the same cuts with the addition of the
VBF cuts of \eqn{CMSVBFCuts}:
in \fig{DiphotonMass-CMSFigure}, the invariant mass distribution of the
photon pair;
in \fig{DiphotonY-CMSFigure}, the absolute value of the rapidity of the
photon pair;
and in \fig{DijetAngle-CMSFigure}, the azimuthal angle difference between
the leading two jets.

The leading-jet transverse-momentum distribution, shown in the left plot
in \fig{Jet1pTFigure}, is fairly typical in many respects.
The upper panel shows the distribution itself.  Because it is steeply falling,
several features are easier to see in the ratio to the central NLO prediction,
shown in the lower panel.  The NLO prediction is somewhat softer than the
LO one; that is, it falls somewhat faster, as seen in the upward slope of the
dashed blue line in the lower panel.  The scale-dependence bands are shown in
hatched orange-brown at LO, and gray at NLO.  The NLO band is narrower than
the LO one throughout, as expected, and is within 10--15\% of the central
value throughout most of the range.  In the lowest $\pT$ bin, the NLO correction
is significant --- the LO prediction is about 30\% lower than the NLO one.
This is accompanied by a wider scale-dependence band in this bin.  The VBF
cuts push the peak of the NLO distribution to around 70~GeV, as shown in the
right plot of \fig{Jet1pTFigure}, from the cut value of 50~GeV.  The lower
bins have larger NLO corrections, and correspondingly larger scale dependence.

We show the same distribution with ATLAS cuts in \fig{Jet1pT-ATLASFigure}.
These cuts flatten the distribution somewhat (note that the plots cut off
at a lower transverse momentum than in \fig{Jet1pTFigure}).  The shape
corrections are again more noticeable after VBF cuts, and the scale
dependence remains large even at NLO in the lowest bins, where the LO
prediction is nearly 50\% lower than the NLO one.

The transverse-momentum distributions of the second jet and of the
leading photon are shown in \fig{Jet2pTFigure} and \fig{Photon1pTFigure},
respectively.  The VBF cuts do not alter the shape of the photon $\pT$
distribution much.  The NLO corrections soften the distribution in a manner
typical for $\pT$ distributions.  The softening is particularly pronounced
for the second-jet $\pT$ distribution after VBF cuts.

The distribution of the dijet invariant mass, shown in the left plot
of \fig{DijetMassFigure}, has a peak around 100~GeV sculpted by the cuts
of \eqn{BasicCuts}.  If we impose VBF cuts, the lower part of the distribution
is cut out, and we are left only with the high-mass tail shown in the right
plot of \fig{DijetMassFigure}.  In the latter case, the LO and NLO
distributions are similar in shape.  The same is true for the distribution
without VBF cuts, in the peak region and above. At low invariant mass,
in contrast, the NLO corrections are large and the NLO scale dependence
remains substantial.

This can be understood as follows. Given the minimum $\pT$s imposed on the
jets, small dijet invariant masses arise primarily from a small angular
separation between the jets.  In this region, the LO matrix element
approaches a collinear factorization limit, where it becomes a lower-point
matrix element, with only three massless objects (two photons and one parton)
in the final state.  Let us consider the real-emission corrections to the LO
process in this region, compared to the real-emission corrections at a
generic point in the LO process's phase space.  The phase space for three
massless final-state objects is more constrained than the one for four
massless final-state objects; and the additional constraints are more
significant than in comparing the phase space for four massless
final-state objects with one for five objects. Accordingly, the additional
emission of a gluon has a relatively larger phase space to fill, so the
additional emission relaxes kinematic constraints in a more substantial
way than at a generic point in phase space.  This is similar to the larger
corrections seen in three-jet production compared to four-jet production,
or in \Wjj-jet production compared to \Wjjj-jet production.  

The distribution of the photon pair's invariant mass, shown in
\fig{DiphotonMassFigure}, has a peak around 100~GeV sculpted by the
cuts of \eqn{BasicCuts}.  In the peak region and above, the distribution
has modest NLO corrections, and its shape is somewhat hardened by VBF cuts.
In these regions, the scale dependence narrows significantly at NLO.
At low invariant mass, in contrast, the NLO corrections are again large
and the NLO scale dependence remains substantial, even more so than for the
dijet mass distribution.  Here this is true whether VBF cuts are imposed or not.
Once again, small invariant masses arise from a small angular separation,
in this case of the photons instead of the jets. In this region, while the
LO matrix element does not factorize (there is no collinear singularity
for a photon pair), the kinematics again resembles that of a lower-point
matrix element, with only three massless objects in the final state.
Once again the kinematic relaxation in the real-emission corrections
is more significant than at a generic point in phase space.
The CMS cuts~(\ref{CMSCuts}) restrict attention to sufficiently large values
of the di-photon invariant mass, shown in \fig{DiphotonMass-CMSFigure},
that the NLO corrections remain modest in magnitude, and do not
alter the shape of the distribution, both before and after VBF cuts.

If we had no jets, or only one jet, in the final state, 
in addition to the pair of photons, then restricting the photon-pair
transverse momentum to small values would impose a strong constraint
on additional radiation; we would expect to see large corrections from a
mismatch between virtual and real-emission contributions there.  With
two jets in the final state, however, such a restriction imposes no
constraint on additional radiation, and the corrections should be small.
This is what we see if we examine the transverse momentum distribution
of the photon pair, shown in \fig{DiphotonPT-ATLASFigure}, both before
and after VBF cuts.  The shape of this distribution has only small corrections
at NLO.  It is influenced by the restriction (suggested by ATLAS)
to a photon-pair mass window around the mass of the Higgs-like boson.
The shape of the photon-pair rapidity distribution, shown subject to
ATLAS cuts and folded over to positive values in \fig{DiphotonY-ATLASFigure},
is similarly unaffected by NLO corrections before VBF cuts; the photon pair
tends to be produced centrally. In contrast, after VBF cuts are applied
the NLO distribution becomes somewhat more central than the LO prediction.
With the CMS cuts, shown in \fig{DiphotonY-CMSFigure}, the corrections
to the shape are similarly modest before VBF cuts, but even more significant
after VBF cuts than for the corresponding ATLAS cuts.

The distribution with respect to the Collins-Soper angle 
$|\cos\theta^*|$ defined in \eqn{CosTheta} is shown in
\fig{CosTheta-ATLASFigure}.  The shapes of these distributions are also
similar at NLO and LO, both before and after VBF cuts.

We show the distribution of the azimuthal angle separation between the two
leading jets, using ATLAS cuts, in \fig{DijetAngle-ATLASFigure}.
The jets are somewhat more decorrelated at NLO, as might be expected
from the addition of radiation.  This effect is much stronger after VBF cuts,
so that at smaller angles ($< 1$~radian), the LO prediction is only half
of the NLO one, and the NLO scale dependence is correspondingly larger. 
The effects are similar when applying CMS cuts, as shown in
\fig{DijetAngle-CMSFigure}, again with a stronger effect after applying
VBF cuts.
\FloatBarrier

\section{Conclusions}
\label{ConclusionSection}

In this paper, we have studied the inclusive production of a photon pair
in association with two jets, at NLO in perturbative QCD.  This final state
is an important background to the study of the Higgs-like
boson~\cite{HiggsBosonExpt} decaying into a photon pair
in the vector-boson fusion production channel.
We have employed a Frixione-style isolation criterion for the photon.  
While this criterion does not correspond precisely to experimental practice, 
given current practice and various uncertainties in traditional cone isolation, 
it is likely to be useful as a theoretical prediction. We have examined the
cross section and a variety of distributions under three different pairs of
cuts. Each pair contains a `standard' set of cuts, corresponding to generic
production of this final state, and an additional set of cuts
restricting the phase space to that corresponding to searches
for Higgs boson production via vector-boson fusion.
We have made the \ntuple{} files~\cite{NTuples} used publicly available,
in process directories {\tt YY2j\/} and {\tt YY2j\_VBF\/},
as explained in \sect{KinematicsSection}.  
(The location of these directories may
be found in {\tt http://blackhat.hepforge.org/trac/wiki/Location\/}.)
One pair of cuts uses fairly generic jet and photon transverse momentum cuts,
while the other two pairs use cuts suggested by CMS and ATLAS, which also
restrict attention to a window in the photon-pair invariant mass surrounding
the Higgs-like boson mass.  In the total cross section and in most parts
of distributions, we find that the NLO scale dependence is reduced to
10--15\%, so that the NLO prediction should be quantitatively reliable.
In some bins of some distributions, the NLO corrections alter the LO
prediction quite substantially, and in these cases the scale dependence
at NLO remains substantially larger.  These features suggest that the
NLO corrections will play an important role in upcoming experimental
analyses of data from the next run of the LHC.

\section*{Acknowledgments}

We thank Joey Huston for discussions about photon isolation and photon
physics.  We also thank Nicolas Greiner and Marco Pieri for
discussions.  FFC and NALP thank the LPNHE for its hospitality, and
FFC acknowledges partial support by ECOS Nord, while the research
reported here was being completed.  This research was supported by the
US Department of Energy under contracts DE--SC0009937 and
DE--AC02--76SF00515.  DAK and NALP's research is supported by the
European Research Council under Advanced Investigator Grant
ERC--AdG--228301.  DM's work was supported by the Research Executive
Agency (REA) of the European Union under the Grant Agreement number
PITN--GA--2010--264564 (LHCPhenoNet). ZB, LJD, SH, HI, DAK and NALP would
also like to thank the Simons Center Geometry and Physics for
hospitality, where this work was finalized.  This research used
resources of Academic Technology Services at UCLA.

\appendix

\section{Tables of Distributions for Basic and VBF Cuts}
\label{BasicTablesAppendix}

\begin{table}[htb]
\vskip .4 cm
\centering
\begin{tabular}{||c|l|l||l|l||}
\hline
\multirow{2}{*}{$\pT$}  &  \multicolumn{2}{|c||}{LO} &  \multicolumn{2}{|c||}{NLO} \\ \cline{2-5}
& \hfil Basic & \hfil VBF & \hfil Basic & \hfil VBF\\ \hline\hline
$40$--$60$ &~$32.13(0.09)^{+9.27}_{-6.71}$ &~$0.611(0.005)^{+0.251}_{-0.163}$~&~$41.0(0.3)^{+4.9}_{-4.6}$~&~$0.99(0.02)^{+0.24}_{-0.18}$~\\ \hline
$60$--$80$ &~$27.58(0.07)^{+8.14}_{-5.85}$ &~$0.933(0.004)^{+0.370}_{-0.244}$~&~$34.3(0.3)^{+3.7}_{-3.7}$~&~$1.36(0.06)^{+0.25}_{-0.22}$~\\ \hline
$80$--$100$ &~$20.09(0.05)^{+6.05}_{-4.32}$ &~$1.078(0.004)^{+0.416}_{-0.276}$~&~$24.4(0.4)^{+2.3}_{-2.5}$~&~$1.29(0.06)^{+0.14}_{-0.16}$~\\ \hline
$100$--$120$ &~$14.82(0.05)^{+4.52}_{-3.22}$ &~$1.059(0.005)^{+0.402}_{-0.269}$~&~$17.0(0.2)^{+1.3}_{-1.6}$~&~$1.18(0.02)^{+0.10}_{-0.13}$~\\ \hline
$120$--$140$ &~$10.72(0.04)^{+3.31}_{-2.35}$ &~$0.883(0.004)^{+0.334}_{-0.223}$~&~$10.8(0.2)^{+0.2}_{-0.7}$~&~$0.90(0.02)^{+0.02}_{-0.08}$~\\ \hline
$140$--$160$ &~$7.44(0.03)^{+2.33}_{-1.65}$ &~$0.665(0.003)^{+0.251}_{-0.168}$~&~$7.8(0.1)^{+0.3}_{-0.6}$~&~$0.66(0.05)^{+0.02}_{-0.06}$~\\ \hline
$160$--$180$ &~$5.15(0.02)^{+1.64}_{-1.15}$ &~$0.471(0.003)^{+0.179}_{-0.120}$~&~$5.09(0.09)^{+0.10}_{-0.34}$~&~$0.42(0.01)^{+0.00}_{-0.03}$~\\ \hline
$180$--$200$ &~$3.56(0.02)^{+1.14}_{-0.80}$ &~$0.324(0.002)^{+0.124}_{-0.083}$~&~$3.51(0.08)^{+0.06}_{-0.24}$~&~$0.31(0.01)^{+0.00}_{-0.02}$~\\ \hline
$200$--$220$ &~$2.51(0.01)^{+0.82}_{-0.57}$ &~$0.224(0.002)^{+0.086}_{-0.058}$~&~$2.52(0.04)^{+0.04}_{-0.17}$~&~$0.188(0.009)^{+0.000}_{-0.012}$~\\ \hline
$220$--$240$ &~$1.81(0.01)^{+0.59}_{-0.42}$ &~$0.156(0.002)^{+0.060}_{-0.040}$~&~$1.72(0.04)^{+0.02}_{-0.11}$~&~$0.146(0.006)^{+0.002}_{-0.011}$~\\ \hline
$240$--$260$ &~$1.32(0.01)^{+0.44}_{-0.31}$ &~$0.108(0.001)^{+0.042}_{-0.028}$~&~$1.29(0.04)^{+0.02}_{-0.09}$~&~$0.088(0.005)^{+0.000}_{-0.006}$~\\ \hline

\hline
\end{tabular}
\caption{The leading-jet transverse-momentum distribution, in fb/GeV, in \YYjj-jet production, as shown
in \fig{Jet1pTFigure}.
}
\label{Jet1pTTable}
\end{table}

\begin{table}[htb]
\vskip .4 cm
\centering
\begin{tabular}{||c|l|l||l|l||}
\hline
\multirow{2}{*}{$\pT$}  &  \multicolumn{2}{|c||}{LO} &  \multicolumn{2}{|c||}{NLO} \\ \cline{2-5}
& \hfil Basic & \hfil VBF & \hfil Basic & \hfil VBF\\ \hline\hline
$25$--$45$ &~$71.8(0.1)^{+21.3}_{-15.3}$ &~$2.332(0.009)^{+0.922}_{-0.608}$~&~$83.5(0.6)^{+7.1}_{-8.0}$~&~$2.86(0.06)^{+0.39}_{-0.39}$~\\ \hline
$45$--$65$ &~$28.87(0.06)^{+8.75}_{-6.24}$ &~$1.775(0.005)^{+0.685}_{-0.455}$~&~$35.6(0.3)^{+3.5}_{-3.7}$~&~$2.12(0.03)^{+0.20}_{-0.26}$~\\ \hline
$65$--$85$ &~$12.77(0.03)^{+3.95}_{-2.80}$ &~$1.182(0.004)^{+0.446}_{-0.298}$~&~$15.1(0.1)^{+1.4}_{-1.5}$~&~$1.32(0.02)^{+0.11}_{-0.15}$~\\ \hline
$85$--$105$ &~$6.58(0.02)^{+2.06}_{-1.46}$ &~$0.664(0.002)^{+0.250}_{-0.168}$~&~$7.50(0.07)^{+0.51}_{-0.69}$~&~$0.70(0.01)^{+0.04}_{-0.07}$~\\ \hline
$105$--$125$ &~$3.78(0.02)^{+1.20}_{-0.84}$ &~$0.346(0.002)^{+0.132}_{-0.088}$~&~$4.22(0.04)^{+0.28}_{-0.38}$~&~$0.333(0.009)^{+0.006}_{-0.027}$~\\ \hline
$125$--$145$ &~$2.32(0.01)^{+0.74}_{-0.52}$ &~$0.191(0.001)^{+0.074}_{-0.049}$~&~$2.44(0.04)^{+0.06}_{-0.19}$~&~$0.173(0.005)^{+0.003}_{-0.013}$~\\ \hline
$145$--$165$ &~$1.50(0.01)^{+0.49}_{-0.34}$ &~$0.1119(0.0009)^{+0.0436}_{-0.0290}$~&~$1.58(0.03)^{+0.07}_{-0.13}$~&~$0.100(0.004)^{+0.001}_{-0.007}$~\\ \hline
$165$--$185$ &~$0.996(0.007)^{+0.325}_{-0.227}$ &~$0.0670(0.0007)^{+0.0263}_{-0.0175}$~&~$1.02(0.02)^{+0.03}_{-0.08}$~&~$0.049(0.005)^{+0.001}_{-0.008}$~\\ \hline
$185$--$205$ &~$0.700(0.007)^{+0.230}_{-0.161}$ &~$0.0423(0.0005)^{+0.0167}_{-0.0111}$~&~$0.75(0.03)^{+0.04}_{-0.06}$~&~$0.038(0.002)^{+0.000}_{-0.002}$~\\ \hline
$205$--$225$ &~$0.497(0.005)^{+0.165}_{-0.115}$ &~$0.0274(0.0004)^{+0.0109}_{-0.0072}$~&~$0.46(0.03)^{+0.01}_{-0.03}$~&~$0.022(0.002)^{+0.000}_{-0.002}$~\\ \hline
$225$--$245$ &~$0.356(0.004)^{+0.119}_{-0.083}$ &~$0.0188(0.0004)^{+0.0075}_{-0.0050}$~&~$0.32(0.02)^{+0.00}_{-0.01}$~&~$0.014(0.001)^{+0.000}_{-0.001}$~\\ \hline
$245$--$265$ &~$0.256(0.003)^{+0.086}_{-0.060}$ &~$0.0129(0.0003)^{+0.0052}_{-0.0034}$~&~$0.24(0.02)^{+0.00}_{-0.01}$~&~$0.007(0.001)^{+0.001}_{-0.003}$~\\ \hline

\hline
\end{tabular}
\caption{The second-jet transverse-momentum distribution in fb/GeV in \YYjj-jet production, as shown
in \fig{Jet2pTFigure}.
}
\label{Jet2pTTable}
\end{table}

\begin{table}[ht]
\vskip .4 cm
\centering
\begin{tabular}{||c|l|l||l|l||}
\hline
\multirow{2}{*}{$\pT$}  &  \multicolumn{2}{|c||}{LO} &  \multicolumn{2}{|c||}{NLO} \\ \cline{2-5}
& \hfil Basic & \hfil VBF & \hfil Basic & \hfil VBF\\ \hline\hline
$50$--$70$ &~$49.85(0.09)^{+14.50}_{-10.46}$ &~$2.238(0.006)^{+0.862}_{-0.573}$~&~$60.1(0.3)^{+5.7}_{-6.0}$~&~$2.69(0.03)^{+0.31}_{-0.35}$~\\ \hline
$70$--$90$ &~$29.33(0.07)^{+8.76}_{-6.27}$ &~$1.436(0.005)^{+0.555}_{-0.368}$~&~$35.5(0.3)^{+3.4}_{-3.6}$~&~$1.73(0.04)^{+0.20}_{-0.22}$~\\ \hline
$90$--$110$ &~$18.47(0.05)^{+5.63}_{-4.01}$ &~$0.947(0.004)^{+0.367}_{-0.244}$~&~$21.6(0.4)^{+1.9}_{-2.1}$~&~$1.07(0.02)^{+0.11}_{-0.13}$~\\ \hline
$110$--$130$ &~$11.65(0.04)^{+3.61}_{-2.56}$ &~$0.634(0.003)^{+0.246}_{-0.163}$~&~$13.0(0.2)^{+0.8}_{-1.1}$~&~$0.73(0.02)^{+0.06}_{-0.08}$~\\ \hline
$130$--$150$ &~$7.31(0.04)^{+2.31}_{-1.63}$ &~$0.444(0.003)^{+0.172}_{-0.114}$~&~$8.02(0.09)^{+0.50}_{-0.74}$~&~$0.48(0.01)^{+0.03}_{-0.05}$~\\ \hline
$150$--$170$ &~$4.68(0.02)^{+1.50}_{-1.06}$ &~$0.308(0.003)^{+0.120}_{-0.080}$~&~$5.0(0.1)^{+0.2}_{-0.4}$~&~$0.33(0.02)^{+0.02}_{-0.03}$~\\ \hline
$170$--$190$ &~$3.02(0.02)^{+0.99}_{-0.69}$ &~$0.219(0.002)^{+0.085}_{-0.057}$~&~$3.22(0.05)^{+0.14}_{-0.28}$~&~$0.216(0.008)^{+0.008}_{-0.018}$~\\ \hline
$190$--$210$ &~$2.06(0.02)^{+0.68}_{-0.47}$ &~$0.154(0.002)^{+0.060}_{-0.040}$~&~$2.14(0.03)^{+0.08}_{-0.17}$~&~$0.11(0.03)^{+0.00}_{-0.01}$~\\ \hline
$210$--$230$ &~$1.37(0.01)^{+0.46}_{-0.32}$ &~$0.110(0.001)^{+0.043}_{-0.028}$~&~$1.43(0.03)^{+0.04}_{-0.11}$~&~$0.097(0.006)^{+0.000}_{-0.006}$~\\ \hline

\hline
\end{tabular}
\caption{The leading-photon transverse-momentum distribution in fb/GeV in \YYjj-jet production, as shown 
in \fig{Photon1pTFigure}.
}
\label{Photon1pTTable}
\end{table}

\begin{table}[ht]
\vskip .4 cm
\centering
\begin{tabular}{||c|l|l||l|l||}
\hline
\multirow{2}{*}{$m_{\gamma\gamma}$}  &  \multicolumn{2}{|c||}{LO} &  \multicolumn{2}{|c||}{NLO} \\ \cline{2-5}
& \hfil Basic & \hfil VBF & \hfil Basic & \hfil VBF\\ \hline\hline
$0$--$20$ &~$0.117(0.004)^{+0.034}_{-0.025}$ &~$0.0052(0.0003)^{+0.0020}_{-0.0013}$~&~$0.23(0.01)^{+0.05}_{-0.04}$~&~$0.008(0.002)^{+0.003}_{-0.002}$~\\ \hline
$20$--$40$ &~$4.33(0.03)^{+1.27}_{-0.92}$ &~$0.220(0.002)^{+0.085}_{-0.056}$~&~$7.9(0.4)^{+1.7}_{-1.3}$~&~$0.365(0.007)^{+0.083}_{-0.066}$~\\ \hline
$40$--$60$ &~$7.94(0.03)^{+2.35}_{-1.69}$ &~$0.395(0.003)^{+0.153}_{-0.102}$~&~$11.6(0.2)^{+1.9}_{-1.5}$~&~$0.566(0.009)^{+0.111}_{-0.091}$~\\ \hline
$60$--$80$ &~$13.57(0.05)^{+3.98}_{-2.86}$ &~$0.616(0.003)^{+0.240}_{-0.159}$~&~$17.0(0.1)^{+1.9}_{-1.9}$~&~$0.80(0.01)^{+0.12}_{-0.11}$~\\ \hline
$80$--$100$ &~$22.31(0.07)^{+6.50}_{-4.69}$ &~$0.944(0.004)^{+0.370}_{-0.245}$~&~$24.7(0.3)^{+1.7}_{-2.1}$~&~$1.15(0.03)^{+0.13}_{-0.15}$~\\ \hline
$100$--$120$ &~$20.92(0.06)^{+6.17}_{-4.44}$ &~$0.897(0.004)^{+0.351}_{-0.232}$~&~$23.0(0.2)^{+1.4}_{-1.9}$~&~$1.01(0.02)^{+0.09}_{-0.12}$~\\ \hline
$120$--$140$ &~$15.76(0.05)^{+4.73}_{-3.38}$ &~$0.706(0.004)^{+0.276}_{-0.183}$~&~$17.2(0.2)^{+1.0}_{-1.4}$~&~$0.82(0.01)^{+0.08}_{-0.10}$~\\ \hline
$140$--$160$ &~$11.40(0.04)^{+3.47}_{-2.47}$ &~$0.548(0.003)^{+0.213}_{-0.141}$~&~$12.5(0.1)^{+0.8}_{-1.1}$~&~$0.60(0.01)^{+0.05}_{-0.07}$~\\ \hline
$160$--$180$ &~$8.18(0.03)^{+2.52}_{-1.79}$ &~$0.419(0.003)^{+0.162}_{-0.108}$~&~$8.96(0.09)^{+0.51}_{-0.76}$~&~$0.46(0.02)^{+0.03}_{-0.05}$~\\ \hline
$180$--$200$ &~$5.96(0.03)^{+1.86}_{-1.32}$ &~$0.333(0.003)^{+0.128}_{-0.085}$~&~$6.6(0.2)^{+0.4}_{-0.5}$~&~$0.354(0.009)^{+0.024}_{-0.038}$~\\ \hline
$200$--$220$ &~$4.44(0.02)^{+1.40}_{-0.99}$ &~$0.261(0.003)^{+0.100}_{-0.067}$~&~$4.97(0.07)^{+0.33}_{-0.46}$~&~$0.284(0.006)^{+0.021}_{-0.030}$~\\ \hline
$220$--$240$ &~$3.31(0.02)^{+1.05}_{-0.74}$ &~$0.211(0.002)^{+0.080}_{-0.054}$~&~$3.63(0.04)^{+0.21}_{-0.31}$~&~$0.18(0.03)^{+0.00}_{-0.01}$~\\ \hline
$240$--$260$ &~$2.56(0.02)^{+0.82}_{-0.58}$ &~$0.175(0.002)^{+0.066}_{-0.044}$~&~$2.92(0.04)^{+0.20}_{-0.28}$~&~$0.184(0.008)^{+0.006}_{-0.018}$~\\ \hline
$260$--$280$ &~$1.97(0.01)^{+0.64}_{-0.45}$ &~$0.145(0.001)^{+0.055}_{-0.037}$~&~$2.26(0.03)^{+0.13}_{-0.20}$~&~$0.145(0.005)^{+0.003}_{-0.011}$~\\ \hline
$280$--$300$ &~$1.54(0.01)^{+0.50}_{-0.35}$ &~$0.123(0.001)^{+0.046}_{-0.031}$~&~$1.78(0.03)^{+0.11}_{-0.17}$~&~$0.118(0.005)^{+0.002}_{-0.009}$~\\ \hline

\hline
\end{tabular}
\caption{The photon-pair invariant-mass distribution in fb/GeV in \YYjj-jet production, as shown
in \fig{DiphotonMassFigure}.
}
\label{DiphotonMassTable}
\end{table}

\begin{table}[ht]
\vskip .4 cm
\centering
\begin{tabular}{||c|l||l||}
\hline
$m_{jj}$  &  LO &  NLO \\ \hline\hline
$0$--$20$ &~$0.67(0.02)^{+0.18}_{-0.13}$~&~$0.95(0.08)^{+0.14}_{-0.11}$~\\ \hline
$20$--$40$ &~$5.26(0.04)^{+1.46}_{-1.07}$~&~$8.4(0.1)^{+1.5}_{-1.2}$~\\ \hline
$40$--$60$ &~$6.08(0.03)^{+1.71}_{-1.25}$~&~$8.9(0.1)^{+1.3}_{-1.1}$~\\ \hline
$60$--$80$ &~$9.53(0.04)^{+2.63}_{-1.93}$~&~$12.1(0.2)^{+1.4}_{-1.3}$~\\ \hline
$80$--$100$ &~$12.53(0.04)^{+3.47}_{-2.55}$~&~$15.5(0.3)^{+1.7}_{-1.6}$~\\ \hline
$100$--$120$ &~$12.70(0.04)^{+3.58}_{-2.61}$~&~$14.5(0.3)^{+1.0}_{-1.2}$~\\ \hline
$120$--$140$ &~$11.63(0.04)^{+3.35}_{-2.43}$~&~$12.9(0.1)^{+0.8}_{-1.1}$~\\ \hline
$140$--$160$ &~$10.08(0.04)^{+2.95}_{-2.13}$~&~$11.5(0.1)^{+0.9}_{-1.0}$~\\ \hline
$160$--$180$ &~$8.59(0.03)^{+2.56}_{-1.84}$~&~$9.6(0.1)^{+0.6}_{-0.9}$~\\ \hline
$180$--$200$ &~$7.28(0.03)^{+2.21}_{-1.58}$~&~$8.6(0.3)^{+0.7}_{-0.8}$~\\ \hline
$200$--$220$ &~$6.20(0.03)^{+1.91}_{-1.36}$~&~$6.4(0.3)^{+0.3}_{-0.5}$~\\ \hline
$220$--$240$ &~$5.28(0.03)^{+1.65}_{-1.17}$~&~$5.7(0.1)^{+0.3}_{-0.5}$~\\ \hline
$240$--$260$ &~$4.41(0.02)^{+1.39}_{-0.98}$~&~$5.0(0.1)^{+0.4}_{-0.5}$~\\ \hline
$260$--$280$ &~$3.80(0.02)^{+1.21}_{-0.85}$~&~$4.2(0.1)^{+0.3}_{-0.4}$~\\ \hline
$280$--$300$ &~$3.27(0.02)^{+1.06}_{-0.74}$~&~$3.3(0.1)^{+0.1}_{-0.3}$~\\ \hline

\hline
\end{tabular}
\caption{The dijet invariant-mass distribution in fb/GeV for standard cuts in \YYjj-jet production, as shown
in the left plot in \fig{DijetMassFigure}.
}
\label{DijetMassTable}
\end{table}

\begin{table}[ht]
\vskip .4 cm
\centering
\begin{tabular}{||c|l||l||}
\hline
$m_{jj}$  &  LO &  NLO \\ \hline\hline
$400$--$420$ &~$0.603(0.004)^{+0.220}_{-0.149}$~&~$0.67(0.06)^{+0.05}_{-0.07}$~\\ \hline
$420$--$440$ &~$0.545(0.003)^{+0.200}_{-0.135}$~&~$0.68(0.04)^{+0.08}_{-0.09}$~\\ \hline
$440$--$460$ &~$0.490(0.003)^{+0.181}_{-0.122}$~&~$0.55(0.03)^{+0.04}_{-0.06}$~\\ \hline
$460$--$480$ &~$0.453(0.003)^{+0.169}_{-0.114}$~&~$0.55(0.03)^{+0.07}_{-0.07}$~\\ \hline
$480$--$500$ &~$0.405(0.003)^{+0.152}_{-0.102}$~&~$0.43(0.03)^{+0.02}_{-0.04}$~\\ \hline
$500$--$520$ &~$0.363(0.002)^{+0.137}_{-0.092}$~&~$0.43(0.02)^{+0.04}_{-0.05}$~\\ \hline
$520$--$540$ &~$0.331(0.002)^{+0.125}_{-0.084}$~&~$0.35(0.02)^{+0.01}_{-0.03}$~\\ \hline
$540$--$560$ &~$0.303(0.002)^{+0.115}_{-0.077}$~&~$0.36(0.02)^{+0.03}_{-0.04}$~\\ \hline
$560$--$580$ &~$0.275(0.002)^{+0.106}_{-0.070}$~&~$0.33(0.01)^{+0.04}_{-0.04}$~\\ \hline
$580$--$600$ &~$0.247(0.002)^{+0.095}_{-0.063}$~&~$0.27(0.01)^{+0.01}_{-0.03}$~\\ \hline
$600$--$620$ &~$0.228(0.002)^{+0.088}_{-0.059}$~&~$0.28(0.02)^{+0.03}_{-0.03}$~\\ \hline

\hline
\end{tabular}
\caption{The dijet invariant-mass distribution in fb/GeV for VBF cuts in \YYjj-jet production, as shown
in the right plot in \fig{DijetMassFigure}.
}
\label{DijetMassVBFTable}
\end{table}

In this appendix, we provide tables for the kinematical
distributions displayed and discussed in \sect{ResultsSection}.
All tables show differential cross sections at both LO and NLO, with numerical 
integration uncertainties given in parentheses, and scale-dependence bands
indicated by super- and subscripts.  For distributions in dimensionful
variables, the variables are given in GeV, and the units for the
distributions are femtobarns per GeV.  We display results both with the
cuts of \eqn{BasicCuts}, shown in columns marked with `Basic', and these cuts
supplemented by the VBF cuts of \eqn{BasicVBFCuts}, shown in columns
marked with `VBF'.  In \Tab{Jet1pTTable}, we display the leading-jet
transverse-momentum distribution; in \Tab{Jet2pTTable}, the second-jet
transverse-momentum distribution;
in \Tab{Photon1pTTable}, the leading-photon transverse-momentum distribution;
and in \Tab{DiphotonMassTable}, the photon-pair invariant-mass distribution.

The dijet invariant-mass distribution has different ranges for the standard
and VBF cuts; we display the results for the two sets of cuts in separate
tables:  in \Tab{DijetMassTable}, the distribution for the cuts of
\eqn{BasicCuts}, and in \Tab{DijetMassVBFTable}, the distribution with these
cuts supplemented by the VBF cuts of \eqn{BasicVBFCuts}.
\FloatBarrier

\section{Tables of Distributions for ATLAS and VBF Cuts}
\label{ATLASTablesAppendix}

\begin{table}[ht]
\vskip .4 cm
\centering
\begin{tabular}{||c|l|l||l|l||}
\hline
\multirow{2}{*}{$\pT$}  & \multicolumn{2}{|c||}{LO} & \multicolumn{2}{|c||}{NLO}
\\ \cline{2-5}
& \hfil ATLAS & \hfil VBF & \hfil ATLAS & \hfil VBF\\ \hline\hline
$30$--$50$ &~$0.86(0.01)^{+0.24}_{-0.18}$ &~$0.0106(0.0005)^{+0.0044}_{-0.0029}$~&~$0.99(0.04)^{+0.08}_{-0.09}$~&~$0.019(0.002)^{+0.005}_{-0.004}$~\\ \hline
$50$--$70$ &~$1.10(0.01)^{+0.32}_{-0.23}$ &~$0.0275(0.0006)^{+0.0112}_{-0.0073}$~&~$1.31(0.05)^{+0.13}_{-0.14}$~&~$0.041(0.002)^{+0.008}_{-0.007}$~\\ \hline
$70$--$100$ &~$0.743(0.007)^{+0.221}_{-0.158}$ &~$0.0350(0.0005)^{+0.0136}_{-0.0090}$~&~$0.80(0.04)^{+0.04}_{-0.07}$~&~$0.038(0.002)^{+0.003}_{-0.004}$~\\ \hline
$100$--$140$ &~$0.366(0.004)^{+0.111}_{-0.079}$ &~$0.0288(0.0004)^{+0.0110}_{-0.0073}$~&~$0.42(0.01)^{+0.02}_{-0.03}$~&~$0.032(0.001)^{+0.002}_{-0.003}$~\\ \hline

\hline
\end{tabular}
\caption{The leading-jet transverse-momentum distribution in fb/GeV
in \YYjj-jet production, as shown in \fig{Jet1pT-ATLASFigure}.}
\label{Jet1pT-ATLASTable}
\end{table}

\begin{table}[ht]
\vskip .4 cm
\centering
\begin{tabular}{||c|l|l||l|l||}
\hline
\multirow{2}{*}{$\pT$} & \multicolumn{2}{|c||}{LO} & \multicolumn{2}{|c||}{NLO}
 \\ \cline{2-5}
& \hfil ATLAS & \hfil VBF & \hfil ATLAS & \hfil VBF\\ \hline\hline
$0$--$20$ &~$0.465(0.007)^{+0.133}_{-0.096}$ &~$0.0140(0.0003)^{+0.0055}_{-0.0036}$~&~$0.42(0.02)^{+0.00}_{-0.02}$~&~$0.016(0.002)^{+0.002}_{-0.002}$~\\ \hline
$20$--$30$ &~$0.90(0.01)^{+0.26}_{-0.19}$ &~$0.0324(0.0007)^{+0.0127}_{-0.0084}$~&~$0.90(0.06)^{+0.05}_{-0.06}$~&~$0.039(0.003)^{+0.005}_{-0.005}$~\\ \hline
$30$--$40$ &~$0.97(0.02)^{+0.28}_{-0.20}$ &~$0.0363(0.0008)^{+0.0141}_{-0.0093}$~&~$1.09(0.07)^{+0.08}_{-0.10}$~&~$0.039(0.004)^{+0.001}_{-0.004}$~\\ \hline
$40$--$50$ &~$0.97(0.02)^{+0.29}_{-0.21}$ &~$0.0392(0.0009)^{+0.0153}_{-0.0101}$~&~$1.05(0.03)^{+0.07}_{-0.08}$~&~$0.044(0.005)^{+0.005}_{-0.005}$~\\ \hline
$50$--$60$ &~$0.88(0.01)^{+0.26}_{-0.19}$ &~$0.040(0.001)^{+0.016}_{-0.010}$~&~$0.9(0.1)^{+0.0}_{-0.1}$~&~$0.048(0.005)^{+0.003}_{-0.005}$~\\ \hline
$60$--$80$ &~$0.83(0.01)^{+0.25}_{-0.18}$ &~$0.0361(0.0007)^{+0.0141}_{-0.0093}$~&~$0.93(0.03)^{+0.06}_{-0.08}$~&~$0.039(0.002)^{+0.003}_{-0.005}$~\\ \hline
$80$--$100$ &~$0.573(0.008)^{+0.173}_{-0.123}$ &~$0.0275(0.0006)^{+0.0107}_{-0.0071}$~&~$0.70(0.01)^{+0.07}_{-0.07}$~&~$0.033(0.002)^{+0.004}_{-0.004}$~\\ \hline
$100$--$200$ &~$0.133(0.001)^{+0.041}_{-0.029}$ &~$0.0078(0.0001)^{+0.0030}_{-0.0020}$~&~$0.176(0.003)^{+0.023}_{-0.022}$~&~$0.0101(0.0004)^{+0.0016}_{-0.0016}$~\\ \hline

\hline
\end{tabular}
\caption{The photon-pair transverse-momentum distribution, in fb/GeV, 
in \YYjj-jet production, as shown in \fig{DiphotonPT-ATLASFigure}.}
\label{DiphotonPT-ATLASTable}
\end{table}

\begin{table}[ht]
\vskip .4 cm
\centering
\begin{tabular}{||c|l|l||l|l||}
\hline
\multirow{2}{*}{$|y|$} & \multicolumn{2}{|c||}{LO} & \multicolumn{2}{|c||}{NLO}
\\ \cline{2-5}
& \hfil ATLAS & \hfil VBF & \hfil ATLAS & \hfil VBF\\ \hline\hline
$0$--$0.3$ &~$46.0(0.5)^{+13.5}_{-9.7}$ &~$1.44(0.02)^{+0.55}_{-0.37}$~&~$53.8(0.9)^{+4.1}_{-4.9}$~&~$2.04(0.07)^{+0.39}_{-0.31}$~\\ \hline
$0.3$--$0.65$ &~$46.1(0.4)^{+13.5}_{-9.7}$ &~$1.57(0.02)^{+0.60}_{-0.40}$~&~$54(2)^{+4}_{-5}$~&~$2.18(0.08)^{+0.41}_{-0.35}$~\\ \hline
$0.65$--$1$ &~$47.9(0.5)^{+14.1}_{-10.1}$ &~$1.92(0.04)^{+0.73}_{-0.49}$~&~$56(1)^{+5}_{-5}$~&~$2.38(0.09)^{+0.31}_{-0.35}$~\\ \hline
$1$--$1.4$ &~$47.4(0.5)^{+14.1}_{-10.1}$ &~$2.14(0.03)^{+0.83}_{-0.55}$~&~$53(1)^{+4}_{-5}$~&~$2.6(0.1)^{+0.3}_{-0.3}$~\\ \hline
$1.4$--$2.4$ &~$23.6(0.3)^{+7.2}_{-5.2}$ &~$1.41(0.02)^{+0.56}_{-0.37}$~&~$25(1)^{+1}_{-2}$~&~$1.36(0.07)^{+0.03}_{-0.12}$~\\ \hline

\hline
\end{tabular}
\caption{The distribution of the absolute value of the photon-pair rapidity,
in fb/unit rapidity, in \YYjj-jet production, as shown in
\fig{DiphotonY-ATLASFigure}.}
\label{DiphotonY-ATLASTable}
\end{table}

\begin{table}[ht]
\vskip .4 cm
\centering
\begin{tabular}{||c|l|l||l|l||}
\hline
\multirow{2}{*}{$|\cos\theta^*|$} & \multicolumn{2}{|c||}{LO}
& \multicolumn{2}{|c||}{NLO} \\ \cline{2-5}
& \hfil ATLAS & \hfil VBF & \hfil ATLAS & \hfil VBF\\ \hline\hline
$0$--$0.1$ &~$111(1)^{+33}_{-24}$ &~$5.5(0.1)^{+2.2}_{-1.4}$~&~$116(11)^{+4}_{-7}$~&~$5.6(0.4)^{+0.2}_{-0.6}$~\\ \hline
$0.1$--$0.2$ &~$113(2)^{+34}_{-24}$ &~$5.2(0.1)^{+2.0}_{-1.4}$~&~$117(3)^{+5}_{-8}$~&~$5.6(0.4)^{+0.4}_{-0.6}$~\\ \hline
$0.2$--$0.3$ &~$106(1)^{+32}_{-23}$ &~$4.7(0.1)^{+1.9}_{-1.2}$~&~$116(3)^{+6}_{-9}$~&~$5.0(0.3)^{+0.5}_{-0.5}$~\\ \hline
$0.3$--$0.4$ &~$103(1)^{+31}_{-22}$ &~$4.55(0.09)^{+1.78}_{-1.18}$~&~$120(8)^{+10}_{-12}$~&~$5.3(0.4)^{+0.3}_{-0.6}$~\\ \hline
$0.4$--$0.5$ &~$102(2)^{+30}_{-22}$ &~$4.2(0.1)^{+1.7}_{-1.1}$~&~$111(3)^{+8}_{-9}$~&~$5.0(0.3)^{+0.5}_{-0.6}$~\\ \hline
$0.5$--$0.6$ &~$100(1)^{+30}_{-21}$ &~$3.92(0.08)^{+1.53}_{-1.01}$~&~$110(3)^{+7}_{-9}$~&~$4.8(0.3)^{+0.6}_{-0.6}$~\\ \hline
$0.6$--$0.7$ &~$97(1)^{+28}_{-20}$ &~$3.61(0.08)^{+1.40}_{-0.93}$~&~$106(5)^{+8}_{-10}$~&~$4.8(0.3)^{+0.7}_{-0.7}$~\\ \hline
$0.7$--$0.8$ &~$84(2)^{+25}_{-18}$ &~$3.4(0.1)^{+1.3}_{-0.9}$~&~$97(3)^{+9}_{-9}$~&~$4.3(0.2)^{+0.7}_{-0.7}$~\\ \hline
$0.8$--$0.9$ &~$40.5(0.9)^{+12.4}_{-8.8}$ &~$2.01(0.05)^{+0.77}_{-0.51}$~&~$59(3)^{+7}_{-7}$~&~$2.5(0.2)^{+0.5}_{-0.4}$~\\ \hline
$0.9$--$1$ &~$35.8(0.7)^{+10.9}_{-7.7}$ &~$1.91(0.04)^{+0.73}_{-0.49}$~&~$53(1)^{+8}_{-7}$~&~$3.1(0.2)^{+0.7}_{-0.5}$~\\ \hline

\hline
\end{tabular}
\caption{The distribution of the $|\cos\theta^*|$ variable defined in
\eqn{CosTheta}, in fb, in \YYjj-jet production, as shown in
\fig{CosTheta-ATLASFigure}.}
\label{CosTheta-ATLASTable}
\end{table}

\begin{table}[ht]
\vskip .4 cm
\centering
\begin{tabular}{||c|l|l||l|l||}
\hline
\multirow{2}{*}{$\Delta\phi_{jj}$} & \multicolumn{2}{|c||}{LO}
& \multicolumn{2}{|c||}{NLO} \\ \cline{2-5}
& \hfil ATLAS & \hfil VBF & \hfil ATLAS & \hfil VBF\\ \hline\hline
$0 $--$ \frac{\pi }{3}$ &~$10.2(0.2)^{+3.0}_{-2.2}$ &~$0.22(0.01)^{+0.09}_{-0.06}$~&~$15.6(0.4)^{+2.6}_{-2.2}$~&~$0.44(0.03)^{+0.13}_{-0.09}$~\\ \hline
$\frac{\pi }{3} $--$ \frac{2 \pi }{3}$ &~$17.1(0.2)^{+5.1}_{-3.7}$ &~$0.61(0.01)^{+0.24}_{-0.16}$~&~$21.2(0.5)^{+2.3}_{-2.2}$~&~$0.93(0.04)^{+0.20}_{-0.16}$~\\ \hline
$\frac{2 \pi }{3} $--$ \frac{5 \pi }{6}$ &~$41.7(0.5)^{+12.5}_{-8.9}$ &~$1.90(0.03)^{+0.74}_{-0.49}$~&~$45(3)^{+3}_{-4}$~&~$2.3(0.1)^{+0.2}_{-0.3}$~\\ \hline
$\frac{5 \pi }{6} $--$ \pi$ &~$74.1(0.5)^{+22.1}_{-15.8}$ &~$3.90(0.04)^{+1.51}_{-1.00}$~&~$73(3)^{+1}_{-5}$~&~$3.7(0.1)^{+0.1}_{-0.3}$~\\ \hline

\hline
\end{tabular}
\caption{The distribution of the azimuthal angle difference between the
two leading jets, in fb/radian, in \YYjj-jet production, as shown in
\fig{DijetAngle-ATLASFigure}.}
\label{DijetAzimuthalAngle-ATLASTable}
\end{table}

In this appendix, we provide tables for distributions displayed and discussed
in \sect{ResultsSection}. All tables show differential cross sections at both
LO and NLO, with numerical integration uncertainties given in parentheses,
and scale-dependence bands indicated by super- and subscripts. We display
results both with the cuts of \eqn{ATLASCuts}, shown in columns marked with
`ATLAS', and these cuts supplemented by the VBF cuts of \eqn{BasicVBFCuts},
shown in columns marked with 'VBF'.
In \Tab{Jet1pT-ATLASTable}, we display the leading-jet transverse-momentum;
in \Tab{DiphotonPT-ATLASTable}, the photon-pair transverse-momentum
distribution (in fb/GeV); 
in \Tab{DiphotonY-ATLASTable}, the distribution of the absolute value of
the photon-pair rapidity (in fb/unit rapidity);
in \Tab{CosTheta-ATLASTable}, the distribution in $|\cos\theta^*|$ as
defined in \eqn{CosTheta} (in fb);
in \Tab{DijetAzimuthalAngle-ATLASTable}, the distribution of the
dijet azimuthal-angle difference (in fb/radian) between the two leading jets.

\clearpage

\section{Tables of Distributions for CMS and VBF Cuts}
\label{CMSTablesAppendix}

\begin{table}[ht]
\vskip .4 cm
\centering
\begin{tabular}{||c|l|l||l|l||}
\hline
\multirow{2}{*}{$m_{\gamma\gamma}$} & \multicolumn{2}{|c||}{LO}
& \multicolumn{2}{|c||}{NLO} \\ \cline{2-5}
& \hfil CMS & \hfil VBF & \hfil CMS & \hfil VBF\\ \hline\hline
$100$--$110$ &~$13.02(0.05)^{+3.76}_{-2.72}$ &~$0.230(0.002)^{+0.091}_{-0.060}$~&~$13.1(0.1)^{+0.4}_{-0.9}$~&~$0.282(0.008)^{+0.040}_{-0.039}$~\\ \hline
$110$--$120$ &~$10.73(0.05)^{+3.13}_{-2.26}$ &~$0.201(0.002)^{+0.080}_{-0.053}$~&~$10.9(0.1)^{+0.3}_{-0.8}$~&~$0.251(0.007)^{+0.037}_{-0.036}$~\\ \hline
$120$--$130$ &~$8.82(0.04)^{+2.60}_{-1.87}$ &~$0.174(0.002)^{+0.069}_{-0.045}$~&~$9.3(0.1)^{+0.4}_{-0.6}$~&~$0.218(0.006)^{+0.028}_{-0.030}$~\\ \hline
$130$--$140$ &~$7.11(0.03)^{+2.12}_{-1.52}$ &~$0.150(0.002)^{+0.059}_{-0.039}$~&~$7.53(0.08)^{+0.29}_{-0.56}$~&~$0.188(0.005)^{+0.025}_{-0.025}$~\\ \hline
$140$--$150$ &~$5.83(0.03)^{+1.76}_{-1.26}$ &~$0.129(0.001)^{+0.051}_{-0.034}$~&~$6.23(0.07)^{+0.31}_{-0.48}$~&~$0.172(0.004)^{+0.028}_{-0.026}$~\\ \hline
$150$--$160$ &~$4.79(0.03)^{+1.46}_{-1.04}$ &~$0.115(0.001)^{+0.045}_{-0.030}$~&~$4.90(0.07)^{+0.17}_{-0.36}$~&~$0.140(0.009)^{+0.021}_{-0.019}$~\\ \hline
$160$--$170$ &~$3.88(0.02)^{+1.19}_{-0.85}$ &~$0.098(0.001)^{+0.039}_{-0.026}$~&~$4.18(0.05)^{+0.17}_{-0.32}$~&~$0.121(0.006)^{+0.018}_{-0.016}$~\\ \hline
$170$--$180$ &~$3.20(0.02)^{+0.99}_{-0.70}$ &~$0.086(0.001)^{+0.034}_{-0.022}$~&~$3.43(0.05)^{+0.13}_{-0.27}$~&~$0.100(0.004)^{+0.008}_{-0.011}$~\\ \hline

\hline
\end{tabular}
\caption{The distribution of the invariant mass of the photon pair in fb/GeV
in \YYjj-jet production, as shown in \fig{DiphotonMass-CMSFigure}.}
\label{DiphotonMass-CMSTable}
\end{table}

\begin{table}[ht]
\vskip .4 cm
\centering
\begin{tabular}{||c|l|l||l|l||}
\hline
\multirow{2}{*}{$|y|$} & \multicolumn{2}{|c||}{LO} & \multicolumn{2}{|c||}{NLO}
 \\ \cline{2-5}
& \hfil CMS & \hfil VBF & \hfil CMS & \hfil VBF\\ \hline\hline
$0$--$0.3$ &~$320(1)^{+94}_{-68}$ &~$6.40(0.05)^{+2.54}_{-1.68}$~&~$345(2)^{+14}_{-25}$~&~$8.8(0.2)^{+1.5}_{-1.3}$~\\ \hline
$0.3$--$0.65$ &~$307(1)^{+90}_{-65}$ &~$6.42(0.04)^{+2.53}_{-1.67}$~&~$331(3)^{+14}_{-25}$~&~$8.9(0.3)^{+1.6}_{-1.4}$~\\ \hline
$0.65$--$1$ &~$297(1)^{+87}_{-63}$ &~$6.59(0.05)^{+2.58}_{-1.71}$~&~$315(2)^{+13}_{-23}$~&~$8.8(0.2)^{+1.5}_{-1.3}$~\\ \hline
$1$--$1.4$ &~$276(1)^{+81}_{-58}$ &~$6.09(0.05)^{+2.39}_{-1.58}$~&~$285(3)^{+9}_{-20}$~&~$7.4(0.2)^{+1.0}_{-1.0}$~\\ \hline
$1.4$--$2.4$ &~$155.0(0.7)^{+46.7}_{-33.5}$ &~$2.91(0.03)^{+1.16}_{-0.77}$~&~$151(2)^{+4}_{-11}$~&~$2.91(0.09)^{+0.14}_{-0.28}$~\\ \hline

\hline
\end{tabular}
\caption{The distribution of the absolute value of the photon-pair rapidity,
in fb/unit rapidity, in \YYjj-jet production, as shown in
\fig{DiphotonY-CMSFigure}.}
\label{DiphotonY-CMSTable}
\end{table}

\begin{table}[ht]
\vskip .4 cm
\centering
\begin{tabular}{||c|l|l||l|l||}
\hline
\multirow{2}{*}{$\Delta\phi_{jj}$} & \multicolumn{2}{|c||}{LO}
& \multicolumn{2}{|c||}{NLO} \\ \cline{2-5}
& \hfil CMS & \hfil VBF & \hfil CMS & \hfil VBF\\ \hline\hline
$0 $--$ \frac{\pi }{3}$ &~$101.0(0.5)^{+29.7}_{-21.4}$ &~$1.18(0.02)^{+0.48}_{-0.31}$~&~$134(1)^{+17}_{-16}$~&~$2.54(0.06)^{+0.79}_{-0.55}$~\\ \hline
$\frac{\pi }{3} $--$ \frac{2 \pi }{3}$ &~$147.4(0.5)^{+43.5}_{-31.3}$ &~$2.59(0.02)^{+1.04}_{-0.68}$~&~$170(2)^{+13}_{-16}$~&~$4.39(0.09)^{+1.06}_{-0.81}$~\\ \hline
$\frac{2 \pi }{3} $--$ \frac{5 \pi }{6}$ &~$264.2(0.9)^{+78.1}_{-56.2}$ &~$6.30(0.04)^{+2.48}_{-1.64}$~&~$251(4)^{+2}_{-13}$~&~$6.9(0.2)^{+0.6}_{-0.8}$~\\ \hline
$\frac{5 \pi }{6} $--$ \pi$ &~$335.1(0.9)^{+100.2}_{-71.8}$ &~$8.77(0.05)^{+3.43}_{-2.27}$~&~$277(3)^{+0}_{-18}$~&~$7.4(0.2)^{+0.0}_{-0.4}$~\\ \hline

\hline
\end{tabular}
\caption{The distribution of the azimuthal angle difference between the
two leading jets, in fb/radian, in \YYjj-jet production, as shown in
\fig{DijetAngle-CMSFigure}.}
\label{DijetAzimuthalAngle-CMSTable}
\end{table}
\FloatBarrier

In this appendix, we provide tables for distributions displayed and
discussed in \sect{ResultsSection}. All tables show differential cross
sections at both LO and NLO, with numerical integration uncertainties
given in parentheses, and scale-dependence bands indicated by super- and
subscripts.  We display results both with the cuts of \eqn{CMSCuts},
shown in columns marked with `CMS', and these cuts supplemented by the VBF
cuts of \eqn{CMSVBFCuts}, shown in columns marked with `VBF'.
In \Tab{DiphotonMass-CMSTable}, we display the distribution of the invariant
mass of the photon pair, in fb/GeV;
in \Tab{DiphotonY-CMSTable}, the distribution of the absolute value of the
photon-pair rapidity (in fb/unit rapidity);
in \Tab{DijetAzimuthalAngle-CMSTable}, the distribution of the
dijet azimuthal-angle difference (in fb/radian) between the two leading jets.


\section{Virtual Matrix Elements at a Point in Phase Space.}
\label{VirtualMatrixElements}

In this appendix, we provide reference values of virtual matrix elements.
We provide values for the independent matrix elements in \YYjj-jet production
at the same point in phase-space as given in eq.~(9.1) of ref.~\cite{GenHel}
with the scale parameter $\mu=M_Z=91.188 $ GeV for both renormalization and
factorization scales, with $\alpha_s(M_Z)=0.120$ and $\alpha_{EM}(0)=1/137$
to the required precision.  We show these values in
\tab{VirtualMatrixElementsTable}.  All other matrix elements are obtained
from these by crossing, and by adjusting the electromagnetic charges of the
quarks appropriately.

For all matrix elements with non-vanishing tree-level values, we quote the
values for the ratio of the virtual corrections to the tree-level squared
matrix element, following ref.~\cite{W3jDistributions}. We quote the value
of the ratio,
\begin{equation} \hatMNLO \equiv \frac{1}{8\pi\alpha_S \, c_\Gamma(\e)}
    \frac{\MNLO}{\MLO}\,, \label{ME2normalization} \end{equation}
where we have also separated out the dependence on the strong coupling
$\alpha_s$ and the overall factor $c_\Gamma(\e)$, defined by
\begin{equation} 
c_\Gamma(\e) = \frac{1}{(4\pi)^{2-\epsilon}}
    \frac{\Gamma(1+\epsilon)\Gamma^2(1-\epsilon)} {\Gamma(1-2\epsilon)}\,.
    \label{cGammaDef} 
\end{equation} 
In the second column of \tab{VirtualMatrixElementsTable} we give the value of
tree-level matrix element squared for the indicated subprocess.  

For the $(1_g2_g\rightarrow 3_\y4_\y5_g6_g)$ subprocess we give
the finite part of the one-loop squared matrix elements $\MOLS$ dressed
with couplings and factors of $c_\Gamma(\e)$ directly, because the
associated tree-level amplitudes vanish, and poles in $1/\epsilon$ are absent.

\begin{table*}
\vskip .4 cm
\begin{tabular}{||c||c||c|c|c||}
\hline
$\hatMNLO{}$ & Tree-level &  $1/\eps^2$ &  $1/\eps$ & finite  \\
\hline
\hline
$1_g 2_g \rightarrow 3_\y4_\y5_d6_{\bar d} $ &  
    $\;  3.722387496\cdot 10^{-5} $ &
    $\; -8.666666667 \;$ &  
    $\;-30.997687242 \;$ & 
    $\;-29.978172584 \;$ \\
\hline
$1_d 2_{\bar d} \rightarrow 3_\y 4_\y 5_u6_{\bar u} $ & 
    $\;  1.726257408 \cdot 10^{-7} $ & 
    $\; -5.333333333 \;$ &  
    $\;-15.845128704 \;$ &
    $\;-12.304984940 \;$ \\\hline
$1_d 2_{\bar d} \rightarrow 3_\y 4_\y 5_d6_{\bar d} $ & 
    $\;  2.344204568 \cdot 10^{-5} $ & 
    $\; -5.333333333 \;$ &  
    $\;-15.947959115 \;$ &
    $\;  7.0934319706 \;$ \\
\hline
\hline
$\MOLS$ & --- & ---  & --- & finite  \\
\hline
$1_g2_g \rightarrow 3_\y4_\y5_g6_g$ &
    --- & 
    --- &  
    --- & 
    $\; 9.0522165549 \cdot 10^{-4} \; $
\\ \hline 
\end{tabular} \caption{The virtual matrix elements at the point in phase space
    used in ref.~\cite{GenHel}. The first column labels the subprocess and the
    second gives the tree-level matrix element. The third, fourth and fifth
    columns give, respectively, the $1/\eps^2$, $1/\eps$ and finite
    contributions to the normalized virtual matrix element
    $\widehat{d\sigma}_V^{(1)}$, following the conventions in
    ref.~\cite{W3jDistributions}.  For the subprocess 
    $(1_g2_g\rightarrow 3_\y4_\y5_g6_g)$ the finite part of the one-loop
    squared amplitudes $\MOLS$ are given directly.  Our conventions, as well
    as the values of the scale parameters and couplings, are given in the
    main text of this appendix.
    \label{VirtualMatrixElementsTable} } 
\end{table*}

\end{document}